%% file: main.tex
\documentclass[manuscript]{aastex}
\usepackage{lineno}
\usepackage{graphicx, subfigure}

\graphicspath{{./figures/}}

\usepackage{amssymb}
\usepackage{amsmath}
\usepackage{epstopdf}
\usepackage{hyperref}
\usepackage{color}
\usepackage{placeins}
\usepackage[hmargin=1in]{geometry}
\usepackage{natbib}
\usepackage{enumerate}
\usepackage{textcomp}
\usepackage{footmisc}
\usepackage{xspace}
\usepackage{longtable}
\usepackage{pdflscape}
\usepackage{booktabs}
\usepackage{standalone}  
\usepackage{makecell}      

\newcommand{\Fermi}{\emph{Fermi}\xspace}
\newcommand{\fermi}{\emph{Fermi}\xspace}

\def\p0{$\pi^{\rm 0}$}
\def\Angst{$\buildrel _{\circ} \over {\mathrm{A}}$}
\def\de{$^{\circ}$\xspace}

\def\g{\gamma}

\def\like{\mathcal{L}}

\def\p8{\texttt{Pass8}}
\def\p7{\texttt{Pass7REP}}

\def\sc{{\rm sc}}

\newcommand{\SunMon}{\texttt{SunMonitor}\xspace}

\newcommand{\solar}{{solar}\xspace}

\shorttitle{Solar Catalog}

\begin{document}
\title{First \Fermi-LAT Solar flare catalog}
\author{
M.~Ajello\altaffilmark{1}, 
L.~Baldini\altaffilmark{2}, 
D.~Bastieri\altaffilmark{3,4}, 
R.~Bellazzini\altaffilmark{5}, 
A.~Berretta\altaffilmark{6}, 
E.~Bissaldi\altaffilmark{7,8}, 
R.~D.~Blandford\altaffilmark{9}, 
R.~Bonino\altaffilmark{10,11}, 
P.~Bruel\altaffilmark{12}, 
S.~Buson\altaffilmark{56}
R.~A.~Cameron\altaffilmark{9}, 
R.~Caputo\altaffilmark{13}, 
E.~Cavazzuti\altaffilmark{14}, 
C.~C.~Cheung\altaffilmark{15}, 
G.~Chiaro\altaffilmark{16}, 
D.~Costantin\altaffilmark{17}, 
S.~Cutini\altaffilmark{18}, 
F.~D'Ammando\altaffilmark{19}, 
F.~de~Palma\altaffilmark{10}, 
R.~Desiante\altaffilmark{10},
N.~Di~Lalla\altaffilmark{9}, 
L.~Di~Venere\altaffilmark{7,8}, 
F.~Fana~Dirirsa\altaffilmark{20}, 
S.~J.~Fegan\altaffilmark{12}, 
Y.~Fukazawa\altaffilmark{21}, 
S.~Funk\altaffilmark{22}, 
P.~Fusco\altaffilmark{7,8}, 
F.~Gargano\altaffilmark{8}, 
D.~Gasparrini\altaffilmark{23,24}, 
F.~Giordano\altaffilmark{7,8}, 
M.~Giroletti\altaffilmark{19}, 
D.~Green\altaffilmark{25}, 
S.~Guiriec\altaffilmark{26,13}, 
E.~Hays\altaffilmark{13}, 
J.W.~Hewitt\altaffilmark{27}, 
D.~Horan\altaffilmark{12}, 
G.~J\'ohannesson\altaffilmark{28,29}, 
M.~Kovac'evic'\altaffilmark{18}, 
M.~Kuss\altaffilmark{5}, 
S.~Larsson\altaffilmark{30,31,32}, 
L.~Latronico\altaffilmark{10}, 
J.~Li\altaffilmark{33}, 
F.~Longo\altaffilmark{34,35,36}, 
M.~N.~Lovellette\altaffilmark{15}, 
P.~Lubrano\altaffilmark{18}, 
S.~Maldera\altaffilmark{10}, 
A.~Manfreda\altaffilmark{2}, 
G.~Mart\'i-Devesa\altaffilmark{37}, 
M.~N.~Mazziotta\altaffilmark{8}, 
I.Mereu\altaffilmark{6,18}, 
P.~F.~Michelson\altaffilmark{9}, 
T.~Mizuno\altaffilmark{38}, 
M.~E.~Monzani\altaffilmark{9}, 
A.~Morselli\altaffilmark{23}, 
I.~V.~Moskalenko\altaffilmark{9}, 
M.~Negro\altaffilmark{39,40}, 
N.~Omodei\altaffilmark{9,41}, 
M.~Orienti\altaffilmark{19}, 
E.~Orlando\altaffilmark{42,9}, 
D.~Paneque\altaffilmark{25}, 
Z.~Pei\altaffilmark{4}, 
M.~Persic\altaffilmark{34,43}, 
M.~Pesce-Rollins\altaffilmark{5,44}, 
V.~Petrosian\altaffilmark{9,45}, 
F.~Piron\altaffilmark{46}, 
T.~A.~Porter\altaffilmark{9}, 
G.~Principe\altaffilmark{19}, 
J.~L.~Racusin\altaffilmark{13}, 
S.~Rain\`o\altaffilmark{7,8}, 
R.~Rando\altaffilmark{47,3,48}, 
B.~Rani\altaffilmark{49,13}, 
M.~Razzano\altaffilmark{5,50}, 
S.~Razzaque\altaffilmark{20}, 
A.~Reimer\altaffilmark{37,9}, 
O.~Reimer\altaffilmark{37}, 
D.~Serini\altaffilmark{7}, 
C.~Sgr\`o\altaffilmark{5}, 
E.~J.~Siskind\altaffilmark{51}, 
G.~Spandre\altaffilmark{5}, 
P.~Spinelli\altaffilmark{7,8}, 
D.~Tak\altaffilmark{52,13}, 
E.~Troja\altaffilmark{13,53}, 
J.~Valverde\altaffilmark{12}, 
K.~Wood\altaffilmark{54}, 
G.~Zaharijas\altaffilmark{42,55}
}
\altaffiltext{1}{Department of Physics and Astronomy, Clemson University, Kinard Lab of Physics, Clemson, SC 29634-0978, USA}
\altaffiltext{2}{Universit\`a di Pisa and Istituto Nazionale di Fisica Nucleare, Sezione di Pisa I-56127 Pisa, Italy}
\altaffiltext{3}{Istituto Nazionale di Fisica Nucleare, Sezione di Padova, I-35131 Padova, Italy}
\altaffiltext{4}{Dipartimento di Fisica e Astronomia ``G. Galilei'', Universit\`a di Padova, I-35131 Padova, Italy}
\altaffiltext{5}{Istituto Nazionale di Fisica Nucleare, Sezione di Pisa, I-56127 Pisa, Italy}
\altaffiltext{6}{Dipartimento di Fisica, Universit\`a degli Studi di Perugia, I-06123 Perugia, Italy}
\altaffiltext{7}{Dipartimento di Fisica "M. Merlin" dell'Universit\`a e del Politecnico di Bari, via Amendola 173, I-70126 Bari, Italy}
\altaffiltext{8}{Istituto Nazionale di Fisica Nucleare, Sezione di Bari, I-70126 Bari, Italy}
\altaffiltext{9}{W. W. Hansen Experimental Physics Laboratory, Kavli Institute for Particle Astrophysics and Cosmology, Department of Physics and SLAC National Accelerator Laboratory, Stanford University, Stanford, CA 94305, USA}
\altaffiltext{10}{Istituto Nazionale di Fisica Nucleare, Sezione di Torino, I-10125 Torino, Italy}
\altaffiltext{11}{Dipartimento di Fisica, Universit\`a degli Studi di Torino, I-10125 Torino, Italy}
\altaffiltext{12}{Laboratoire Leprince-Ringuet, \'Ecole polytechnique, CNRS/IN2P3, F-91128 Palaiseau, France}
\altaffiltext{13}{NASA Goddard Space Flight Center, Greenbelt, MD 20771, USA}
\altaffiltext{14}{Italian Space Agency, Via del Politecnico snc, 00133 Roma, Italy}
\altaffiltext{15}{Space Science Division, Naval Research Laboratory, Washington, DC 20375-5352, USA}
\altaffiltext{16}{INAF-Istituto di Astrofisica Spaziale e Fisica Cosmica Milano, via E. Bassini 15, I-20133 Milano, Italy}
\altaffiltext{17}{University of Padua, Department of Statistical Science, Via 8 Febbraio, 2, 35122 Padova}
\altaffiltext{18}{Istituto Nazionale di Fisica Nucleare, Sezione di Perugia, I-06123 Perugia, Italy}
\altaffiltext{19}{INAF Istituto di Radioastronomia, I-40129 Bologna, Italy}
\altaffiltext{20}{Department of Physics, University of Johannesburg, PO Box 524, Auckland Park 2006, South Africa}
\altaffiltext{21}{Department of Physical Sciences, Hiroshima University, Higashi-Hiroshima, Hiroshima 739-8526, Japan}
\altaffiltext{22}{Friedrich-Alexander Universit\"at Erlangen-N\"urnberg, Erlangen Centre for Astroparticle Physics, Erwin-Rommel-Str. 1, 91058 Erlangen, Germany}
\altaffiltext{23}{Istituto Nazionale di Fisica Nucleare, Sezione di Roma ``Tor Vergata", I-00133 Roma, Italy}
\altaffiltext{24}{Space Science Data Center - Agenzia Spaziale Italiana, Via del Politecnico, snc, I-00133, Roma, Italy}
\altaffiltext{25}{Max-Planck-Institut f\"ur Physik, D-80805 M\"unchen, Germany}
\altaffiltext{26}{The George Washington University, Department of Physics, 725 21st St, NW, Washington, DC 20052, USA}
\altaffiltext{27}{University of North Florida, Department of Physics, 1 UNF Drive, Jacksonville, FL 32224 , USA}
\altaffiltext{28}{Science Institute, University of Iceland, IS-107 Reykjavik, Iceland}
\altaffiltext{29}{Nordita, Royal Institute of Technology and Stockholm University, Roslagstullsbacken 23, SE-106 91 Stockholm, Sweden}
\altaffiltext{30}{Department of Physics, KTH Royal Institute of Technology, AlbaNova, SE-106 91 Stockholm, Sweden}
\altaffiltext{31}{The Oskar Klein Centre for Cosmoparticle Physics, AlbaNova, SE-106 91 Stockholm, Sweden}
\altaffiltext{32}{School of Education, Health and Social Studies, Natural Science, Dalarna University, SE-791 88 Falun, Sweden}
\altaffiltext{33}{Deutsches Elektronen Synchrotron DESY, D-15738 Zeuthen, Germany}
\altaffiltext{34}{Istituto Nazionale di Fisica Nucleare, Sezione di Trieste, I-34127 Trieste, Italy}
\altaffiltext{35}{Dipartimento di Fisica, Universit\`a di Trieste, I-34127 Trieste, Italy}
\altaffiltext{36}{email: francesco.longo@trieste.infn.it}
\altaffiltext{37}{Institut f\"ur Astro- und Teilchenphysik, Leopold-Franzens-Universit\"at Innsbruck, A-6020 Innsbruck, Austria}
\altaffiltext{38}{Hiroshima Astrophysical Science Center, Hiroshima University, Higashi-Hiroshima, Hiroshima 739-8526, Japan}
\altaffiltext{39}{Center for Research and Exploration in Space Science and Technology (CRESST) and NASA Goddard Space Flight Center, Greenbelt, MD 20771, USA}
\altaffiltext{40}{Department of Physics and Center for Space Sciences and Technology, University of Maryland Baltimore County, Baltimore, MD 21250, USA}
\altaffiltext{41}{email: nicola.omodei@stanford.edu}
\altaffiltext{42}{Istituto Nazionale di Fisica Nucleare, Sezione di Trieste, and Universit\`a di Trieste, I-34127 Trieste, Italy}
\altaffiltext{43}{Osservatorio Astronomico di Trieste, Istituto Nazionale di Astrofisica, I-34143 Trieste, Italy}
\altaffiltext{44}{email: melissa.pesce.rollins@pi.infn.it}
\altaffiltext{45}{email: vahep@stanford.edu}
\altaffiltext{46}{Laboratoire Univers et Particules de Montpellier, Universit\'e Montpellier, CNRS/IN2P3, F-34095 Montpellier, France}
\altaffiltext{47}{Department of Physics and Astronomy, University of Padova, Vicolo Osservatorio 3, I-35122 Padova, Italy}
\altaffiltext{48}{Center for Space Studies and Activities "G. Colombo", University of Padova, Via Venezia 15, I-35131 Padova, Italy}
\altaffiltext{49}{Korea Astronomy and Space Science Institute, 776 Daedeokdae-ro, Yuseong-gu, Daejeon 30455, Korea}
\altaffiltext{50}{Funded by contract FIRB-2012-RBFR12PM1F from the Italian Ministry of Education, University and Research (MIUR)}
\altaffiltext{51}{NYCB Real-Time Computing Inc., Lattingtown, NY 11560-1025, USA}
\altaffiltext{52}{Department of Physics, University of Maryland, College Park, MD 20742, USA}
\altaffiltext{53}{Department of Astronomy, University of Maryland, College Park, MD 20742, USA}
\altaffiltext{54}{Praxis Inc., Alexandria, VA 22303, resident at Naval Research Laboratory, Washington, DC 20375, USA}
\altaffiltext{55}{Center for Astrophysics and Cosmology, University of Nova Gorica, Nova Gorica, Slovenia}
\altaffiltext{56}{Institut f\"ur Theoretische Physik und Astrophysik, Universit\"at W\"urzburg, Campus Hubland Nord Emil-Fischer-Str. 31
97074 W\"urzburg, Germany}

\begin{abstract}
We present the first \Fermi-Large Area Telescope (LAT) \solar flare catalog covering the 24$^{th}$ \solar cycle. This catalog contains 45 \Fermi-LAT \solar flares ({FLSFs}) with emission  in the $\gamma$-ray energy band $(30 $ MeV - $10 ~$GeV$)$ 
detected with a significance $\geq$5$\sigma$  over the years 2010-2018. {A subsample containing  37 of these flares exhibit delayed emission beyond the prompt-impulsive hard X-ray phase with 21 flares showing delayed emission lasting more than two hours. No prompt-impulsive emission is detected in four of these flares.} We also present in this catalog the  observations of GeV emission from 3 flares originating from Active Regions located behind the {limb (BTL) of the} visible \solar disk. We report the light curves, spectra, best proton index and localization (when possible) for all the {FLSFs}. The $\gamma$-ray spectra is consistent with the decay of pions produced by $>$300 MeV protons. 
This work contains the largest sample of high-energy $\gamma$-ray flares ever reported and provides the unique opportunity to perform population studies on the different phases of the flare and thus allowing to open a new window in \solar physics.  

\end{abstract}

\section{Introduction}
\label{sec:intro}
It is generally accepted that the magnetic energy released through reconnection during \solar flares is capable of accelerating electrons and ions to relativistic energies on time scales as short as a few seconds.  Much is known of the electron acceleration during these explosive phenomena thanks to the observations made in Hard X-rays (10 keV - 1 MeV; HXRs; \citep[see, e.g.~][]{Vilmer1987,1988SoPh..118...49D,LIN20031001} and microwaves \citep[see, e.g.~][]{1998A&A...334.1099T}. {The}  observed impulsive phase radiation in \solar flares is dominated by electron emission, however a fair fraction of stronger flares, with longer impulsive phase, show even higher-energy emission at $\gamma$-ray energies ($E > 3$ MeV) by accelerated protons and other ions in the form of  
nuclear de-excitation lines, by $\sim$3 - 50 MeV ions, and $>$100 MeV continuum due to decay of pions produced by $>$ 300 MeV ions
\citep[see, e.g.~][]{2011SSRv..159..167V}. The first reported observation of $\gamma$-rays with energies above 10 MeV was made in 1981 with the Solar Maximum Mission (SMM) spectrometer~\citep{chup82} and throughout the 1980's several other observations were made \citep[see, e.g.~][]{forr85,forr86} providing evidence of pion-decay emission and revealing multiple phases in the flares.

{The first detection of GeV $\gamma$-rays were made by the {\it Energetic Gamma-Ray Experiment Telescope} (EGRET) on board the {\it Compton Gamma-Ray Observatory} (CGRO) \citep[see, e.g.~][]{kanb93,vilm03}.} 
{The majority of the flares observed from 50 MeV to 2 GeV by EGRET had durations lasting tens of minutes but up to several hours {in two flares} leading to a new class of flares initially known as \emph{Long Duration Gamma-Ray Flares} \citep{ryan00,chup09}. This new class of flares presented a challenge to the \emph{classical} magnetic reconnection theory for particle acceleration during flares because the $\gamma$-ray emission persisted beyond any other flare emissions, therefore suggesting the need for an additional} {mechanism and site for acceleration of protons and other ions. However, with only two such detections the search for an additional acceleration mechanism and site was very challenging.}

{Additional cases suggesting the need for a new source of ion acceleration came with the observations of} $\gamma$-ray emission, up to only 100~MeV, from three flares whose host Active Regions (ARs) were located behind the limb (BTL) of the visible \solar disk~\citep{1993ApJ...409L..69V,1999A&A...342..575V,1994ApJ...425L.109B}. 
It is generally believed that the lower energy $\gamma$-rays are produced at the dense footpoints of flare loops by ions accelerated at the reconnection regions near the top of these loops. 
 Thus, observations of BTL flares pose interesting questions regarding the acceleration site and mechanism, of the ions and about their transport  to the high density photospheric regions on the visible disk.
 Although there were 
 some scenarios put forth~\citep{cliv93}, no convincing explanations were given for the acceleration and transport sites and mechanisms of particles responsible for these observations. 

Prior to the launch of the \fermi Gamma-ray Space Telescope in 2008, the understanding of these emission mechanisms was severely limited because {of the limited amount of high-energy $\gamma$-ray flares detected.}

The \fermi-Large Area Telescope \citep[LAT,][]{LATPaper} observations of the flaring Sun over the first 12 years in orbit have revealed an extremely rich and diverse sample of events, spanning from short prompt-impulsive flares~\citep{2012ApJ...745..144A} to the gradual-delayed long-duration phases~\citep{0004-637X-787-1-15} including the longest extended emission ever detected ($\sim$20 hours) from the SOL2012-03-07; a GOES X-class flare~\citep{0004-637X-789-1-20}%
\footnote{{Solar flares observed by the Geostationary Operational Environmental Satellite (GOES)} are classified, on the basis of their peak flux in the soft X-ray range of 0.5 to 10 keV,  as X, M, C and A class with peak fluxes greater than $10^{-4}, 10^{-5}, 10^{-6}$ and $10^{-7}$ 
Watt m$^{-2}$, respectively}. The LAT, thanks to its large field of view (FoV) of 2.4 sr, monitors the entire sky 
every two orbits as an excellent general purpose $\gamma$-ray astrophysics observatory but in doing so it keeps the Sun in the FoV ~40\% of the time.

Nonetheless, thanks to its technology improvements with respect to previous $\gamma$-ray space based missions, the \fermi-LAT has increased the total number of $>$30~MeV detected \solar flares  by almost a factor of 10. 
{More importantly the LAT with its higher spatial resolution than EGRET can localize the centroids of the $\gamma$-ray emissions on the photosphere, which is particulary important for the interpretation of the BTL flares.}

In this \Fermi-LAT Solar Flare (FLSF) catalog we present the observations of 45 flares with  $>$30~MeV emission in the period January 2010 - January 2018 (covering most of the 
24$^{\rm th}$ \solar cycle). From these observations we now know that $>$100 MeV $\gamma$-ray emission from even moderate GOES class flares is fairly common (roughly half of the {FLSFs} in our catalog are associated with M class flares) and that this high-energy emission is not correlated with the intensity of the X-ray flare, as one might expect. Our spectral analysis indicates that the $>$100~MeV emission is due to accelerated ions as opposed to HXR and microwave producing electrons. Based on the timing evolution of the $\gamma$-ray emission we find that there are two main populations of $\gamma$-ray flares: impulsive-prompt (prompt hereafter) and gradual-delayed (delayed hereafter). The prompt flares are those whose emission evolution is 
similar to that of the HXRs, indicating common acceleration sites and mechanism of electrons and ions.
The emission of delayed FLSF flares, which are always (with the exception of FLSF~2012-10-23 and FLSF~2012-11-27) associated with fast Coronal Mass Ejections (CMEs), rises at the end of the impulsive HXR phase  and, like Solar Energetic Particles (SEPs), extends well beyond the end of the HXR emission (for up to tens of hours). This and other observations suggest a different  acceleration site and mechanism.

In section \ref{sec:analysis} we describe the analysis methods and 
procedures used in this work, which includes the description of an automated pipeline (\ref{sec:sunmon}), the LAT Low Energy (LLE) analysis (\ref{sec:LLE}), spectral analysis (\ref{sec:spectral_analysis}), how we perform our localization of the $\gamma$-ray emission (\ref{sec:localization_alg}), and the search for spatial extension in the $\gamma$-ray {emission 
of} the brightest flares (\ref{sec:extension_alg}). Here we also describe the methods used to calculate the total emission, fluence and the total number of accelerated $>$500 MeV protons needed to produce the observed emission (\ref{sec:total_emission}).  In section \ref{sec:classification} we describe how the \solar flares are classified based on the  evolution of their $\gamma$-ray emission. In section \ref{sec:results} we present the results of the catalog. In section \ref{sec:discussion} we discuss the main findings of this work and the theoretical implications of our results. The tables and figures for each individual flare in this catalog are reported in~\cite{ajello_m_2020_4311157}.

\section{Analysis methods and procedures}
\label{sec:analysis}
The LAT is sensitive to $\gamma$-rays in the energy range {between $30$ MeV and $>$300 GeV}~\citep{Pass8}. 
The LAT {registers} energy, direction and time information for each {detected particle}. 
Each {such ``event"} is classified {by on-ground processing as a photon or other  particle based on the consistency of its interaction with those expected from energetic $\gamma$ rays}. 

 Event classes correspond to different levels of purity tolerance of the $\gamma$-ray sample appropriate for use in different types of analyses.
For each event {class there is a corresponding set of Instrument Response Functions (IRFs) describing the performance of the instrument}. The standard analysis and software are described at the \Fermi Science Support Center (FSSC) web site\footnote{\url{http://fermi.gsfc.nasa.gov/ssc/}} and, in great detail, in \citet{2012ApJS..203....4A}.

For the FLSF catalog we developed two analysis chains, the first one, that we call \emph{standard}, uses data with energies between 
 60\,MeV and 10\,GeV from two sets of event classes, \texttt{P8R3\_SOURCE} and the \solar flare Transient class \texttt{P8R3\_TRANSIENT015s} (S15)\footnote{{Events belonging to the \texttt{P8R\_TRANSIENT015s} class are} available in the extended photon data through the \Fermi Science Support Center}. The \texttt{P8R3\_SOURCE}\citep{bruel2018fermilat} class is the event class recommended for the standard \Fermi-LAT source analysis, while the S15 class was specifically developed to be insensitive to the potential pulse pile-up in the anti-coincidence detector (ACD) scintillators of the LAT resulting from the intense flux of X-rays during the prompt phase of \solar flares. {Pile-up of X-rays during the readout integration time of the ACD coincident with the entry of a $\gamma$-ray into the LAT can cause the otherwise good $\gamma$-ray to be misidentified as a charge particle} by the instrument flight software or event-classification ground software and thereby mistakenly vetoed. The \Fermi-LAT instrument team closely monitors this
effect and tags {time intervals with particularly high activity in the sunward ACD tiles as ``bad time intervals" (BTI) in} the public
data archive\footnote{\url{http://fermi.gsfc.nasa.gov/ssc/data/access/}}. 
The S15 event class is robust against these spurious vetoes because it is defined using selections that exclude variables associated with the Anti-Coincidence Detector and are therefore less susceptible to X-ray pile-up activity which can occur during the impulsive phase of \solar flares; thus all analysis in this catalog during a BTI used the S15 event class.

Additionally, a subset of results on short duration prompt \solar flares was obtained using the second chain based on LLE analysis methods. 
The LLE technique is an analysis method designed to study bright transient phenomena, such as Gamma Ray Bursts and \solar flares, in the 30 MeV--1 GeV energy range. The LAT collaboration developed this analysis using a different approach from that used in the standard photon analysis. The idea behind LLE is to maximize the effective area below $\sim$ 1 GeV by relaxing the standard analysis requirement on background rejection; see \cite{0004-637X-789-1-20} for a full description of the LLE method. The LAT collaboration has already used the LLE technique to analyze \solar flares, in particular FLSF 2010-06-12 \citep[the first flare detected by the LAT, see][]{2012ApJ...745..144A} and the prompt phase of the FLSF 2012-03-07 flares~\citep{0004-637X-789-1-20}.
In this FLSF catalog we used the LLE selection to study the short prompt phase of 14 \solar flares.

{These two approaches are complementary: the LLE method suffers from large background contamination and is effective only for short transients but, because it is much less restrictive than the \texttt{P8R3\_SOURCE} event class, the LLE class has much larger effective area and has significantly greater sensitivity at high incidence angles.}

Indeed, the FLSF 2010-06-12 was detected with the LLE approach when the Sun was more than 75$^{\circ}$ off-axis \citep{2012ApJ...745..144A}.

\subsection{The \Fermi-LAT \SunMon}
\label{sec:sunmon}
We have created an automated data analysis
pipeline, the \fermi-LAT \SunMon, to monitor the high-energy
$\gamma$-ray flux from the Sun throughout the {\fermi} mission. 
\footnote{Results from this pipeline are available online at 
\url{https://hesperia.gsfc.nasa.gov/fermi_solar/}.} 
The time intervals during which we run the analysis are when the Sun is
$<$ 70{\de} from the LAT boresight. 

\begin{figure}
\begin{center}
\includegraphics[width=\linewidth]{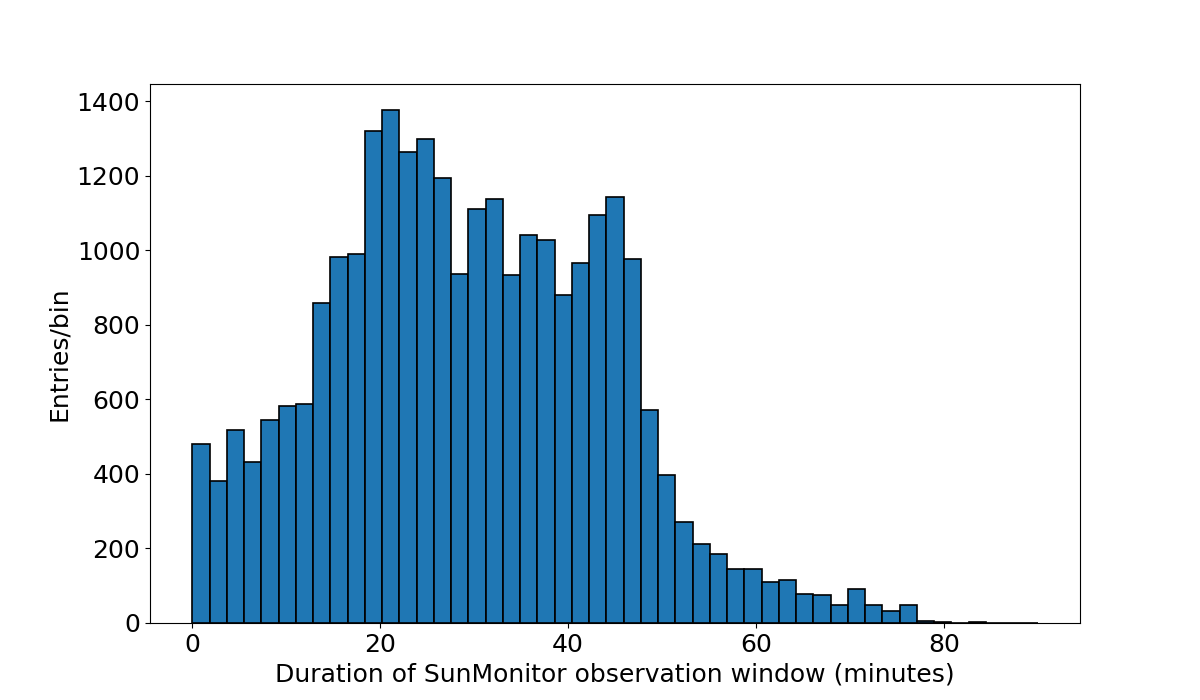}
\caption{Duration of the \Fermi \texttt{SunMonitor} observation windows. The duration varies from 5 to 80 minutes with an average duration of 30 minutes.}
\label{fig:sunmon_windows}
\end{center} 
\end{figure}

The effective area of the LAT decreases significantly for sources at incidence angles larger than 60{\de}, so only very bright
transients are detectable past this limit. 
Selecting a maximum off-axis angle of 70{\de} extends the window of continuous Sun exposure for the brightest flares.  The
duration of these windows vary (ranging from 5 to 80 minutes, with an average duration of 30 minutes, as is shown in Figure~\ref{fig:sunmon_windows}) as the Sun advances along
the ecliptic and as the orbit of {\it Fermi} precesses.  
Contamination from $\g$~rays produced by cosmic-ray interactions with the
Earth's atmosphere is reduced by selecting only events arriving within 100{\de}
of the zenith\footnote{We used the \texttt{gtmktime} filter cut = \texttt{(DATA\_QUAL$>$0||DATA\_QUAL==-1) LAT\_CONFIG==1 angsep(RA\_ZENITH,DEC\_ZENITH,RA,DEC)$<$ (zmax-rad)} where RA and DEC are those of the postion of the Sun at the time of the flare, zmax= 100{\de} and rad is the radius of the RoI used for the analysis.}.

Each interval is analyzed using a Region of Interest (RoI) of
10{\de} radius, centered on the position of the Sun at the central time of the
interval. On average the duration of a \SunMon interval is 30 {minutes. During this time, the
maximum deviation of the true position of the Sun from the RoI center due to its
apparent motion is $\sim$ 0\fdg02.} 
This is
smaller than the typical angular resolution of the instrument: the 68\%
containment angle of the reconstructed incoming $\gamma$-ray direction for
normal incidence at 1 GeV is 0\fdg8 and at 100 MeV is 5{\de}. Furthermore, the statistical uncertainty on the measured centroid of the $>$100~MeV emission is always larger than 0\fdg03, even for the brightest \solar flares. 
It is therefore not necessary to apply a correction to account for the motion of the Sun from the center of the RoI. 
In each \SunMon interval we perform an unbinned maximum likelihood analysis  using the
tools in the
\Fermi ScienceTools software package\footnote{We used version 11-05-03 available at \url{http://fermi.gsfc.nasa.gov/ssc/}}. 
The unbinned analysis computes the log-likelihood
of the data using the reconstructed direction and energy of each individual $\gamma$-ray
and  the  assumed  sky  model  folded through the instrument response {functions} corresponding to {the selected} event class.

The likelihood analysis consists of
maximizing the probability of obtaining the data given an input model {as well as deriving error estimates}.
The RoI is modeled with a \solar component and two templates for diffuse
$\gamma$-ray background emission: a galactic component produced by the interaction
of cosmic rays with the gas and interstellar radiation fields of the Milky Way,
and an isotropic component that includes both the contribution of the
extragalactic diffuse emission and the residual cosmic rays that passed
the $\gamma$-ray classification\footnote{The models used for this analysis,
\texttt{gll\_iem\_v07.fits} and \texttt{iso\_P8R3\_SOURCE\_V2\_v1.txt},
are available at \url{http://fermi.gsfc.nasa.gov/ssc/data/access/lat/BackgroundModels.html}}.
We fix the normalization of the galactic component but leave the normalization
of the isotropic background as a free parameter to account for variable
fluxes of residual cosmic rays.

When the Sun is not flaring, it is a steady, faint source of $\gamma$ rays. This emission consists of two components: {a disk emission originating} from hadronic cosmic-ray cascades in the \solar atmosphere and a spatially extended emission from the inverse Compton scattering of  cosmic-ray  electrons  on  \solar  photons  in  the  heliosphere. The  {disk}  emission  was  first mentioned by \citet{doi:10.1029/RG003i002p00319} and \citet{1991ApJ...382..652S} and the existence of an additional, spatially extended component was not realized until recently \citep{2006ApJ...652L..65M,2007Ap&SS.309..359O,PhysRevLett.121.131103,PhysRevD.101.083011}.  The quiet Sun was detected for the first time in $\gamma$ rays in the EGRET data~\citep{2008A&A...480..847O}.
We also include the quiet Sun emission disk component as a point source in our RoI; however, we did not include the extended Inverse Compton (IC) component described in \citep{2011ApJ...734..116A} because it is too faint to be detected during these time intervals.
The $>$100 MeV  flux of the \solar disk component used in the FLSF catalog, obtained during the first 18 months of \Fermi-LAT observations \citep{2011ApJ...734..116A}, is 4.6 ($\pm$0.2$^{stat}\pm$ 1.0$^{syst})\times 10^{-7}$ {ph} cm$^{-2}$ s$^{-1}$.

We rely on the likelihood ratio test and the associated test statistic (TS)~\citep[]{Mattox:96} to estimate the significance of the detection. Here we define TS as twice the increment of the logarithm of the likelihood obtained by fitting the data with the source and background model component simultaneously with respect to a fit with only the background. Note that the significance in $\sigma$ {for the 68\% confidence interval} can be roughly approximated as $\sqrt{\rm TS}$.

{With a pipeline testing for detection in so many time windows (33511 total over the period of this work), we need to account for the trials factor to
understand the statistical significance of a $\gamma$-ray source detected in the \SunMon
with a particular value of TS.}

Assuming {each window is} independent, a TS of 20, which would
otherwise correspond to a confidence of about 4.5$\sigma$, corresponds to 1.38$\sigma$ post trials.
In order to have a detection significance of $\geq$5$\sigma$ we must impose a cut on the TS
with a minimum of 30. This corresponds to a selection of 133 time windows, some of them
consecutive in time for \solar flares lasting more than an hour.
Following this systematic sweep with \SunMon, a detailed analysis is performed on
those windows with a TS above 30.

From January 2010 to the end of {January 2018}, 
{we applied the \SunMon pipeline analysis to} 33511 intervals of duration longer than 5 minutes. The cases when the duration is less than 5 minutes {are likely due to} the RoI {being close to} the maximum zenith angle, or cut short by a passage of the satellite into the South Atlantic Anomaly (SAA). These are generally not long enough to yield a reliable point source likelihood detection and constrain the background. Overall the Sun was observable for an average duty cycle of 28\% for the entire timespan of the FLSF catalog.

Note that outside the time interval considered here, since April 2018, the LAT has been operating with a modified observing profile due to a failure of one of the \solar array drive assemblies that reduces its exposure to the Sun \footnote{see: \url{https://fermi.gsfc.nasa.gov/ssc/observations/types/post_anomaly/} for more information.}. This change in observing strategy results in an average 45\% {reduction in \solar exposure for the standard event classes (22\% reduction for LLE)} and consequently in the potential for \solar physics science with the LAT.

\subsection{LAT Low Energy Spectral Analysis}
\label{sec:LLE}
The LAT Low Energy (LLE) technique is designed to study bright transient phenomena, such as \solar flares, in the 30 MeV--1 GeV energy range. In this catalog we used the LLE selection to study the prompt phase of 14 \solar FLSFs. To obtain the LLE spectral data we used the {\tt gtburst} package, available in the Fermitools distribution from the FSSC. The LLE data are divided by {\tt gtburst} in 50 logarithmically spaced energy bins from 10 MeV to 10 GeV. For the spectral analysis we used only the bins in the energy range optimised for the LLE selection. 

A spectral fit was then performed using the {\tt XSPEC}~\citep{XSPEC} package following an approach similar to the one previously adopted for the analysis of the prompt phase of SOL2012-03-07~\citep{0004-637X-789-1-20}. The results of the joint analysis with \Fermi-Gamma-ray Burst Monitor (GBM) Bismute-Germanate (BGO) data ($300$ keV - $20$ MeV) will be reported in a forthcoming publication.

\subsection{Spectral Analysis}
\label{sec:spectral_analysis}
We fit three models to the \Fermi-LAT $\gamma$-ray \solar spectral data. The first two, a simple power law (PL) and a power-law with an exponential cut-off (PLEXP), are phenomenological functions that may describe bremsstrahlung emission from relativistic electrons. {The parameters of these models are varied to obtain the best fit to the data.} When the PLEXP provides a significantly better fit than the PL, we also fit the data with a third model consisting {of a pion-decay emission templates\footnote{We are using only pion-production emission neglecting other (minor) components that contribute to the $\gamma$-ray emission}. This third model uses a series of $\g$-ray spectral templates derived from a detailed study of $\g$ rays from the decay of pions
produced by interactions of accelerated protons and ions with background protons and ions. 
The accelerated particles are assumed to have a power-law} {energy} spectrum {(dN/dE $\propto$ E$^{-\beta}$), where E is the kinetic energy of the protons} with index $\beta$ and an isotropic pitch angle distribution, injected into {a thick target with a coronal composition~\citep{ream95} taking He/H = 0.1}~\citep[updated from][]{murp87}.

\begin{figure}
\begin{center}
\includegraphics[width=0.7\linewidth]{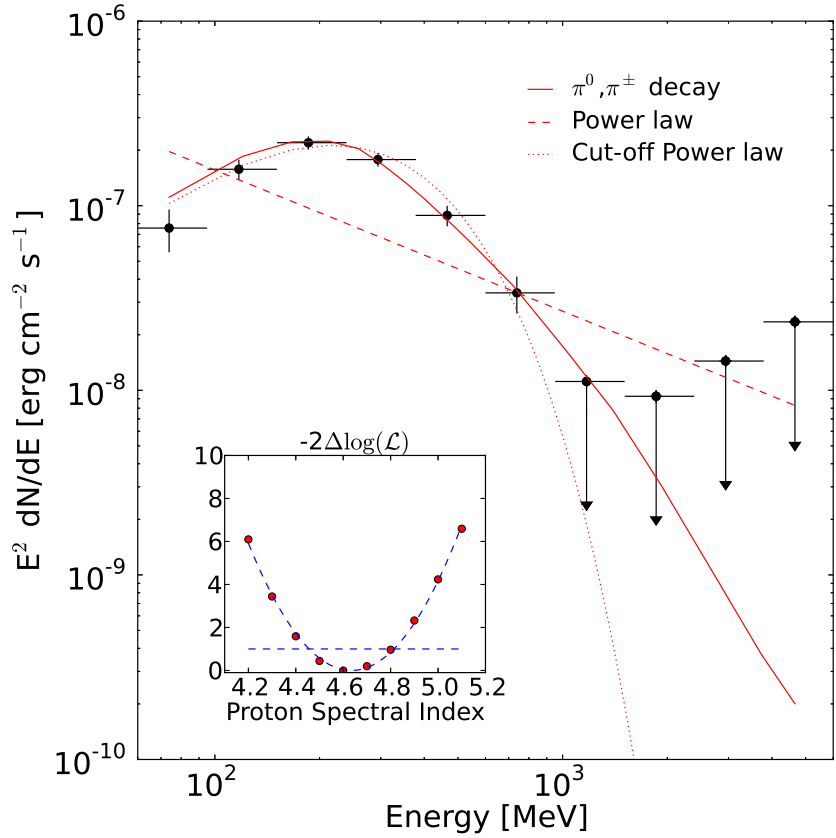}
\caption{Example {$\gamma$-ray spectra} for SOL2012-03-07. The data {were} fit with three models (PL,PLEXP and pion templates) and when the curved model (PLEXP) is preferred to the PL model we perform a scan over the pion templates to search for the best proton index. In the insert we show the fit to the log-likelihood values with a parabola and the  68\% confidence level is indicated by the straight line at $-2$\,$\Delta\log(\like_{\rm min})+1$.}
\label{fig:sed_sample}
\end{center} 
\end{figure}

{When the PLEXP provides a significantly better fit than the PL, we fit the data with the pion-templates }to determine the proton index that best fits the data. To do this, 
we calculate the variation of the log-likelihood 
with the proton spectral index and fit it with a parabola. {We run the likelihood analysis for each of the 41 proton spectral 
indices available from our templates (2.0 - 6.0 in steps of 0.1).} The minimum of 
this distribution ($\like_{\rm min}$) 
gives the best fit spectral index,
and the corresponding value $s_0$ as the maximum likelihood.
Figure~\ref{fig:sed_sample} shows an example of a spectral energy distribution of SOL2012-03-07 obtained following this procedure. 

Once we have found the proton index corresponding to the best fit and the value of the observed $\gamma$-ray emission we can estimate the total number of $>$500~MeV accelerated protons (N500, hereafter) needed to produce the observed $\gamma$-ray emission over a given time following the prescription of \citet{murp87}.

To compute the photon spectral energy distribution we divide the data into ten energy bins {(in the energy range 60 MeV - 10 GeV)} and determine the source flux using the unbinned maximum likelihood algorithm \texttt{gtlike} keeping the normalization of the background constant at the best fit value and assuming that the spectrum of the point source is an $E^{-2}$ power law. For non-detections (TS$<$9), we compute  95\% CL upper limits.

\subsection{Localizing the emission from \Fermi LAT Solar Flares }
\label{sec:localization_alg}

\begin{figure*}[ht!]   
\begin{center}   
\includegraphics[width=0.8\textwidth]{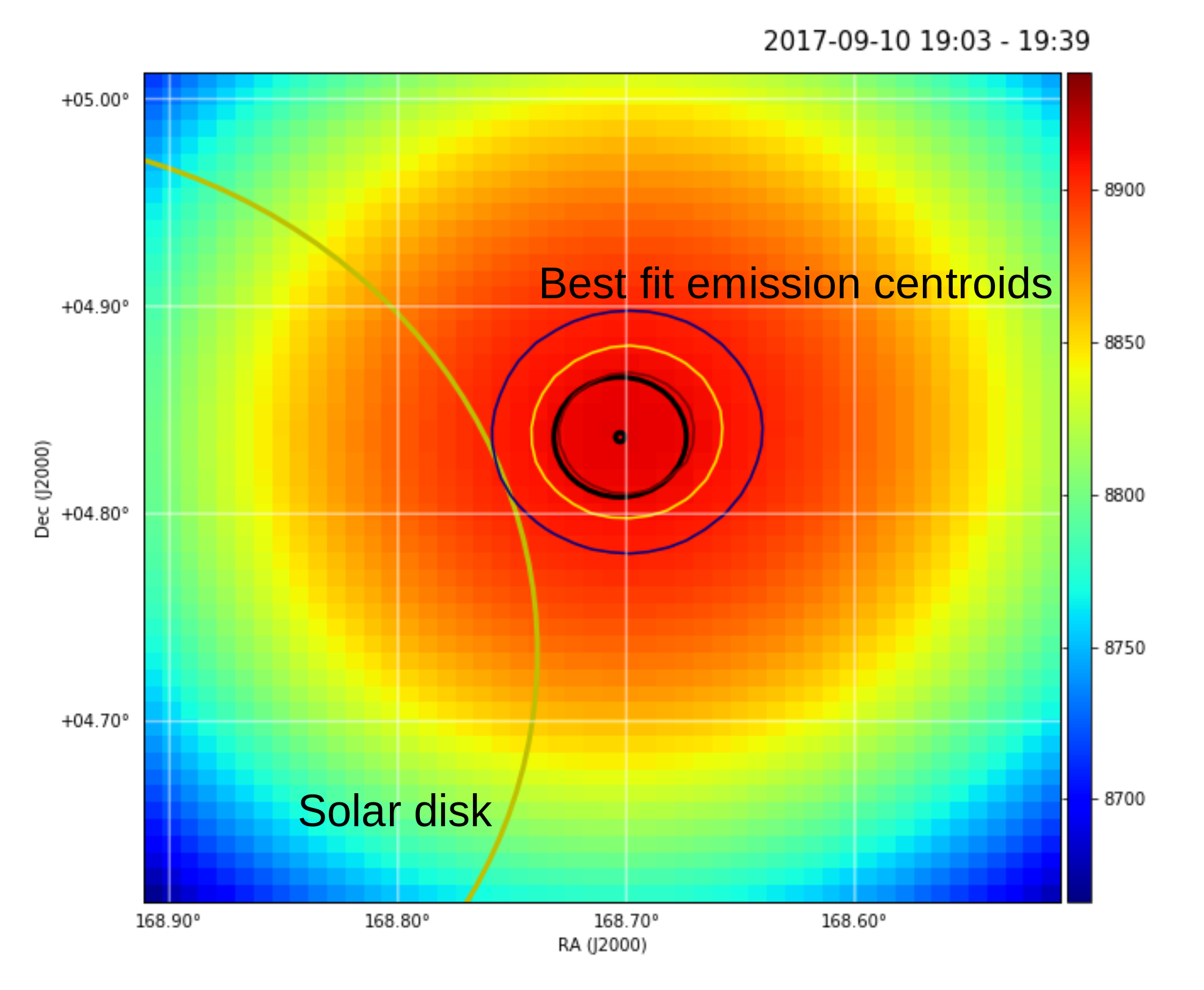} 
\caption{TS map for the observation of FLSF~2017-09-10 in the time interval of 19:03--19:39 UT. {The large yellow circle represents the \solar disk, solid black circle represents the 68\% statistical error. The thin red, yellow and blue lines track the 1, 2 and 3 sigma contours on the TS map.} These are not always perfectly circular, but a circular
error containment region (black circle) provides a good approximation.}
\label{fig:tsmap_example}   
\end{center}   
\end{figure*}

The standard tool to study the localization of $\g$-ray sources with an unbinned likelihood analysis is the \texttt{gtfindsrc} algorithm from the ScienceTools\footnote{Available at \url{http://fermi.gsfc.nasa.gov/ssc/data/analysis/software/}}. The likelihood analysis is based on sky models with background sources at fixed spatial positions and the best spectral fit for the source of interest. \texttt{gtfindsrc} uses a multidimensional minimization of the unbinned likelihood for a grid of positions around an initial guess until the convergence tolerance for a positional fit is reached.
However, the Sun is in the FoV of the LAT for relatively short timescales which can result in inhomogeneous exposure across the FoV. For this reason, we relied on the \texttt{gttsmap} algorithm to study the localization for the {FLSFs} of the catalog. The 
TS maps are created by moving a putative point source through a grid of locations on the
sky and maximizing -log(likelihood) at each grid point, with any other well-identified sources within the RoI included in each fit. The \solar flare source is then identified at local maximum of the TS map. The 68\% containment radius (or 1 $\sigma$ statistical localization error) on the position corresponds to a drop in the TS value of 2.30 (4.61 and 9.21 correspond to 2 and 3 $\sigma$ respectively). See Figure~\ref{fig:tsmap_example} for an example TS map of FLSF~2017-09-10.

\begin{figure*}[ht!]   
\begin{center}   
\includegraphics[height=4.5 in]{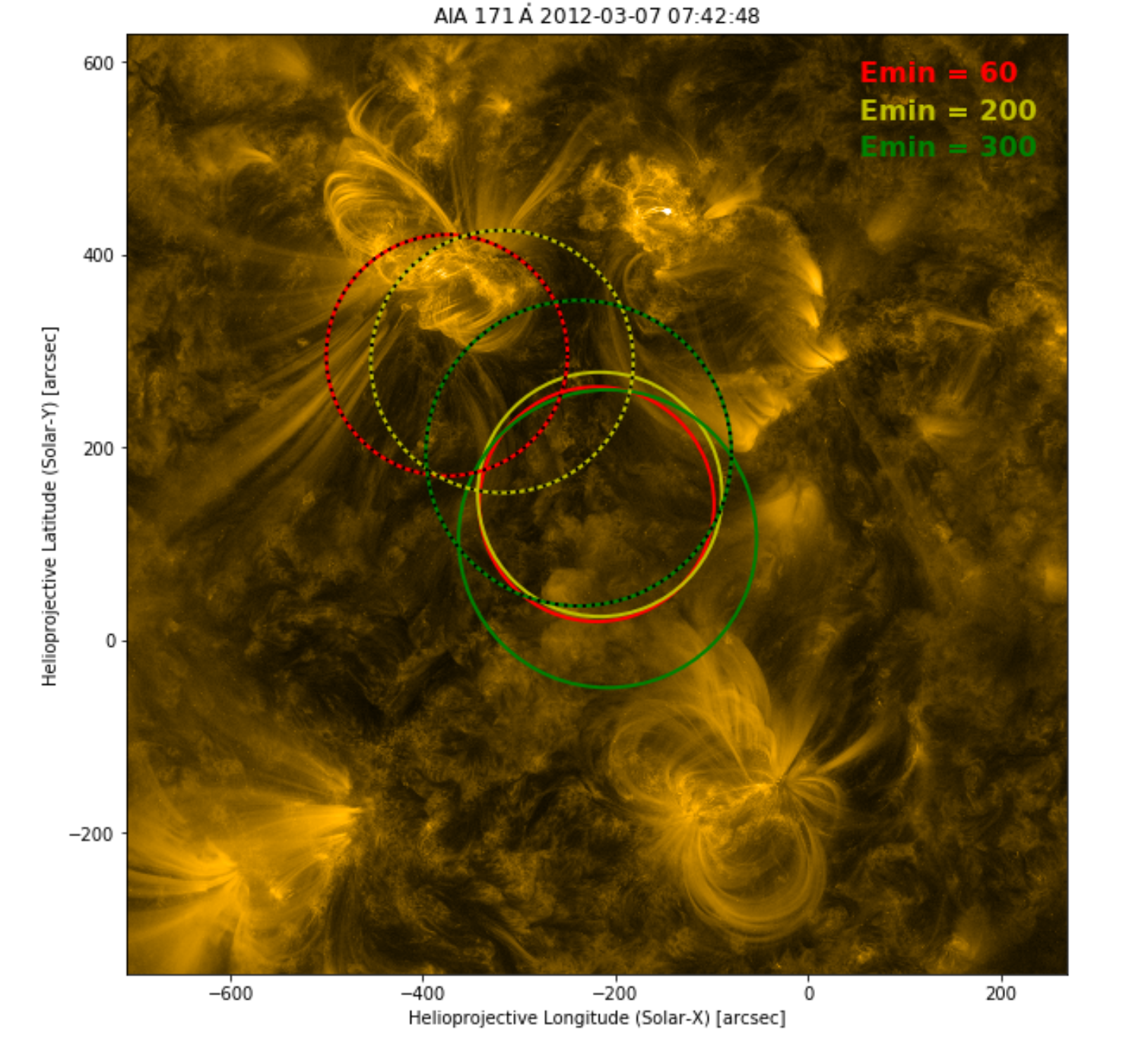}  
\caption{Comparison of the localization of the bright FLSF 2012-03-07 between fish-eye corrected (solid line) and not corrected (dashed line) with 60 (red), 200 (yellow) and 300 (green)~MeV energy thresholds. Each circle marks the 68\% statistical containment radius. The background is an Atmospheric Imaging Assembly (AIA) 171 \Angst\ image taken at 2012-03-07 07:42:48 UT by the Solar Dynamics Observatory (SDO)}. 
\label{march2012_spatial_errors}   
\end{center}   
\end{figure*}

When performing the localization of the \fermi-LAT data of the Sun it is necessary to also take into account for the \emph{fish-eye} effect. The fish-eye effect is a selection bias in the LAT trigger and reconstruction algorithms. At low energies and high incidence angles, particles that scatter toward the LAT boresight (having a smaller apparent incidence angle) are reconstructed with higher efficiency than particles that scatter away from the LAT boresight (having a larger apparent incidence angle). The reconstructed position of the source is biased and ends up appearing closer to the boresight axis than its true position.

The fish-eye effect can be quantified on an event-by-event basis using Monte-Carlo simulations. The correction depends both on the true incidence {angle} and the energy of the particle. The correction becomes dramatic at energies below 100 MeV and incidence angle greater than 70$^{\circ}$, reaching several degrees shift \citep[see][for a detailed description of the fish-eye effect]{2012ApJS..203....4A}. 

The correction of the fish-eye effect is crucial particularly for bright flares, when the statistical error on the position becomes smaller than 0\fdg1 and the uncertainty becomes dominated by systematics. 
We 
investigated 
the effect of the fish-eye 
correction on two bright 
\solar flares (FLSF 2012-03-07 and FLSF 2017-09-10). We varied the value of the minimum energy threshold to quantify the amplitude of the correction and the systematic error it induces. 
The amplitude of the fish-eye correction decreases with energy so we expect the distance between the corrected and uncorrected positions to decrease with energy. This is indeed what we observe in Figure~\ref{march2012_spatial_errors}: the correction is largest above a 60 MeV minimum energy, and above 300 MeV the two positions are consistent.

Solar flares generally have soft $\g$-ray spectra, 
 cutting off at energies just above 100 MeV, so that the localization error (statistical) does not really improve as the threshold energy is increased, as can be seen in an example in Figure~\ref{march2012_spatial_errors}, where the statistical error on the localization above 300 MeV (green) is larger than the one above 60 MeV (red). Due to this, we use only photons with measured energies above 100 MeV  when performing the localization study. Note that, although the localization uncertainties at 60 MeV and 100 MeV are very similar, the fish-eye correction that we had to apply to the events between 60 MeV and 100 MeV is larger than the one for the events above 100 MeV; therefore in order to minimize the systematic uncertainty, we use only events with energy $>$100 MeV to estimate the localization of the emission.
 
\subsubsection{Localization of BTL FLSF~2014-09-01}
\label{sec:loc_20140901}
The emission centroid for the other {FLSFs} previously published {all remained within the 68\% error radius} with the new analysis tool, the FLSF~2014-09-01 is the only exception that we found during the analysis performed for this work.

As mentioned in section~\ref{sec:localization_alg} the tool used to perform localization studies for the FLSF catalog to compensate for the potential systematic errors tied to inhomogeneous exposures across the FoV for short detections is \texttt{gttsmap} and no longer the \texttt{gtfindsrc} tool. We also reported (in section~\ref{sec:localization_alg}) the study performed to quantify the impact on the localization results due to the fish-eye effect and showed that it depends on the energy and incidence angle of the source. For this reason, in the FLSF catalog we have decided to perform localization studies using \texttt{gttsmap} on bright flares with exposure times longer than 20 minutes, with incidence angles smaller than 60$^{\circ}$ and with energies greater than 100 MeV in order to avoid potentially large systematic effects in the resulting emission centroids.

\begin{figure*}[ht!]   
\begin{center}   
\includegraphics[width=0.7\textwidth]{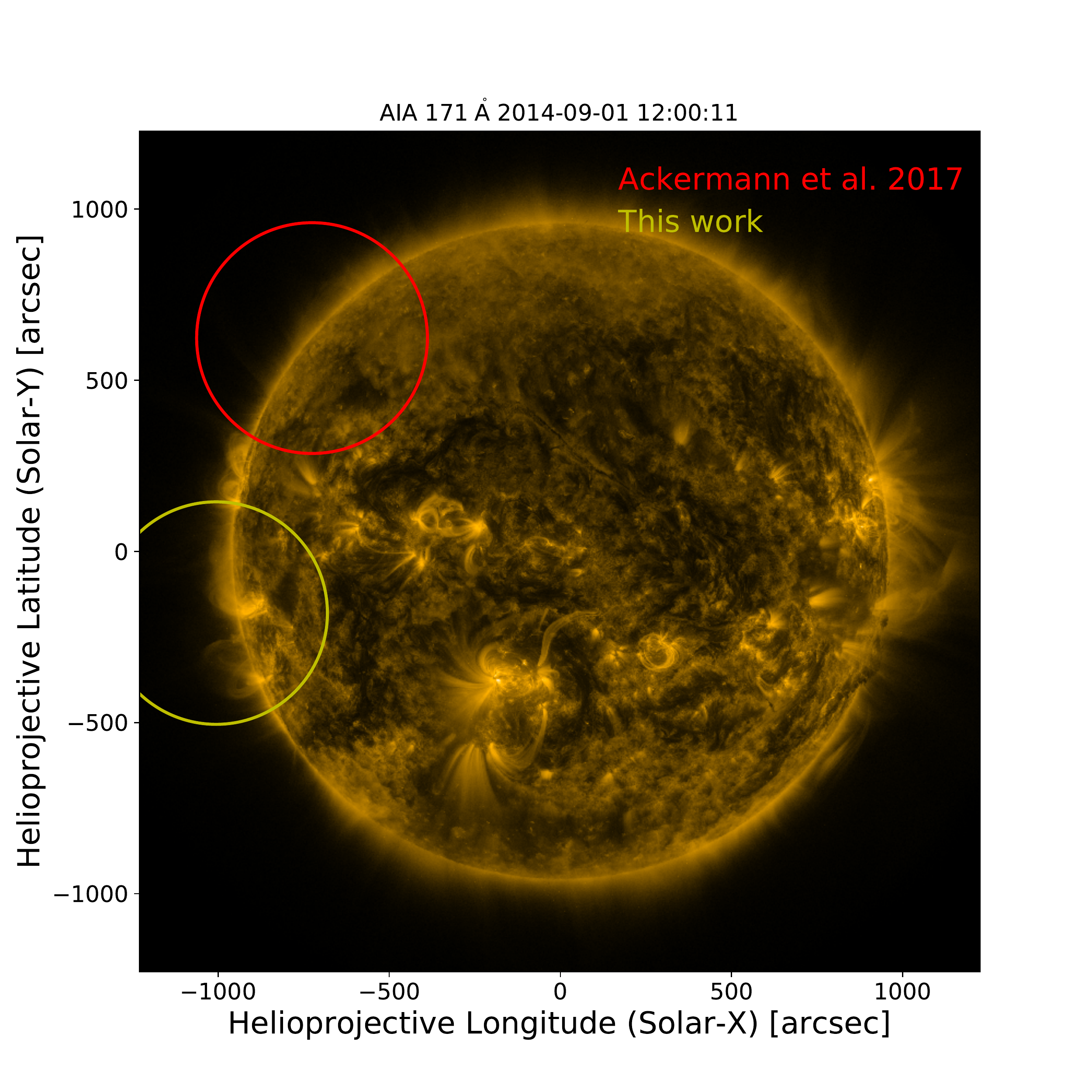}  
\caption{Emission centroid for FLSF~2014-09-01 for energies greater than 100~MeV with 95\% uncertainty error radius using the \texttt{gttsmap} tool and the fish-eye correction in yellow and the previously published position is shown in red (with the 95\% uncertainty error radius). The new position is centered at helioprojective coordinates X,Y=[-1105$\arcsec$,-128$\arcsec$] with a 95\% uncertainty error radius of 643$\arcsec$.}   
\label{fig:FLFSF20140901_localization_old_new}   
\end{center}   
\end{figure*}

The first detection window of the BTL FLSF~2014-09-01 unfortunately occurred when the Sun was at an angle of 67$^{\circ}$ from the LAT boresight and lasted for only 16 minutes and the emission centroid published in~\cite{2017ApJ...835..219A} was obtained using the \texttt{gtfindsrc} tool. After a careful re-analysis of this flare with the new localization tool and the knowledge obtained from the fish-eye systematic study we find that the emission centroid for FLSF~2014-09-01 has moved with respect to the previously published value as can be seen in Figure~\ref{fig:FLFSF20140901_localization_old_new}.  

\subsection{Test for spatial extension}
\label{sec:extension_alg}

We test the possibility to measure spatial extension in the localization results of the bright FLSF~2012-03-07 and FLSF~2017-09-10 by using \texttt{fermipy} \citep{2017ICRC...35..824W}. 
This tool has been used in several \Fermi-LAT publications \citep{2018ApJ...856..106D,2018ApJ...866...24A,2018ApJS..237...32A,2019MNRAS.485..356A}. It is based on a binned likelihood analysis and, although not optimal for low counting statistics\footnote{Both FLSF~2012-03-07 and FLSF~2017-09-10 are very bright and a binned likelihood analysis is appropriate.}, presents the advantage of being very fast and allows to study the extension of $\g$-ray emission by comparing a model with a source with a radial extension ({uniform} disk or gaussian) with the data, and profiling the value of the $\log(\like)$ by varying the extension radius. 

\begin{figure*}[ht!]   
\begin{center}   
\includegraphics[width=0.45\textwidth,trim=0 52 10 30,clip]{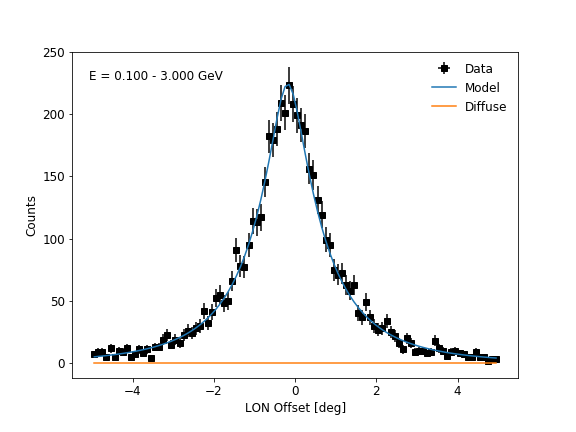}
\includegraphics[width=0.45\textwidth,trim=0 52 10 30,clip]{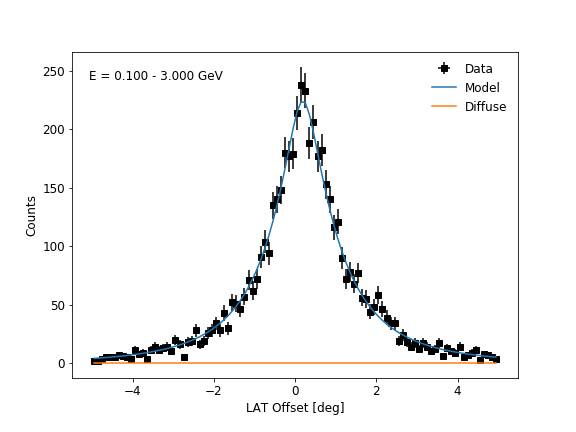}\\
\includegraphics[width=0.45\textwidth,trim=0 0 10 52,clip]{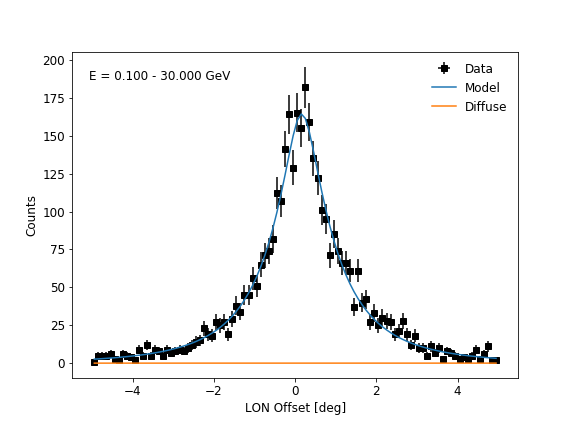}
\includegraphics[width=0.45\textwidth,trim=0 0 10 50,clip]{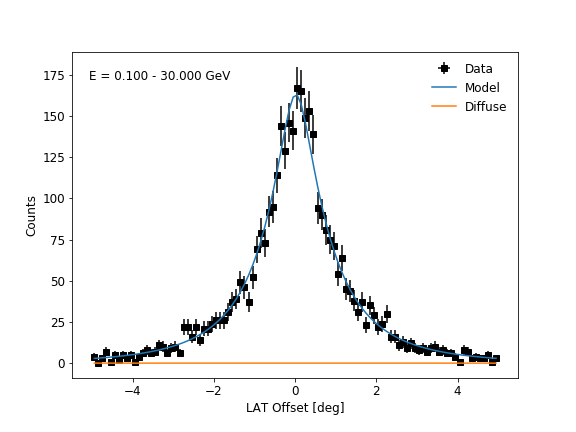}
\caption{Longitude (left) and latitude (right) radial profile for FLSF~2012-03-07 (top row) and for FLSF~2017-09-10 (bottom row). The x-axis shows the offset with respect to the optimized localization.}   
\label{fig:pontsource_extension} 
\end{center}   
\end{figure*}

For FLSF~2012-03-07 we use the same time window used in \citep{0004-637X-789-1-20}, namely from 2012-03-07 02:27:00 UT to 2012-03-07 10:14:32 UT, thus avoiding the time interval affected by ACD pile up. For FLSF~2017-09-10 we use the time window from 2017-09-10 15:56:55 UT to 2017-09-11 02:00:21 UT and SOURCE class events with energies greater than 100 MeV. The RoI is 10\degr\ wide. In this analysis the spectra of the {FLSFs} are described by a power law with exponential cut off, and the model is re-optimized during the fit procedure.
For convenience, we use \texttt{ThreeML} \citep{2015arXiv150708343V} as an interface to \texttt{fermipy}. It allows us to perform the fit to the LAT data using the \texttt{fermipy} plugin, providing, at the same time, an easy interface to download the data and build the model to be fitted.
In Figure~\ref{fig:pontsource_extension} we show the radial profile of a point source model compared {to} the data, for the best fit model. The model (which is convolved with the IRFs of the instrument), matches very well the radial profile of the counts in both directions, and no residual counts that could suggest the presence of a spatially extended emission are visible. Note that in our analysis we first optimize the localization of the source (hence the offset in Figure~\ref{fig:pontsource_extension}) and then we test for an extension. The optimized locations are 
{at helioprojective coordinates X,Y=[-400$\arcsec$,400$\arcsec$] with a 68\% uncertainty error radius of 100$\arcsec$}
for FLSF 2012-03-07, and {X,Y=[600$\arcsec$,-60$\arcsec$] with an uncertainty of 70$\arcsec$ for FLSF~2017-09-10}.

\begin{figure*}[ht!]   
\begin{center}   
\includegraphics[width=0.49\textwidth,trim=0 0 0 0,clip]{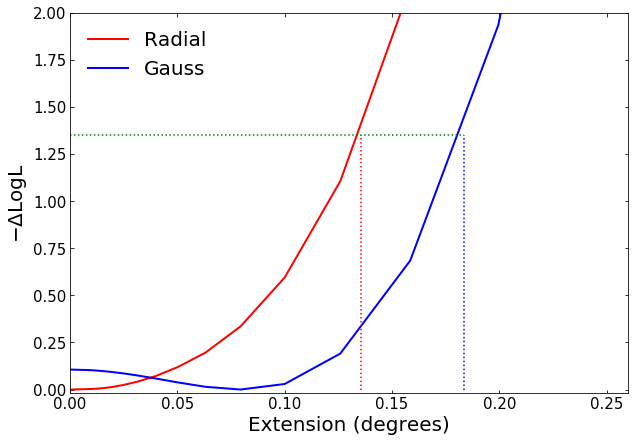}
\includegraphics[width=0.49\textwidth,trim=0 0 0 0,clip]{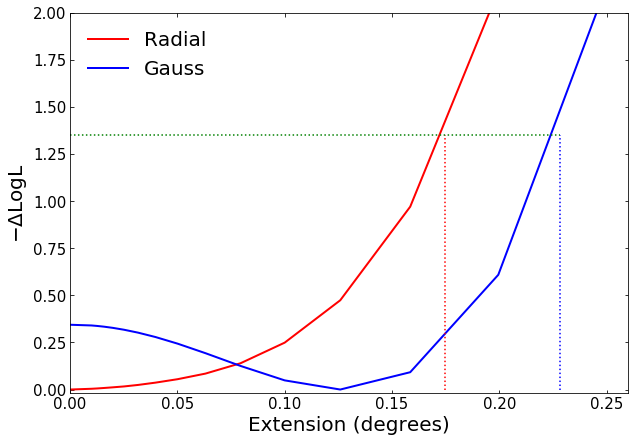}
\caption{Likelihood profile of FLSF 2012-03-07 (left) and FLSF 2017-09-10 (right) as a function of a spatial profile for a Gaussian profile (Gauss) and a Radial profile (Radial). The horizontal green dotted lines show the increment of the $-\Delta\log(\like)\approx$1.35, corresponding to a C.L. of 95\%. The blue and red dotted lines are the estimated values for the upper limits on the radius. }   
\label{fig:likelihood_extension} 
\end{center}   
\end{figure*}

Finally, in Figure~\ref{fig:likelihood_extension} we show the profile of the likelihood as a function of the radial extension for two different spatial templates, for the two flares. 
The improvement with respect to the point source hypothesis is very small ($\Delta$TS$<$1.5 in both cases), and only an upper limit of the radius can be placed. The 95\% confidence level upper limits (corresponding to a $-\Delta\log(\like)\approx$1.35) are, 0\fdg18, for the gaussian disk and 0\fdg14 for the radial disk for FLSF~2012-03-07, and 0\fdg23 (gaussian) and 0\fdg17 (radial) for FLSF~2017-09-10.
These two events are the only two flares detected by the LAT that are bright enough to allow a dedicated spatial extension analysis. Even so, we  can only set an upper limit on the extension that is smaller than the \solar radius.

\subsection{Total emission duration, fluence and total number of greater than 500~MeV protons}
\label{sec:total_emission}
With the Sun being observable by the LAT for only 20 to 40 minutes every 1.5 to 3 hours, it can be challenging to reconstruct the complete light curve and to estimate the true duration of the $\g$-ray emission. In order to overcome the issues caused by the observational gaps, we are forced to make some assumptions on the behavior of the emission when the Sun is outside of the {FoV} of the LAT. To identify the start of the FLSF we rely on the timing of the associated GOES X-ray flare. For example, when the GOES X-ray flare occurs during a LAT data gap and the start of the LAT detection window ($t_{\rm start}$) occurs after the end of the GOES X-ray flare we take the end of the GOES X-ray flare as the start of the $\g$-ray emission. For the cases where the GOES X-ray flare occurs within the detection window, and the LAT statistics are not sufficient to perform a fine time binning analysis, we take $t_{\rm start}$ to be the start of the detection window. The end time of the FLSF ($t_{\rm stop}$) is taken as the midpoint between the end of the last detection window and the start of the following observational window (with an upper limit on the $\g$-ray emission from the Sun). The total duration of the FLSF is then simply $\Delta t=t_{\rm stop}-t_{\rm start}$. These assumptions on the start and stop of the FLSF are not needed for the short prompt FLSF flares where the true start/stop of the $\gamma$-ray emission can be identified within the observational window.

\begin{figure*}[ht!]   
\begin{center}   
\includegraphics[width=\textwidth,trim=10 50 50 50,clip]{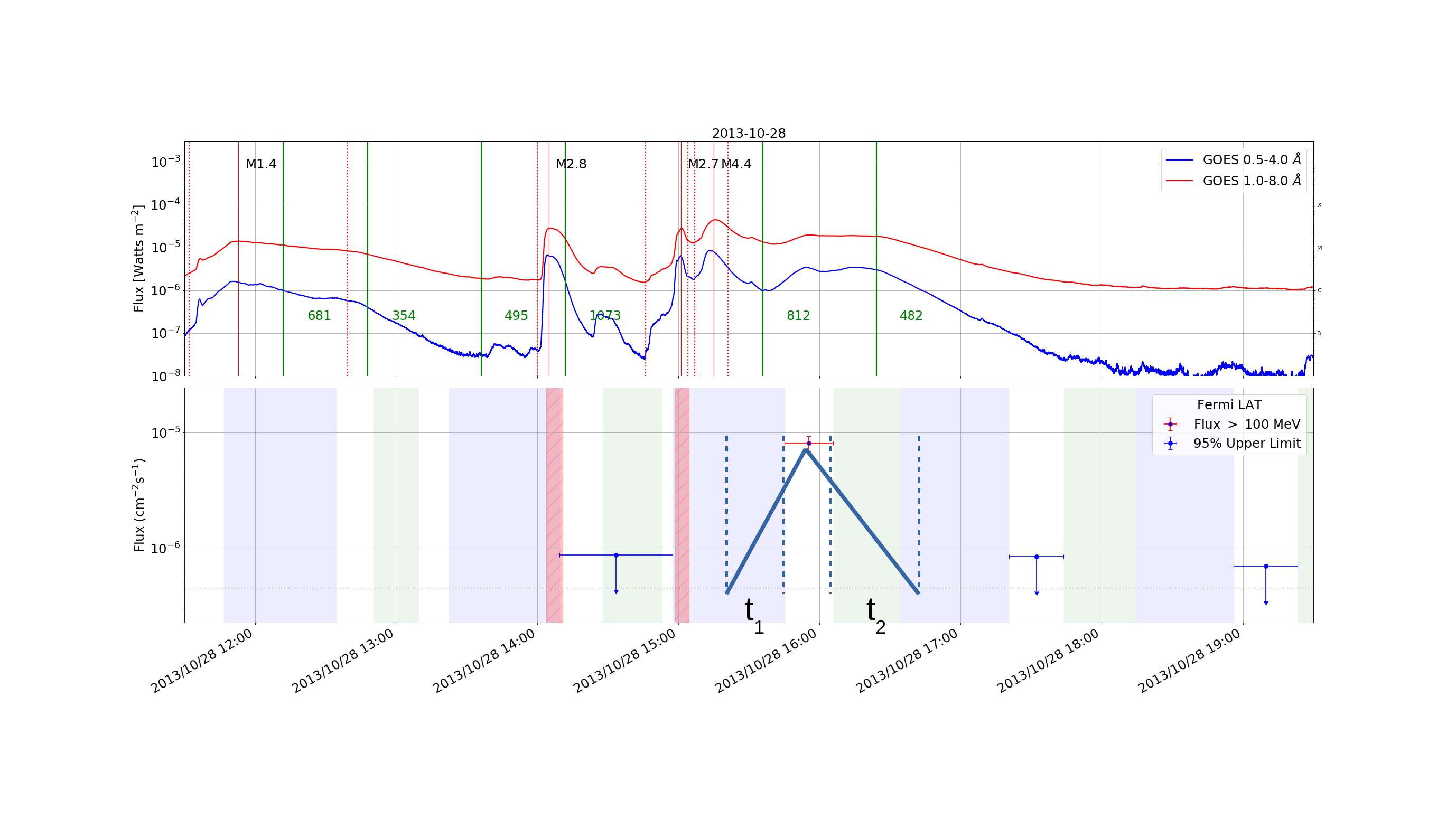} 
\caption{Lightcurve of the {$>$100 MeV emission from} FLSF 2013-10-28 with multiple flaring episodes
prior to the start of the $\g$ rays. The M2.7 and M4.4 and 812 km\,s$^{-1}$  CME all from the same Active Region  (AR) are likely associated with the $\gamma$-ray emission, although it is possible that the activity from another AR (M2.8 flare and 1073 km\,s$^{-1}$  CME) may contribute to the $\gamma$-rays. The solid green lines represent the {Large Angle and Spectrometric Coronagraph (LASCO)} CME C2 first appearance, the linear speed value is annotated next to the line (also in green). The dashed/solid red lines represent the start(stop)/peak of the GOES X-ray flare, the GOES class is also annotated next to the solid red line.  In the lower panel the vertical dashed lines denote the t$_{1}$ and t$_2$ quantities, where t$_1$ is defined as the time between the assumed start of the emission and the start of the detection window and t$_2$ is the time between the end of the detection window and the assumed end of the emission. For further details on how we use the t$_1$ and t$_2$ quantities to determine the uncertainties on the total fluence and total N500 see text in Section~\ref{sec:total_emission}. The solid triangle represents the assumed lightcurve for this flare. The {light green} bands indicate when the \Fermi satellite was in the South Atlantic Anomaly (SAA), the blue bands indicate when the Sun was outside of the {FoV} of the LAT and the pink bands indicate the presence of potential pile-up in the data.}   
\label{fig:Flares_association_SOL20131028} 
\end{center}   
\end{figure*}

Once we have estimated the start and stop of the FLSF, we can build a functional shape\footnote{We use \texttt{scipy} splines to build the functional shape of the $\gamma$-ray lightcurve.} to describe the lightcurve of the FLSF even in the cases where we only have one detection point (see  Figure~\ref{fig:Flares_association_SOL20131028}). Having a full description of the lightcurve of the FLSF emission it is possible to evaluate the total $\gamma$-ray fluence by simply integrating {the lightcurve over the estimated duration of the flare}. When integrating we assume that the flux values at the start and end of the FLSF are equal to 4.6$\times$10$^{-7}$ {ph} cm$^{-2}$s$^{-1}$ which corresponds to the $>$100~MeV quiet Sun emission.

For every FLSF that is best described by the pion template model we {provide} an estimate of {N500} 
needed to produce the $\gamma$-ray emission detected in the observational time window. However, if we want to know the total N500 needed to produce the total $\gamma$-ray emission over the full duration then we need to build a functional form {(just as was done for the lightcurve)} also for the temporal evolution of N500. The start and stop of the FLSF remain the same as described above; the main challenge lies in estimating the value for N500 at t$_{\rm start}$ and t$_{\rm stop}$. The value of N500 depends on two parameters, the normalization of the spectral function used to fit the data and the best proton index resulting from the spectral analysis (as described in \ref{sec:spectral_analysis}). We therefore find the best value for the N500 corresponding to the quiet Sun flux level by performing a scan over all the possible proton indices (ranging from 2 to 6 with the same gradation as used during the likelihood analysis) and used the average value of 6$\times$10$^{22}$. Finally, as in the case for the fluence, we integrate the functional form to find {N500}
needed to produce the total emission of the FLSF. The values for the total fluence and total N500 with their associated uncertainties for all the {FLSFs} in the catalog are listed in Table~\ref{tab:time_integrated_main_table}.

The main uncertainties on the fluence and total N500 are due to the values of $t_{1}$ and $t_{2}$, where t$_1$ is defined as the {duration} between the assumed start of the emission ($t_{\rm start}$) and the start of the detection window and t$_2$ is the {duration} between the end of the detection window and the assumed end of the emission ($t_{\rm stop}$). See Figure~\ref{fig:Flares_association_SOL20131028} for an illustration of t$_1$ and t$_2$ for the case of the single point detection of FLSF 2013-10-28. To estimate this uncertainty we vary the value of t$_1$ and t$_2$ by $\pm$50\% and repeat the integral over the flux and N500, the error is then found by taking the difference between this value and the nominal one. 


\section{FLSF classification}
\label{sec:classification}
We associate each significant detection of $\g$-ray emission from \solar flares with \solar events as seen by other instruments. For most cases the association of the $\g$-ray emission to a specific GOES flare or CME is straightforward: linking the FLSF to a single flare or CME within an hour of the start of the $\g$-rays. In some cases however, the association to a single GOES flare or a single CME is not 
obvious when several events happen within a short time frame. {In these cases, we tend to pick the GOES flare or the CME closest in time to the $\gamma$-ray emission.} For example, the FLSF 2013-10-28  (shown in Figure~\ref{fig:Flares_association_SOL20131028}) a series of three M-class flares occurred, accompanied by two CMEs, all prior to the $\gamma$-ray detection. In this case the $\gamma$-ray emission is likely associated with the pair of flares M2.7 and M4.4 {(both of which started within an hour of the start of the FLSF)} from the same 
AR and the associated CME with speed 812 km\,s$^{-1}$ {(LASCO first appearance occurred $\approx$15 minutes prior to the start of the FLSF)}.

In the cases of the BTL {FLSFs},  the soft X-ray emission detected by GOES is either absent or biased toward lower fluxes than would have been the case {if it was} a disk flare. For those, the STEREO satellites provide the direct {Extreme Ultra Violet} (EUV) observation of the flare, which allows us to estimate 
the peak soft X-ray flux 
(for a detailed description of this procedure, see 
\citealp{2017ApJ...835..219A}).

Once we have found a GOES X-ray flare 
associated with the FLSF then we can begin to classify the 
flares in the catalog. In the attempt to better {characterize the features present in each of the FLSFs and hopefully to also} understand the underlying acceleration mechanisms at work during the  flares in the FLSF catalog, we compare the $\g$-ray timing evolution with that in Hard X-Rays. This is because HXR emission traces the high-energy electron population accelerated during the flare energy release and $\g$-ray signatures of protons accelerated by the same processes and on the same time-scales have been observed in the past by SMM and EGRET~\citep{thompson93}. 

\begin{figure*}[ht!]   
\begin{center}   
\includegraphics[width=\textwidth,trim=0 50 50 50,clip]{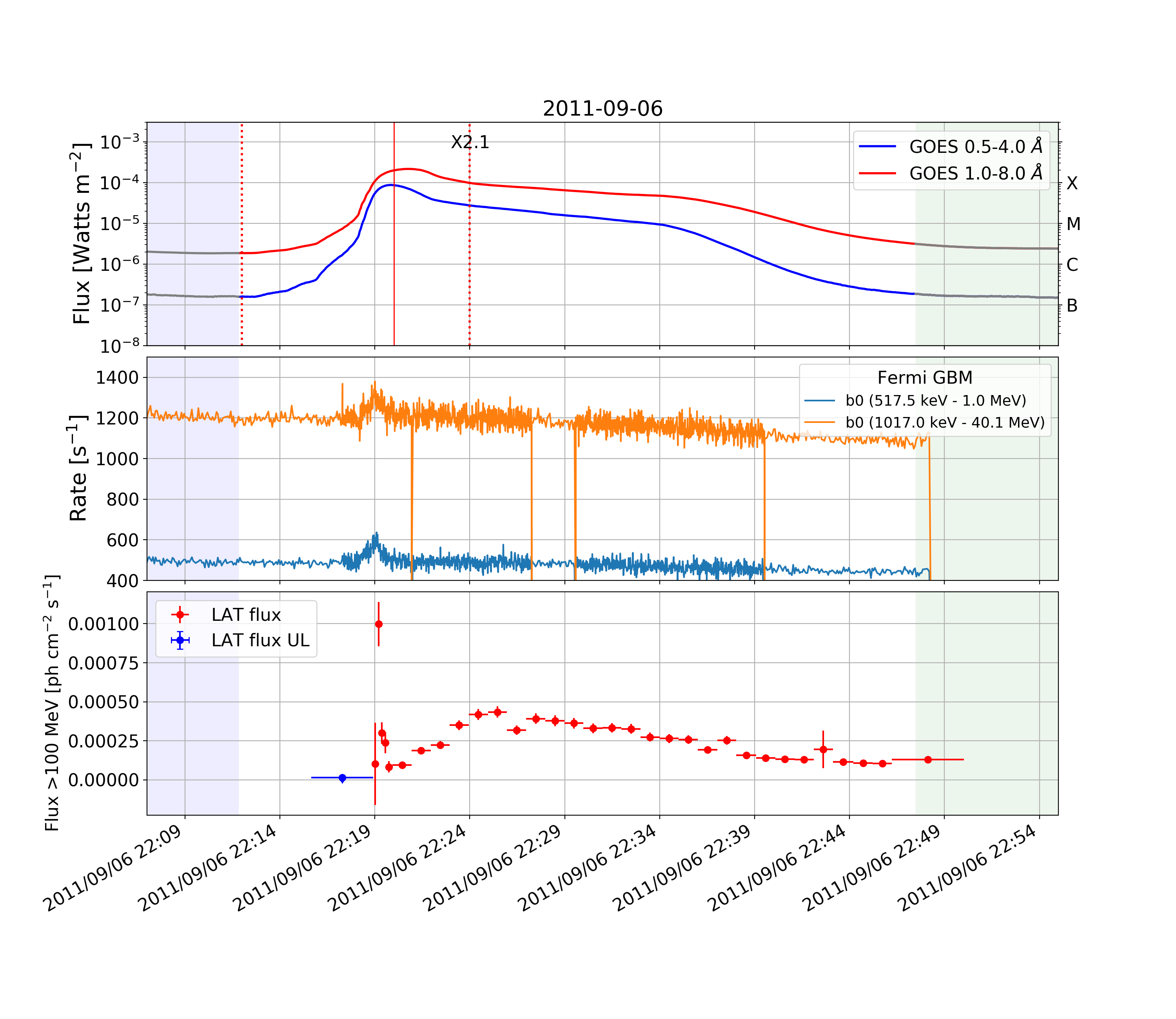} 
\caption{Example of a flare with a prompt component coincident with the bright HXR peak followed by a $\gamma$-ray delayed emission; that occurred on 2011-09-06. From top to bottom, the GOES X-ray flux in two energy bands, the \Fermi-GBM X-ray lightcurve and the \Fermi-LAT {$>$100 MeV }flux using the standard likelihood analysis with a fine time binning to reveal the \emph{prompt} component. The dashed/solid red lines represent the start(stop)/peak of the GOES X-ray flare, the GOES class is also annotated next to the solid red line.}
\label{fig:Flares_categories_example_prompt_short_delay} 
\end{center}   
\end{figure*}

The \Fermi-Gamma-Ray Burst Monitor~\citep[GBM,][]{meeg09} on-board the \Fermi satellite consists of twelve NaI detectors and two BGO detectors covering an energy range 8~keV to 40~MeV. Thanks to the fact that the \Fermi-GBM  continuously monitors the non occulted sky, it provides excellent HXR coverage of the FLSFs in this catalog. For each FLSF in the catalog with a time window coincident with the prompt phase of an X-ray \solar flare, we compare the HXR evolution observed by the two instruments of the \Fermi-GBM to a finely time-resolved $\g$-ray lightcurve as shown in Figure~\ref{fig:Flares_categories_example_prompt_short_delay} for the FLSF~2011-09-06. If we find that the $\g$-ray emission evolution is synchronous to the HXR evolution we classify it as a {\it prompt} flare.

When performing these finely time-resolved lightcurves different patterns emerge revealing a more complex picture of the $\gamma$-ray \solar flares. This can be seen again for the FLSF 2011-09-06  (Figure~\ref{fig:Flares_categories_example_prompt_short_delay}). A prompt component coincident with the bright HXR peak appears in $\g$-rays and is immediately followed by a second phase lasting for more than 20 minutes after the start of the flare. This phase consists of a second less bright peak with a longer rise and fall timescales, but there is no sign of such behavior in the HXRs. The Sun passed in the FoV two hours later and no $\g$-rays were detected. Cases such as the FLSF~2011-09-06 are classified as \emph{prompt short-delayed}.

A flare is {\it prompt-only} if the $\gamma$-ray emission does not extend beyond the HXR duration, as was the case for the flare detected on 2010-06-12~\citep{2012ApJ...745..144A}. All flares detected through the LLE method are associated with prompt emission, but some exhibit delayed emission as well. The fine time-resolved lightcurves for all the {FLSFs} classified as \emph{prompt} are reported in~\cite{ajello_m_2020_4311157}.  

\begin{figure*}[ht!]   
\begin{center}    
\includegraphics[width=0.83\textwidth,trim=30 150 50 100,clip]{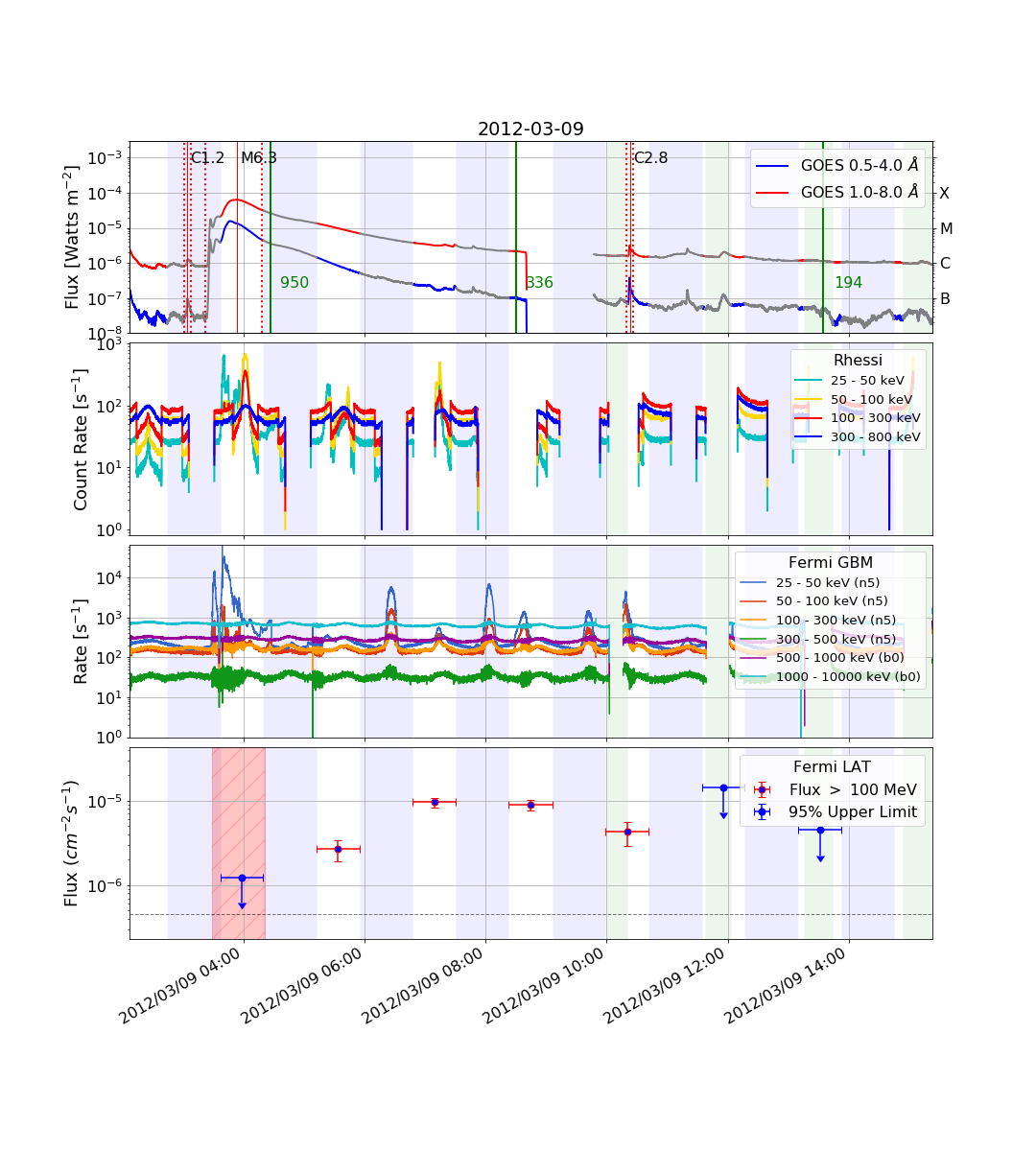} 
\caption{{Lightcurve of the $>$100 MeV emission from} FLSF 2012-03-09 lasting more than 6 hours but with no detectable high-energy $\gamma$-ray emission in the impulsive phase, classified as {\it delayed-only}. The four panels report the light curve measured by GOES, RHESSI, \Fermi/GBM and \Fermi/LAT in various energy ranges. The solid green lines represent the LASCO CME C2 first appearance, the linear speed value is annotated next to the line (also in green). The dashed/solid red lines represent the start(stop)/peak of the GOES X-ray flare, the GOES class is also annotated next to the solid red line. The {light green} bands indicate when the \Fermi satellite was in the SAA and the blue bands indicate when the Sun was outside of the {FoV} of the LAT. Pink bands indicate the time interval over which potential pile-up effects could be present.}   
\label{fig:Flares_delayed_1}   
\end{center}   
\end{figure*}

\begin{figure*}[ht!]   
\begin{center}    
\includegraphics[width=0.83\textwidth,trim=30 150 50 100,clip]{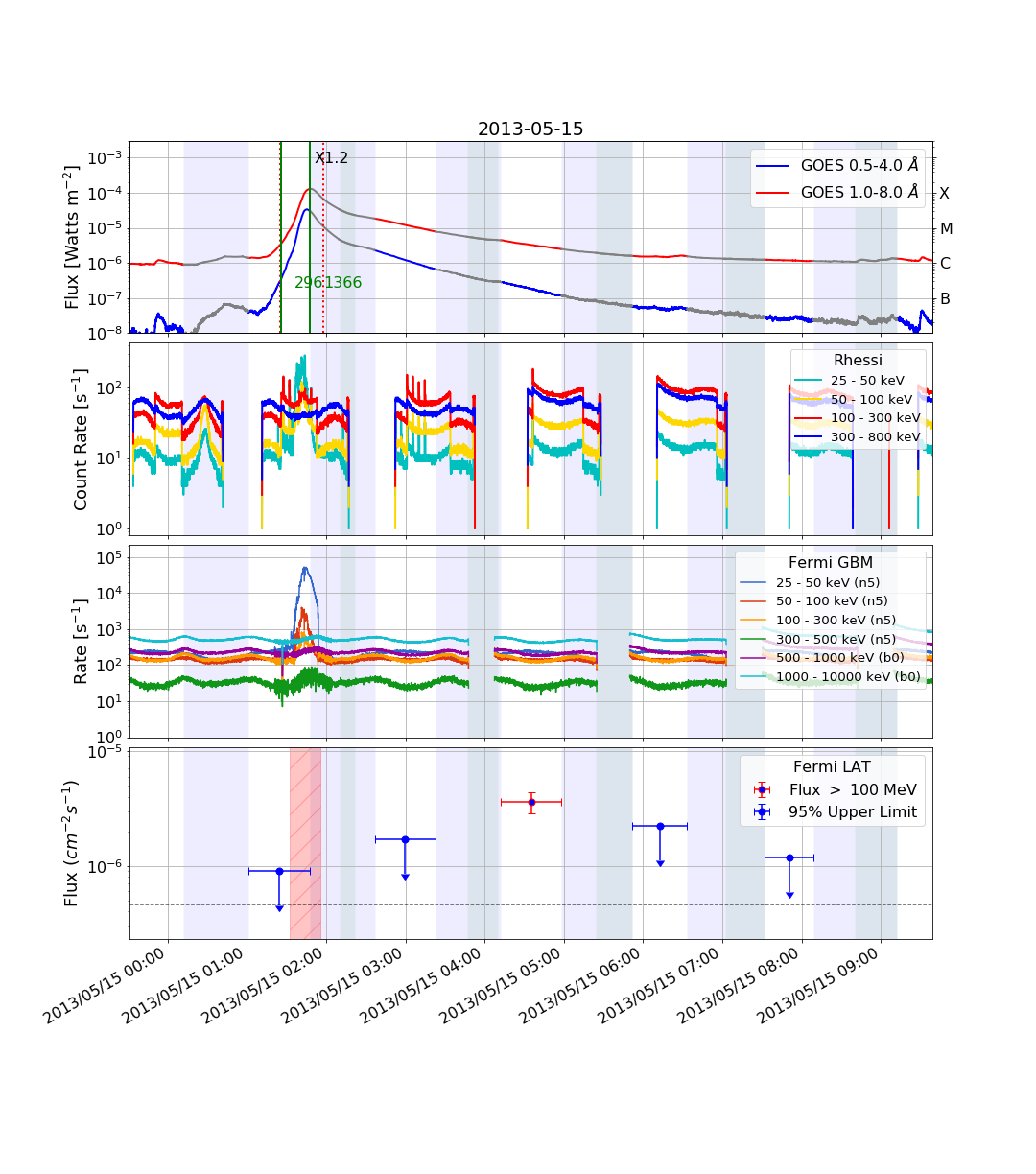} 
\caption{The {\it delayed-only} {lightcurve of the $>$100 MeV emission from} FLSF 2013-05-15 flare with no detectable high-energy $\gamma$-ray emission in the impulsive phase, or the following time window. The four panels report the same quantities as those in Figure~\ref{fig:Flares_delayed_1}.}   
\label{fig:Flares_delayed_2}   
\end{center}   
\end{figure*}

A large number of \solar flares observed by the \Fermi-LAT do not fall in the \emph{prompt} category: $\g$-ray emission is detected beyond the end of the HXR emission and even the end of the SXR seen by GOES. We refer to that general category as {\it delayed} emission.
The subset of flares classified as \emph{delayed} also exhibit a wide variety of behaviors. For example there are cases where no significant $\g$-rays are detected during the prompt phase of the flare in X-rays, but $\g$-ray emission seen rising and falling later on. We refer to these flares as being {\it delayed-only}.

One of the most interesting results of the \Fermi-LAT observations of \solar flares are the events {with detectable emission lasting several hours.} As already discussed in Section~\ref{sec:intro}, the LAT has the Sun in its {FoV} on average only 40\% {of its orbit, greatly limiting the coverage} of these  \emph{delayed} $\gamma$-ray flares. As a result it is difficult to study the time profiles of these flares throughout the entire duration of the emission. 

This is the case of the FLSF~2012-03-09, which is associated with a GOES M6.3 flare with HXR extending up to the GBM NaI 100--300~KeV channel. 
Most of the prompt phase was observable by the \Fermi-LAT and the bright SXR affected the instrument response ({bad time interval} in red in Figure~\ref{fig:Flares_delayed_1}). No $\g$-ray emission was detected during the peak of the prompt phase using the S15 event class or the LLE analysis method.
Yet $\g$-ray emission was detected when the Sun came back in the FoV, almost two hours after the start of the flare in X-rays, and lasted for four orbits. It followed a rise and fall pattern reaching its peak after four hours 
and ending 7 hours after the start of the flare in X-rays.

Similarly, the FLSF~2013-05-15  with no significant emission detected during either the impulsive phase or in the first time window following the flare, but significant emission detected in the following time window (Figure~\ref{fig:Flares_delayed_2}). 
 In itself, it might not be a new type of behavior, as it can be seen as a rise-and-fall pattern with the starting flux being just below the \Fermi-LAT sensitivity but the peak flux being high enough to be detected.

These behaviors highlight the possibility that high-energy emission above 100 MeV can arise at later times, even if the prompt phase itself did not show a strong non-thermal component  (almost no HXR above 300 keV and no $\g$-rays below 30 MeV).
Although these cases are rare {(only four cases in the catalog)}, they are particularly interesting in understanding whether the acceleration of high energy particles is solely due to the prompt phase of \solar flares or due to a separate mechanism entirely.

There are also {FLSFs} with both a clear prompt and a long duration delayed component present, these flares are classified as \emph{prompt-delayed}. An example of this class of flares is the {FLSF~2017-09-10~\citep{2018ApJ...865L...7O} that exhibited a very bright prompt phase and almost 14 hours of delayed $\gamma$-ray emission.} In the FLSF catalog we were able to classify the flares into six different categories: \emph{prompt}, \emph{prompt-only}, \emph{delayed}, \emph{delayed-only}, \emph{prompt short-delayed}, \emph{prompt-delayed}. All the lightcurves and categories of the FLSFs are reported in~\cite{ajello_m_2020_4311157}.

\section{Results}
\label{sec:results}

Continuous monitoring of the Sun has led to the high-confidence (TS$\geq$30) detection of 45 \solar flares with $\g$-ray emission above 60 MeV. {For 39 of these flares $\gamma$-ray emission was significant in 92 \SunMon time windows. 
The remaining 6} {flares were detected with LLE analysis only.}
Of these 45 flares, 6 are classified as \emph{prompt-only}, 4 are classified as \emph{delayed-only} and for 10 flares both the \emph{prompt} and \emph{delayed} emission was clearly observed by \Fermi-LAT. For the remaining cases, we cannot exclude the presence of a \emph{prompt} emission because the Sun was not in the {FoV} of the LAT during the HXR activity. Because of the observing strategy of the \Fermi-LAT, more than half of the \solar flares detected are only detected in a single time window, whereas 16 are detected in more than one window. Of the 16 flares detected in multiple time windows 5
are detected in only 2 time windows, and 11 are detected in 3 or more (up to 11) time windows well beyond the HXR signatures of the high-energy electrons. Seven flares in the latter group show a well-defined pattern of rise and decay phases after the end of the HXR  and 2 show a decay phase only. All 5 flares detected in 2 time windows show a decay between the two points. {Some of these may represent a rise and fall case 
with a peak occurring in between the two time windows. However, this is unlikely because statistically one would expect 2 or 3 of these flares showing rise instead of decay, and because this would imply faster rise and fall than seen in the flares with more than three windows of observation.}


In Table~\ref{tab:time_integrated_main_table} we show the time integrated results for the {FLSFs} detected with the \SunMon. The columns report the {report the LAT detection start date and time, the GOES soft X-ray start and end times, the LAT detection duration,} the total duration of the FLSF%
\footnote{The detection duration is simply the sum of the \SunMon detection windows duration while the total duration is that found using the approach described in section \ref{sec:total_emission}.}, 
{the fluence, namely the time-integrated flux over the total duration}, the FLSF flare type, and the total number of accelerated $>$500 MeV protons (N500). {The GOES classes for the three}  {BTL} flares (identified by  *) {are} estimated based on STEREO UV fluxes as described in~\cite{2015ApJ...805L..15P}.

The characteristics of the $\gamma$-ray emission in each \SunMon time window are listed in Table~\ref{tab:gamma_ray_flare_windows}. {Results from flares detected in more than one time window are listed together.}  The columns of  Table~\ref{tab:gamma_ray_flare_windows} are the {time of each {detection} 
window, the duration of the 
window, the $>$100 MeV flux, TS, and the spectral parameters (power-law indices and cutoff energies) of the best fitting photon model.} For the cases where the $\Delta$TS$>$9 we give the proton index based on pion-decay model {in the last column}. The fluxes are given in $10^{-5}$~{ph} cm$^{-2}$s$^{-1}$ and calculated for the emission between 100 MeV and 10 GeV. {The LAT emission in all \SunMon time windows with TS larger than 70 shows significant spectral curvature and can be well described with the exponential cutoff model. This does not mean that all fainter $\gamma$-ray flares are only consistent with a power law model, but rather that the lower statistics make it impossible to distinguish between the two.}%
\footnote{The FLSF of 2013-10-28 is the only exception, having a TS of 120 and the exponential cutoff model is not preferred ($\Delta$TS=8).}

 {We retract the LAT  detection of the C-class flare on 2011-06-02 reported in \cite{2014ApJ...787...15A}, because  
 during} {the month of June, the Sun passes through the galactic plane, and a higher background flux of photons enters into the RoI around the Sun relative to other periods in the year. After careful re-analysis of this event, we found that the reported detection was not statistically significant.} 

The FLSF LLE catalog results are reported in Table~\ref{tab:lle_main_flare_table}. {Three of the flares detected with LLE were outside the nominal LAT FoV.} For the eleven flares in the FoV,
five were not detected above 60 MeV by the \SunMon { analysis, and an upper limit} was obtained for the time window when the flare happened. For the six flares detected
{with both analyses in the same time window, the $>$100 MeV fluxes reported} in the \SunMon results (Table~\ref{tab:time_integrated_main_table}) {are the average over the time window, and the $>$100 MeV fluxes obtained through} the LLE approach are listed in Table~\ref{tab:lle_main_flare_table}.

\begin{figure*}[ht!]
\begin{center}
\includegraphics[width=0.45\textwidth,trim=50 10 100 50, clip=true]{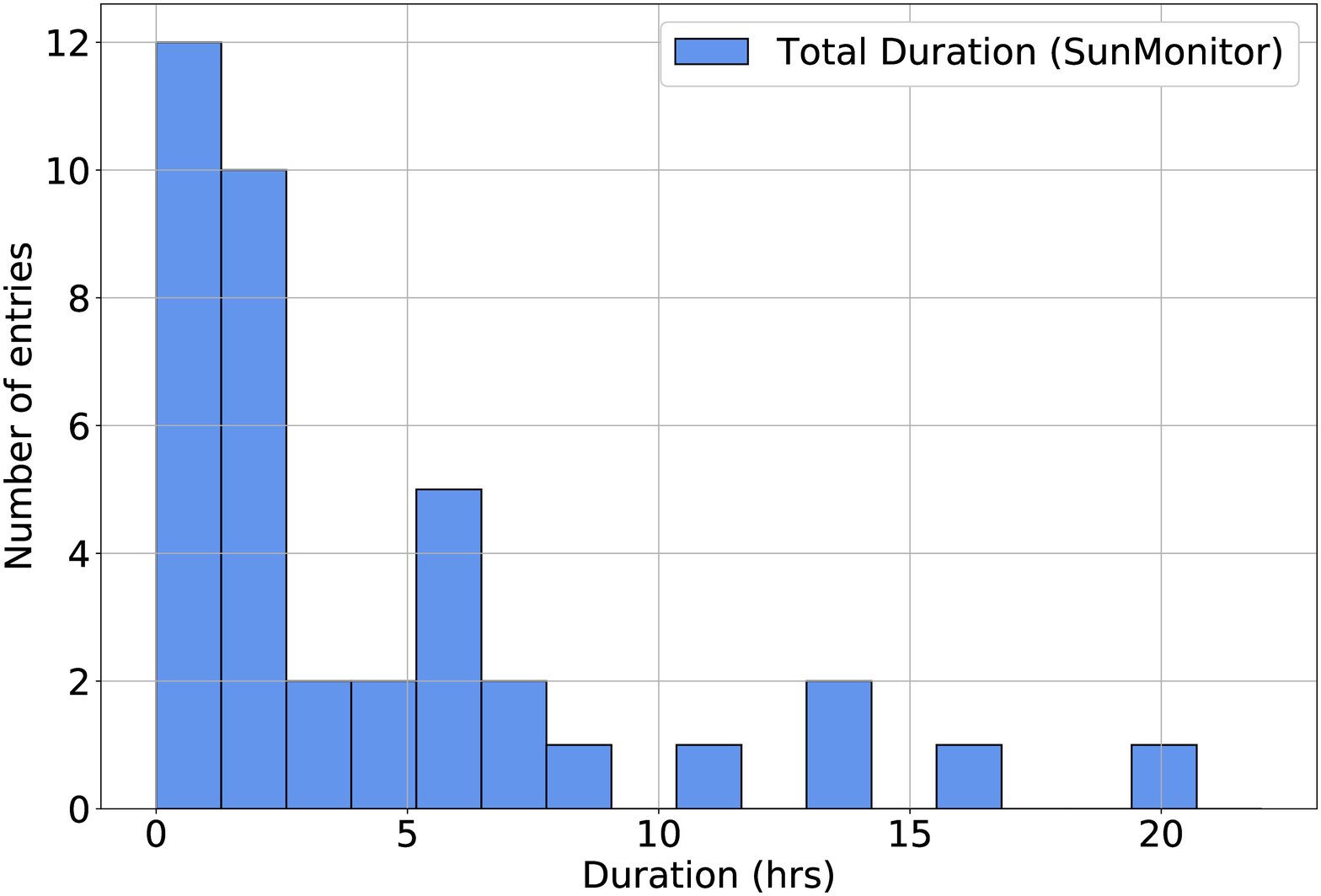}
\includegraphics[width=0.45\textwidth,trim=50 10 100 50, clip=true]{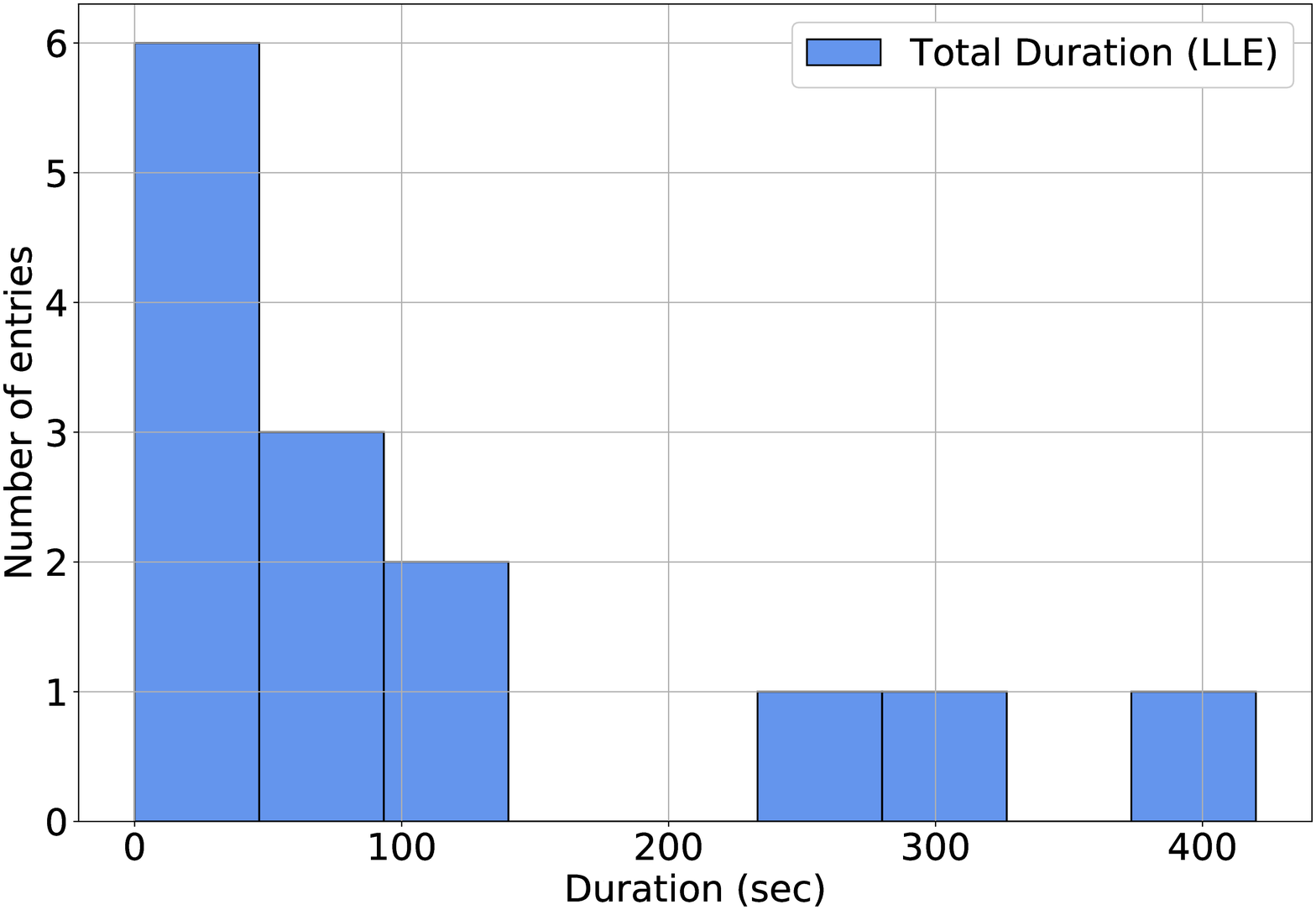}
\end{center}
\begin{center}
 \caption{Distribution of the total duration for all of the \SunMon detected flares (in hours, left panel) and the LLE detected flares (in seconds, right panel) in the FLSF catalog.}
\label{fig:Catalog_durations} 
\end{center}   
\end{figure*}

\begin{figure*}[ht!]   
\begin{center}   
\includegraphics[width=0.45\textwidth,trim=50 10 100 50, clip=true]{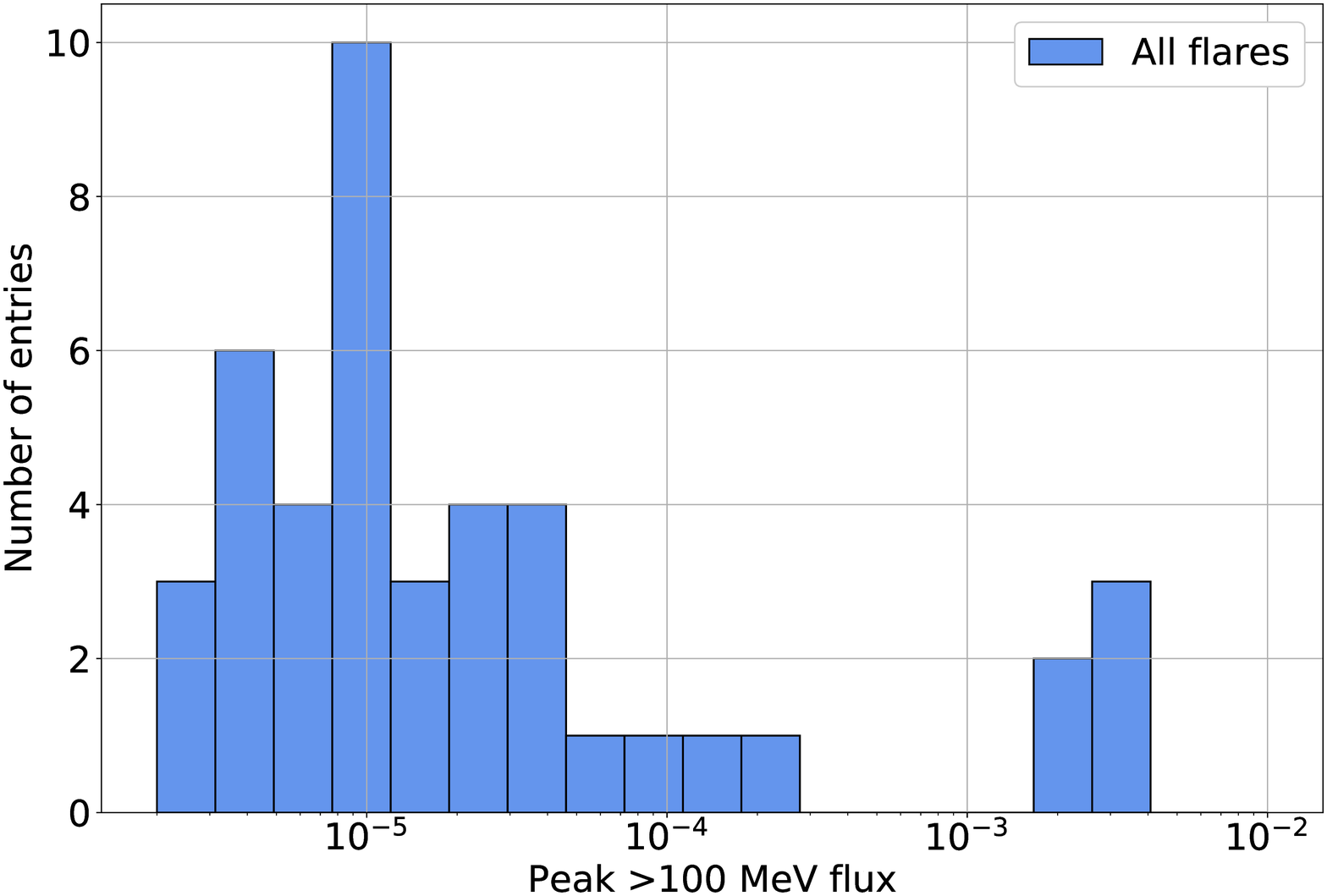}
\includegraphics[width=0.45\textwidth,trim=50 10 100 50, clip=true]{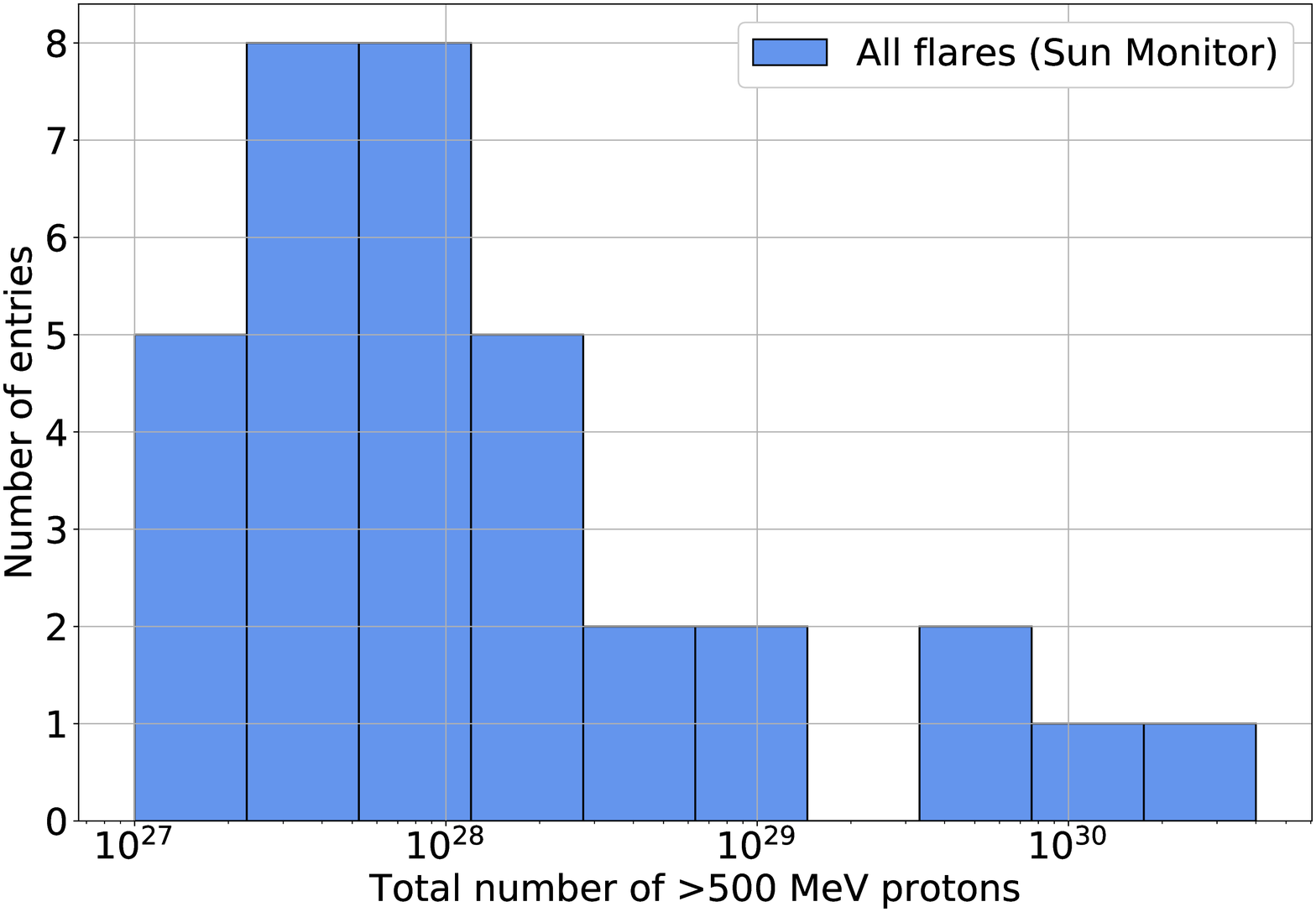}
\end{center}
\begin{center}
\caption{Distributions of the peak $>$100 MeV flux ({in ph cm$^{-2}$ s$^{-1}$}; left panel) for all  {FLSFs} in the catalog, and the {total number of accelerated $>$500 MeV protons needed to} produce the detected $\gamma$-ray emission for each of the \SunMon detected {FLSFs} (right panel).}
\label{fig:Catalog_flux_protons} 
\end{center}   
\end{figure*}

The {durations for the flares detected with the \SunMon range from 0.6 to 20.3 hours, whereas the} LLE detected 
flares have durations ranging from 10 to 400 seconds (see Figure~\ref{fig:Catalog_durations}). Both the $>$100~MeV peak $\gamma$-ray fluxes and total number of $>$500~MeV protons needed to produce the observed $\gamma$-ray emission for all of the {FLSFs} in the catalog span over four orders of magnitude (see Figure~\ref{fig:Catalog_flux_protons}). 

\begin{figure*}[ht!]   
\begin{center}   
\includegraphics[width=\textwidth,trim=50 0 50 50,clip]{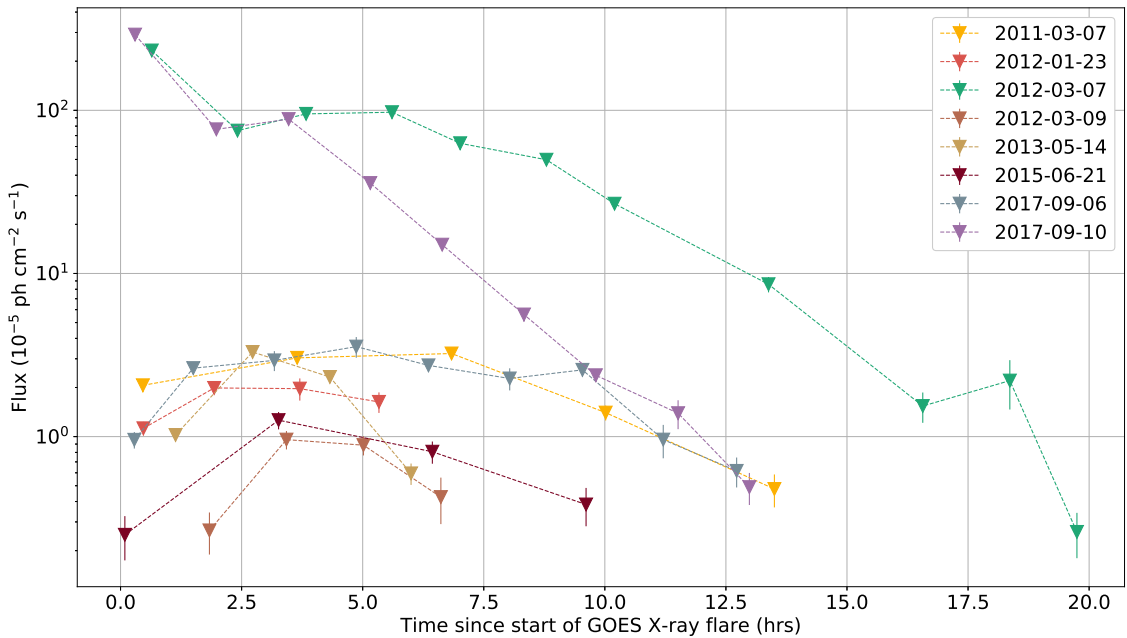}
\caption{{The time profiles of} flux between 0.1 - 10 GeV 
for {for each FLSF lasting two or more hours versus the time} since the start of the GOES X-ray flare. The typical rise and fall behavior of the $\gamma$-ray emission during the \emph{delayed} phase is most evident for the cases where no \emph{prompt} emission was present during the detection.}
\label{fig:flux_time_since_goes} 
\end{center}   
\end{figure*}

Eight of the 45 FLSFs have durations of two hours or more. Their $>$100 MeV fluxes as a function of time (since the start of the associated GOES X-ray flare) are shown in Figure~\ref{fig:flux_time_since_goes}. The time profiles of all these \emph{delayed} {FLSFs}  follow  a rise-and-fall behavior. However, the rise times to reach the peak flux and the fall times vary significantly 
from flare to flare. For example, the FLSF~2017-09-10 has a rise time of $\approx$1.5 hours while the FLSF~2017-09-06 takes $\approx$4.5 hours to reach its peak. The peak flux values also vary from flare to flare by up to two orders of magnitude, emphasizing the wide variety of these \emph{delayed} flares. The two brightest flares in Figure~\ref{fig:flux_time_since_goes} were coincident with very strong SEP events; Ground Level Enhancement (GLE)\#72 in the case of the FLSF~2017-09-10 and a sub-GLE event in the case of the FLSF~2012-03-07\footnote{GLEs are sudden increases in the cosmic ray intensity recorded by ground based detectors. The number following the GLE indicates the number of GLEs that have been observed since 1956, see GLE database \url{http://gle.oulu.fi} for more details.}.  {Coincidentally,} the $\gamma$-ray fluxes for these {two} flares are more than an order of magnitude higher than the other events. In Table~\ref{tab:time_integrated_mw_main_table} we list some multiwavelength associations with the FLSFs presented in this work. In particular, we include GOES X-ray flares, CMEs, SEPs and Hard X-ray counterparts to the gamma-ray flares.

\begin{figure*}[ht!]   
\begin{center}   
\includegraphics[width=\textwidth,trim=50 0 50 50,clip]{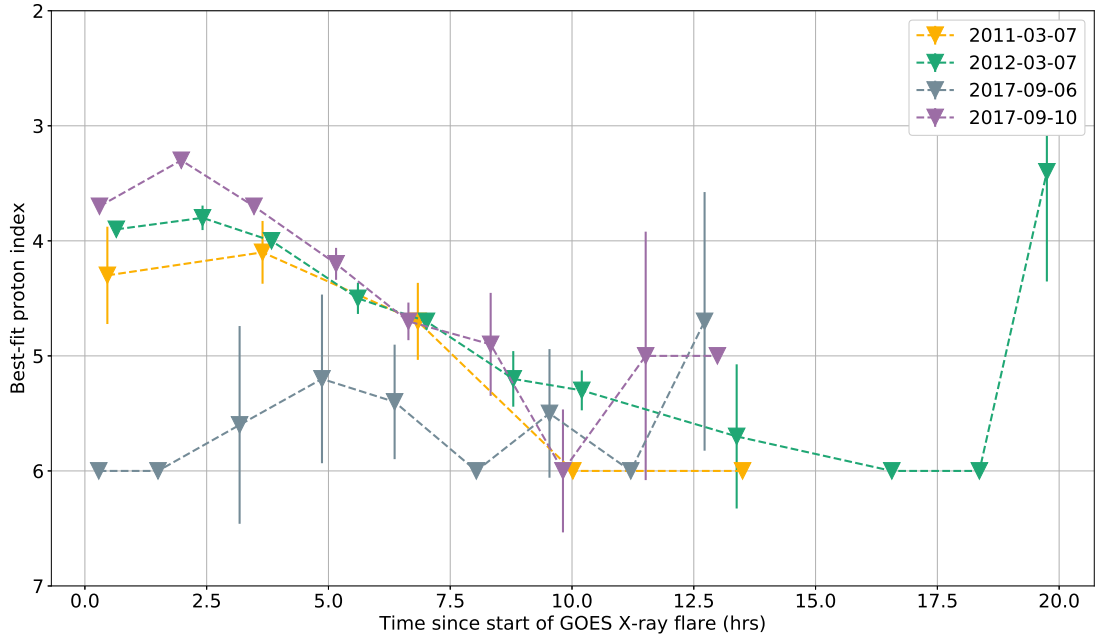}
\caption{{Variation with time (since start of the GOES X-ray flare)} 
of best-fit proton spectral index for the  four FLSFs for which a statistically meaningful measurement can be made.}
\label{fig:Catalog_proton_indexes_variations} 
\end{center}   
\end{figure*}

For the {FLSFs} with more than 4 \SunMon detection windows, it is possible to study the variation of the proton index with time. {In Figure~\ref{fig:Catalog_proton_indexes_variations} we show the accelerated proton spectral index as a function of time since} the start of the GOES X-ray flare {(assuming that the $\gamma$-rays emission is due to pion decay)}. The statistical uncertainties limit the amount of information available from the time variation of the proton indices. {However, the data suggest that 
the proton spectra  tend to gradually steepen (get softer), following a trend similar to the $\gamma$-ray fluxes 
for these delayed flares.} 

\begin{figure*}[ht!]   
\begin{center}   
\includegraphics[width=0.8\textwidth,trim=10 100 50 50,clip]{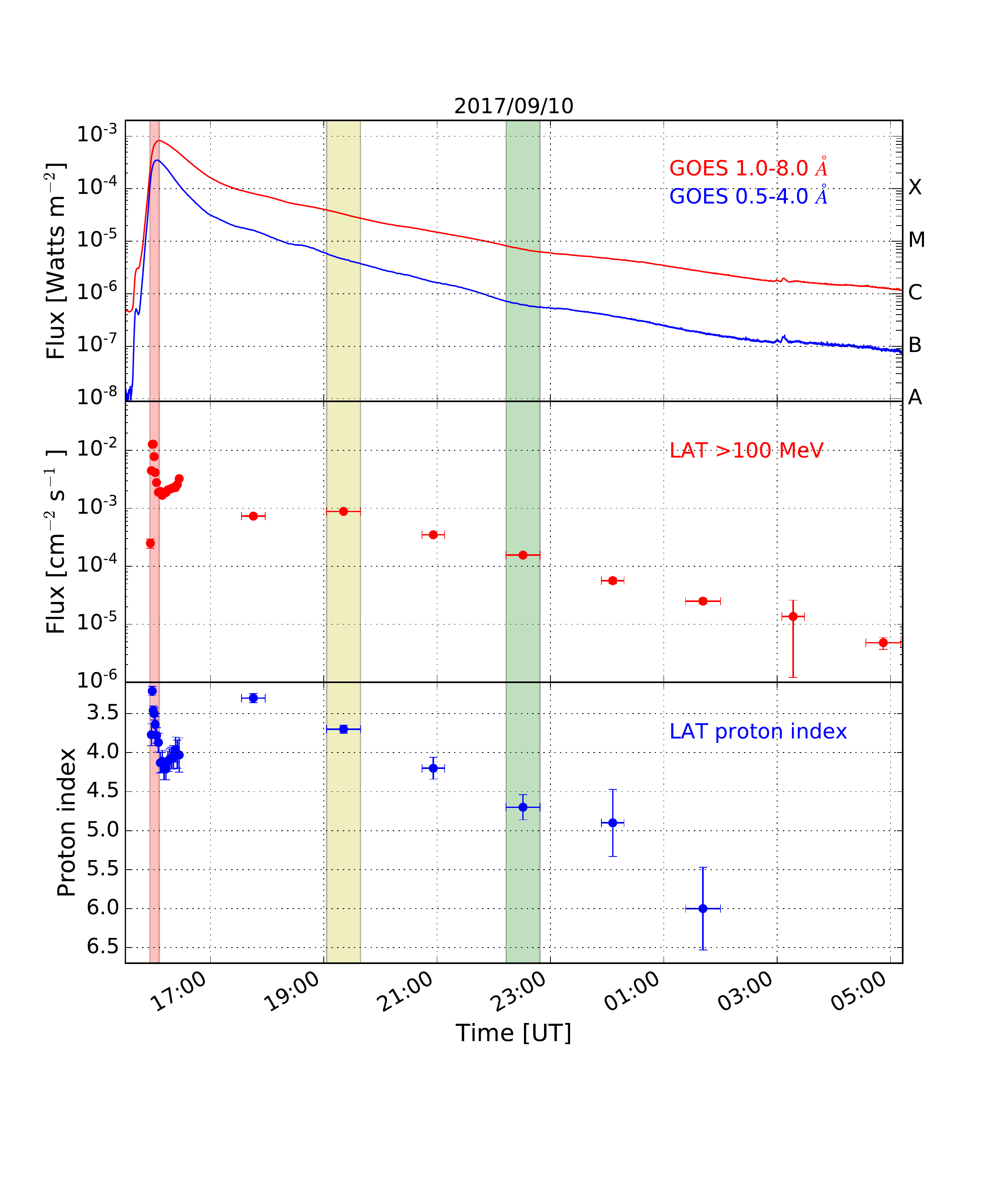}
\caption{Composite light curve for the FLSF~2017-09-10 with data from GOES X-rays, \Fermi-LAT $>$100 MeV flux and the best proton index inferred from the LAT $\gamma$-ray data. The figure is taken from~\cite{2018ApJ...865L...7O}.  The evolution of the proton index shows three distinct phases, {a softening during the prompt-impulsive phase, a plateau and another softening during the decay phase.} 
The three color bands represent the time windows over which we performed the localization of the emission.}
\label{fig:SOL20170910_protons_lc} 
\end{center}   
\end{figure*}

For the extremely bright FLSF~2017-09-10, both the prompt and delayed phases were well observed by the {LAT, and we are} not limited by statistics. The data from this flare show three phases in  the evolution of the proton index over the almost two hours of $\gamma$-ray emission (see Figure~\ref{fig:SOL20170910_protons_lc}). This flare was also associated with GLE $\#$72. and \cite{Kocharov_2020} show that these phases correspond to separate components of the GLE.

Solar cycle 24 has been particularly poor in {GLE} events. 
Only two have been firmly
identified: GLE $\#$71 and $\#$72, which occurred on 2012-05-17 and 2017-09-10. Both events were detected with the \Fermi-LAT. In addition to GLEs, five
{``sub-GLE"} events have been identified. Sub-GLE events are those detected only by high-elevation neutron monitors and correspond to less energetic events, extending to a few hundred
MeV~\citep{Poluianov2017}. They occurred on 2012-01-27, 2012-03-07, 2014-01-06, 2015-06-07, 2015-10-29 at levels of relative increase {in neutron flux} of 5\%, 5\%, 4\%, 8\% and 7\%, respectively
(smaller than the relative increase of 17\% for GLE\#71). The first three correspond to flares in the FLSF catalog, but no emission was detected  
for the last two.

\begin{figure*}[ht!]   
\begin{center}   
\includegraphics[width=0.7\textwidth,trim=10 10 100 50,clip]{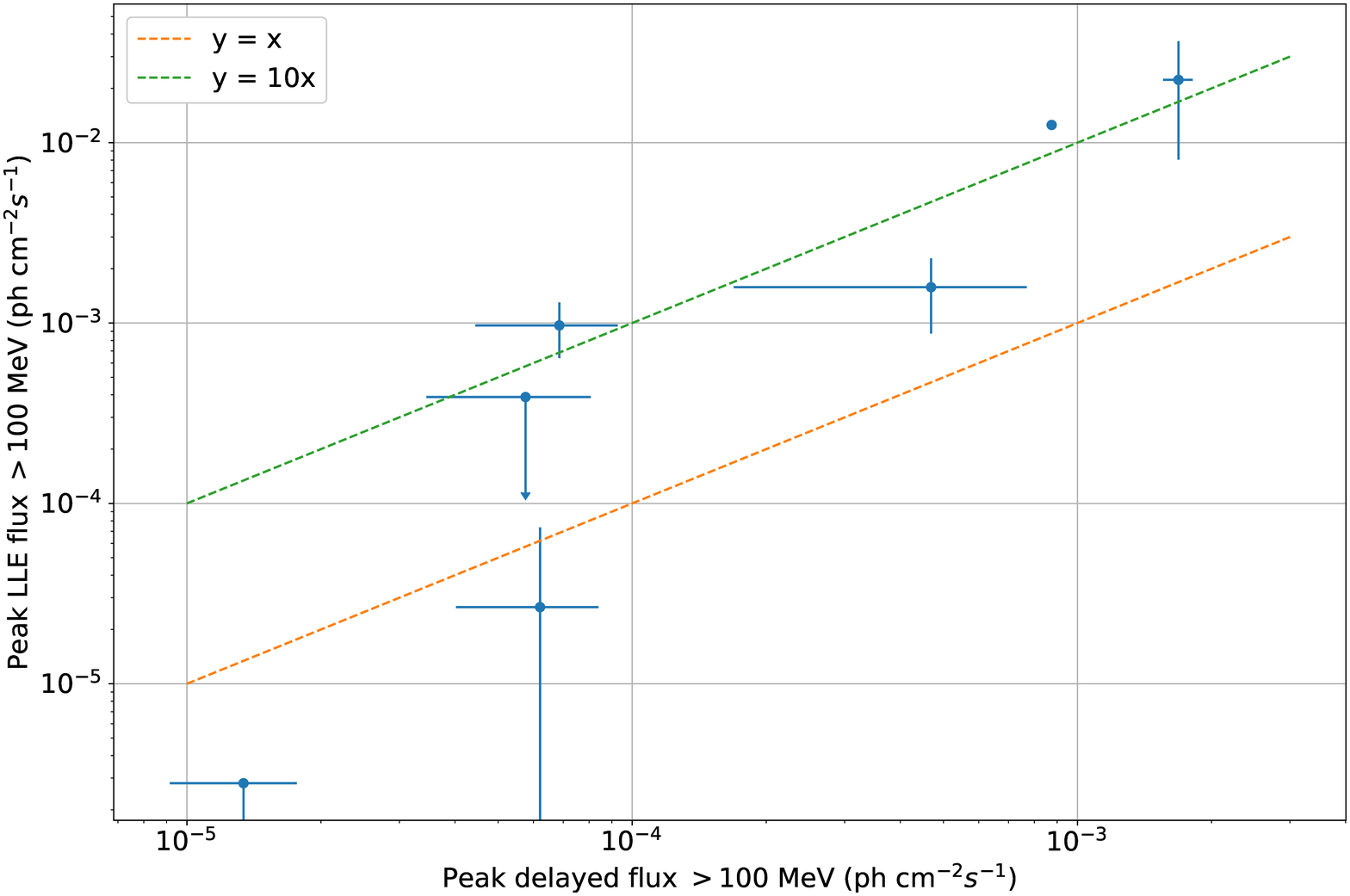}
\includegraphics[width=0.7\textwidth,trim=10 10 100 50,clip]{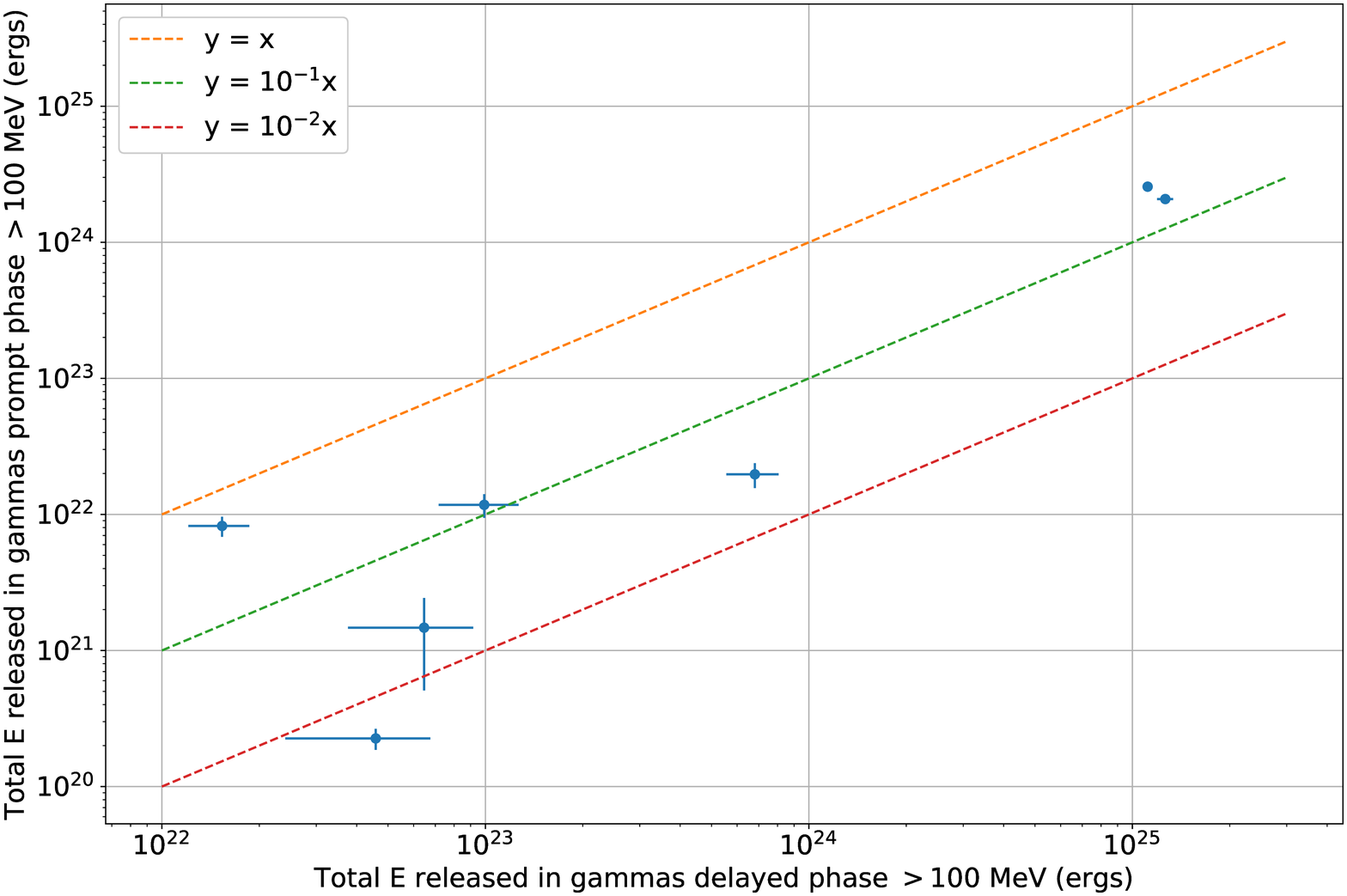}
\caption{Scatter plot of the peak flux during the prompt phase versus the peak flux during the delayed phase for the 7 {FLSFs} with both the prompt a delayed phases observed fully.  The prompt peak fluxes {tend to be higher than those during the delayed phase, in some cases up to more than 10 times}. Bottom panel: Scatter plot of the total energy released in $\gamma$-rays above 100 MeV during the prompt and delayed phases. The total energy released during the delayed phase is {on average about 10 times} larger than the prompt phase.} 
\label{fig:prompt_vs_delayed} 
\end{center}   
\end{figure*}

Flares with both the LLE-prompt and delayed phases detected by the LAT allow a comparison of the  prompt and delayed emission characteristics within the same flare. Seven flares in the catalog (2011-09-06, 2011-09-24, 2012-06-03, 2012-10-23, 2014-02-25, 2014-06-10 and 2017-09-10)
satisfy this {criterion.  For these flares we found the peak flux value for the prompt phase by fitting the LLE data at the peak of the lightcurve with two models: a simple power-law  or a power-law with an exponential cutoff using the \texttt{xspec} analysis package%
\footnote{\texttt{xspec} model \texttt{pegpwrlw} and \texttt{pegpwrlw*highecut}}.} The correlation between the peak fluxes of the prompt and delayed phases are shown in the top panel of Figure~\ref{fig:prompt_vs_delayed}  illustrating that, on average, the prompt peak flux is {up to} 10 times higher than the {peak of the delayed emission}. 
 The bottom panel of this figure shows the correlation between the total {$\gamma$-ray} energies ($>$~100 MeV), showing a larger dispersion and {a total} 
energy released during the delayed phase that, {on average  about 10  times larger} than {that in the prompt phase}.

\begin{figure*}[ht!]   
\begin{center}   
\includegraphics[width=0.7\textwidth,trim=70 10 100 80,clip]{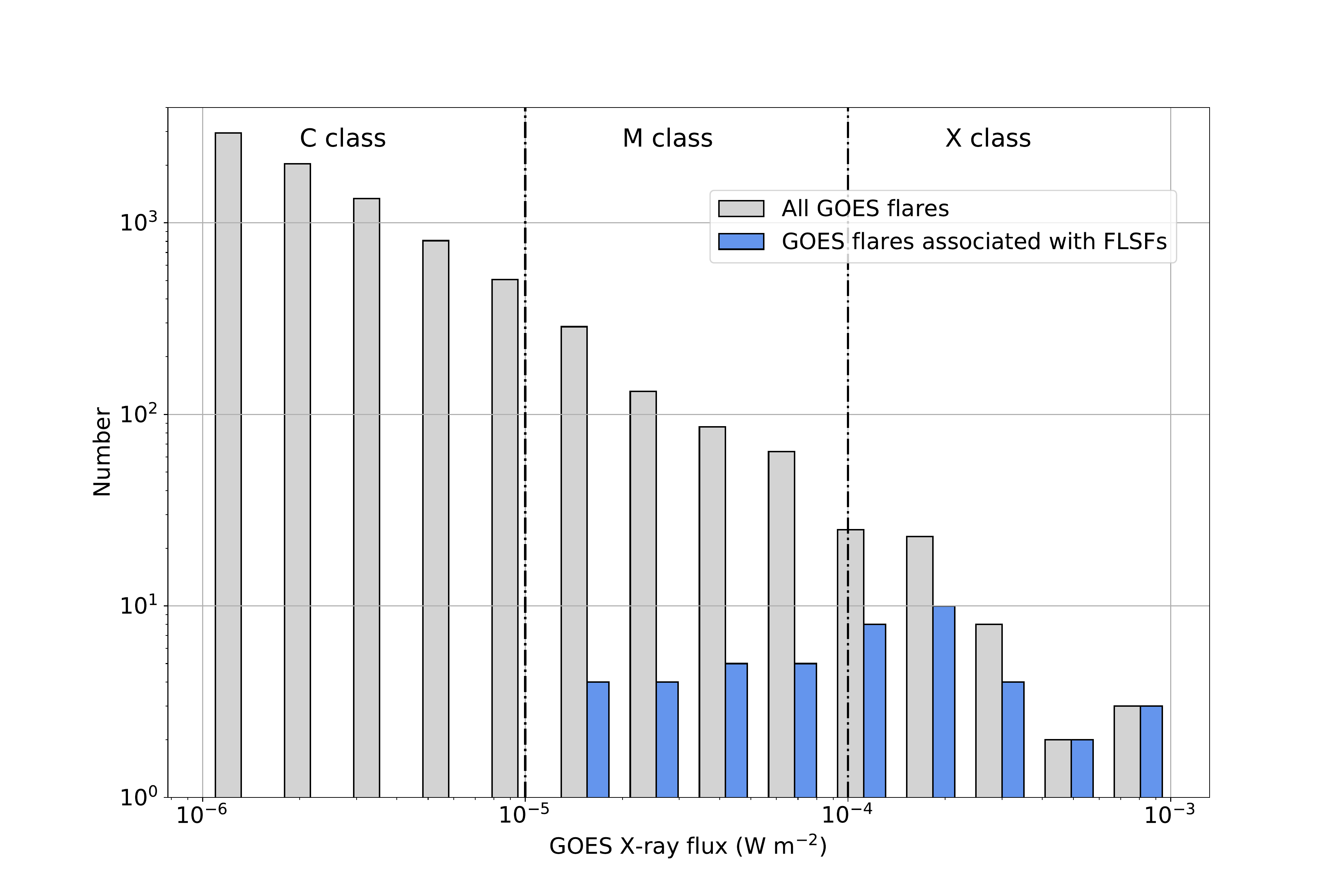}
\includegraphics[width=0.7\textwidth,trim=70 10 100 80,clip]{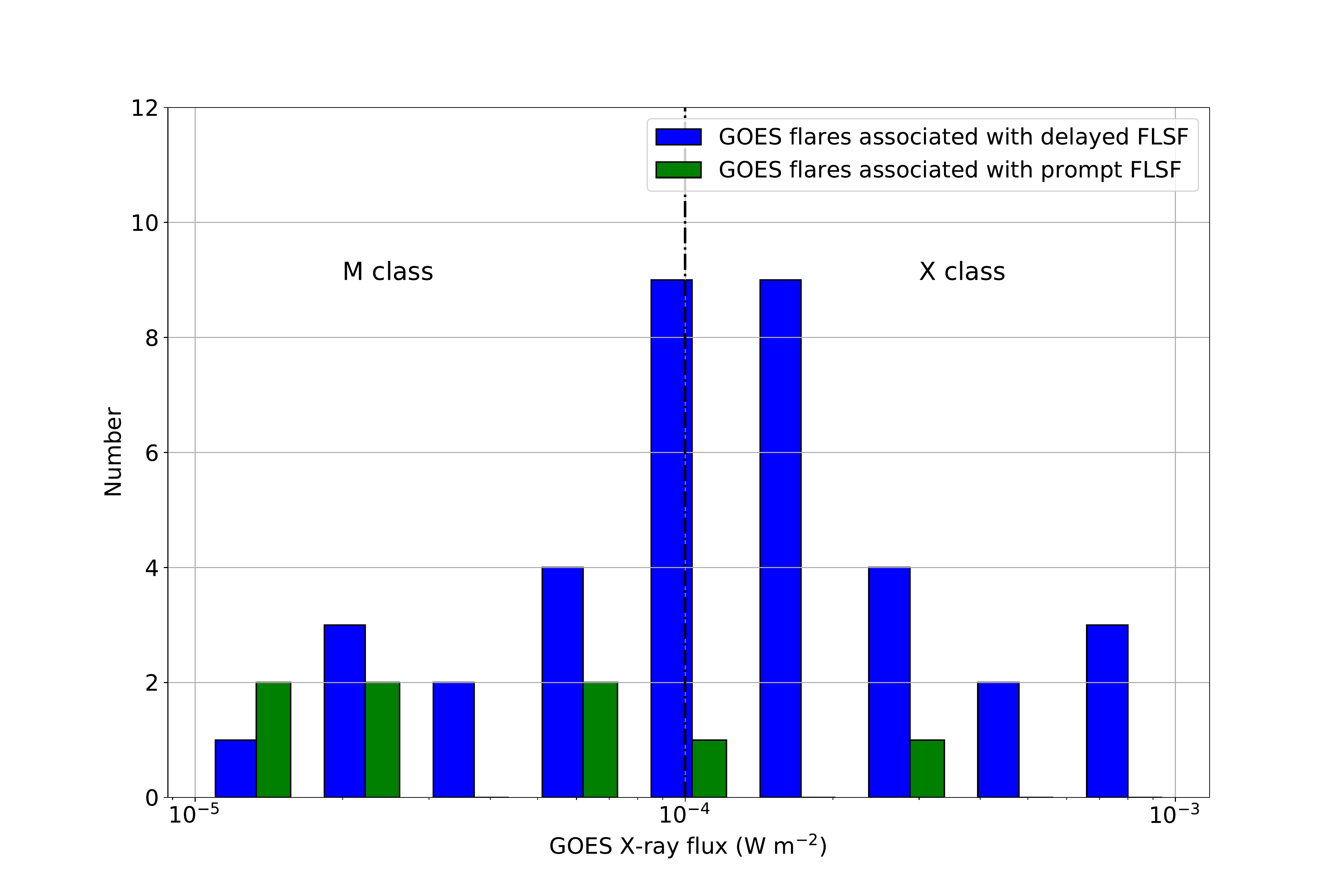}
\caption{{Top panel: Distribution of the GOES class for all of the X-ray flares of \solar cycle 24 (in gray) and for the FLSFs  (light blue). Bottom panel: Distribution of the GOES class for the FLSFs separated by type \emph{delayed} (blue) and \emph{prompt} flares (green).}}
\label{fig:Catalog_goes_flares} 
\end{center}   
\end{figure*}

\begin{figure*}[ht!]   
\begin{center}   
\includegraphics[width=0.7\textwidth,trim=10 10 100 50,clip]{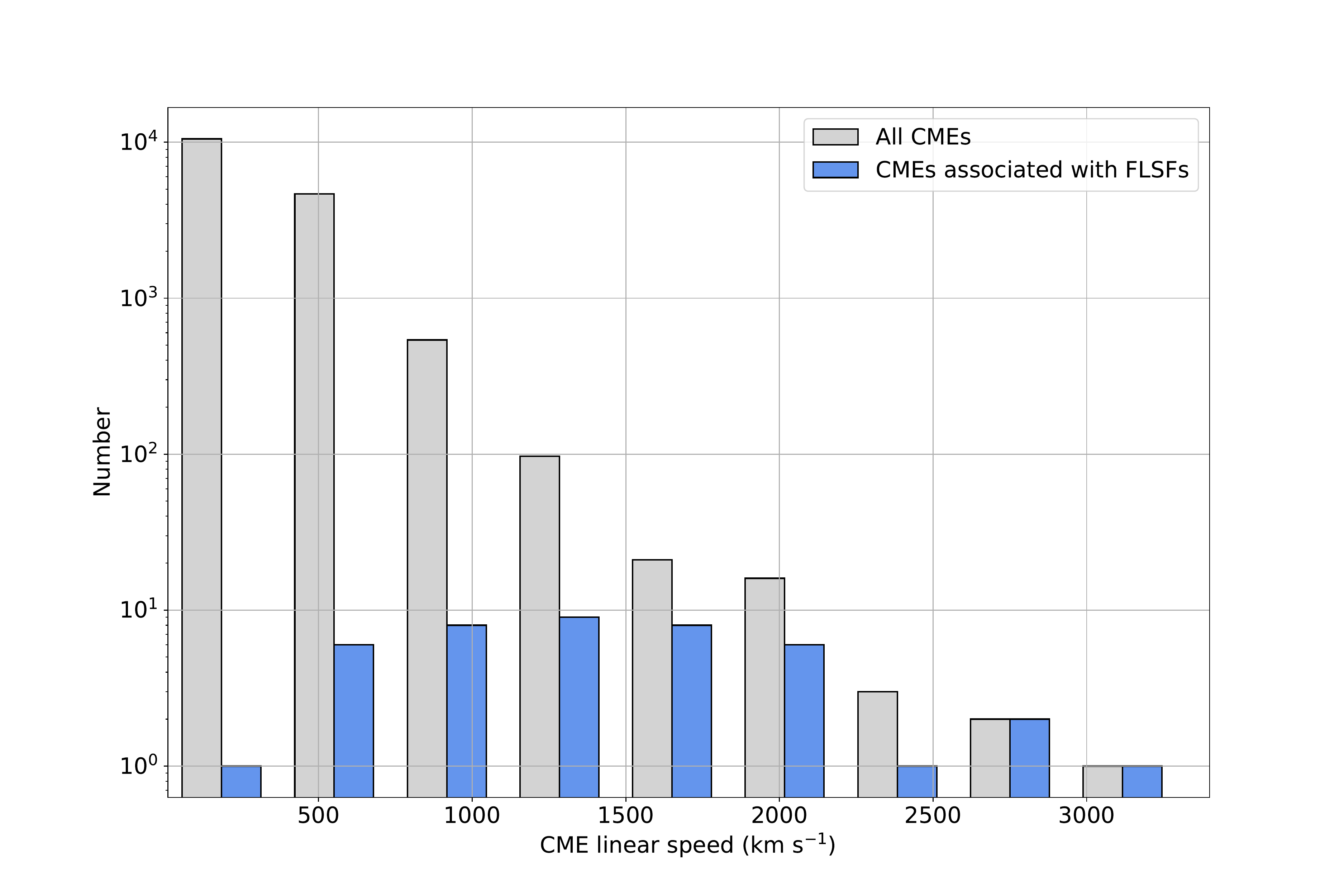}
 \includegraphics[width=0.7\textwidth,trim=10 10 100 50,clip]{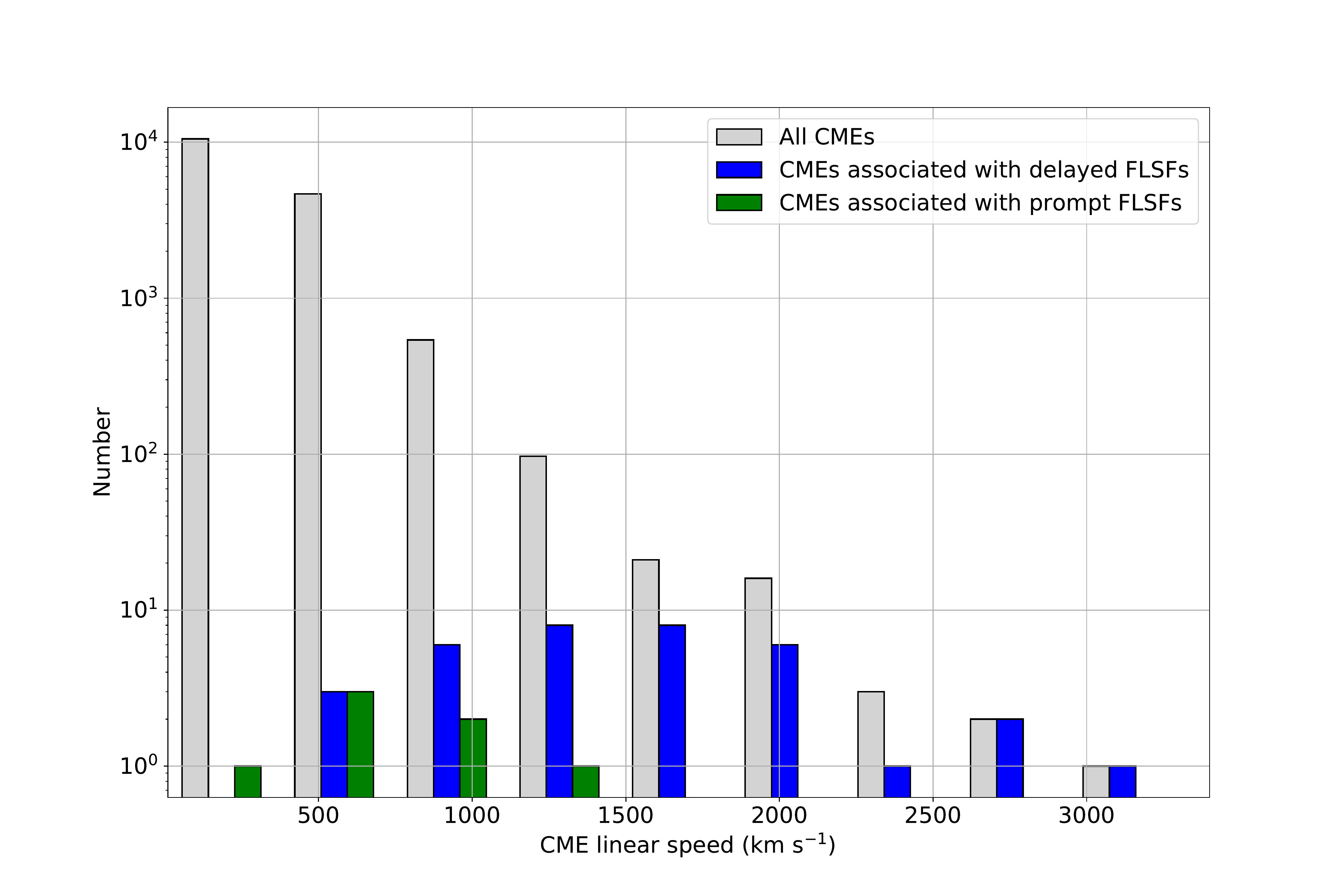}
\caption{{Top panel: Distribution of the CME linear speed for all of \solar cycle 24 (in gray) and for all the {FLSFs} in this work (light blue). Bottom panel: Distribution of the CME linear speed for FLSFs classified as delayed (blue) and FLSFs classified as prompt (green). The mean speed for the \emph{delayed} flares is 1535 km\,s$^{-1}$  and for the \emph{prompt} {flares} is 656 km\,s$^{-1}$. As in top panel, the gray histogram represents the CME linear speed for all of the CMEs of \solar cycle 24 (whose mean speed is 342 km\,s$^{-1}$).}}
\label{fig:Catalog_cme_speed_delayed_prompt} 
\end{center}   
\end{figure*}

The FLSFs in the catalog are almost evenly distributed between GOES M and X-class flares (in the 0.5 to 10 keV energy range), with 25 flare associated with X-class and 20 associated with M-class (see top panel of Figure~\ref{fig:Catalog_goes_flares}, where the gray distribution represents all of the M and X-class GOES flares that occurred during the time period considered in this paper). 
{As can be seen in the bottom panel of Figure~\ref{fig:Catalog_goes_flares}, the FLSFs of \emph{delayed} type are evenly distributed between the M and X-class flares while the \emph{prompt} type flares are mostly associated with M GOES class flares (75\% of the flares are M class).
These distributions also illustrate how the increase in sensitivity of the LAT with respect the previous $\gamma$-ray detectors has allowed to detect $>$100~MeV emission over a wider range of GOES X-ray flares. Furthermore, when combining the information from Figures \ref{fig:Catalog_goes_flares} and \ref{fig:Catalog_cme_speed_delayed_prompt} it appears that the presence of a fast CME is more relevant for the \emph{delayed} type flares than the brightness of the associated X-ray flare. }

\begin{figure*}[ht!]   
\begin{center}    
\includegraphics[width=\textwidth,trim=50 10 80 80,clip]{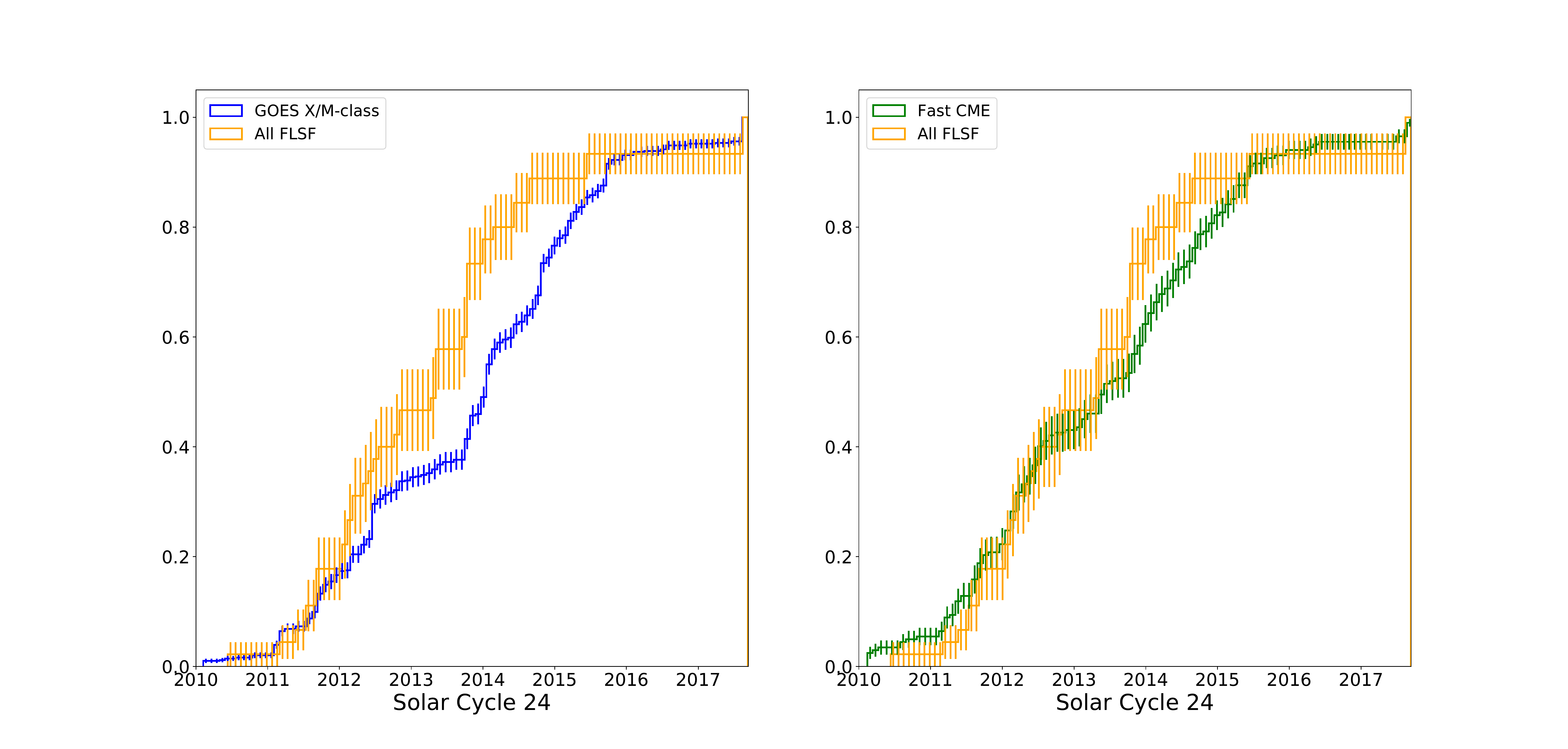} 
\caption{Cumulative number of FLSFs as a function of time compared with the distribution for M/X class GOES flares (left) and fast CME (linear speed $>$1000 km s$^{-1}$) events (right).}
\label{fig:KS_ARposition}   
\end{center}   
\end{figure*}

{During Cycle 24, the number of GOES M-class and X-class flares {in the period covered by this catalog (January 2010 - January, 2018)} was approximately the same in the first half as in the second (384 and 389, respectively), while the majority of fast CME events (those with speed $>$1200 km\,s$^{-1}$) happened in the earlier half 
(January 2010 -- January 2014) (61 vs 35). 
Similar behaviour was observed for major SEP events (30 in the first half and 12 in the second half of the Cycle). 
Interestingly, the number of {FLSFs} is also larger in the first half of the Cycle, with 33 flares, while only 12 occurred in the second half. To quantify this behavior we show in Figure~\ref{fig:KS_ARposition} the cumulative distributions of XRT flares and fast CME (linear speed $>$1000 km s$^{-1}$) events compared with the distribution of {FLSFs}. The latter seems to be in much better agreement with the distribution of fast CME events, with a  Kolmogorov-Smirnov test p-value of 0.15, while the comparison of XRT flares with FLSFs gives a p-value of 4.6$\times10^{-4}$. This result is also suggesting that high-energy \solar flares have a stronger association with fast CME rather than with bright X-ray flares.}

\subsection{FLSF Active Region Positions}
\label{ARpositions}

\begin{figure*}[ht!]   
\begin{center}    
\includegraphics[width=\textwidth,trim=10 50 100 50,clip]{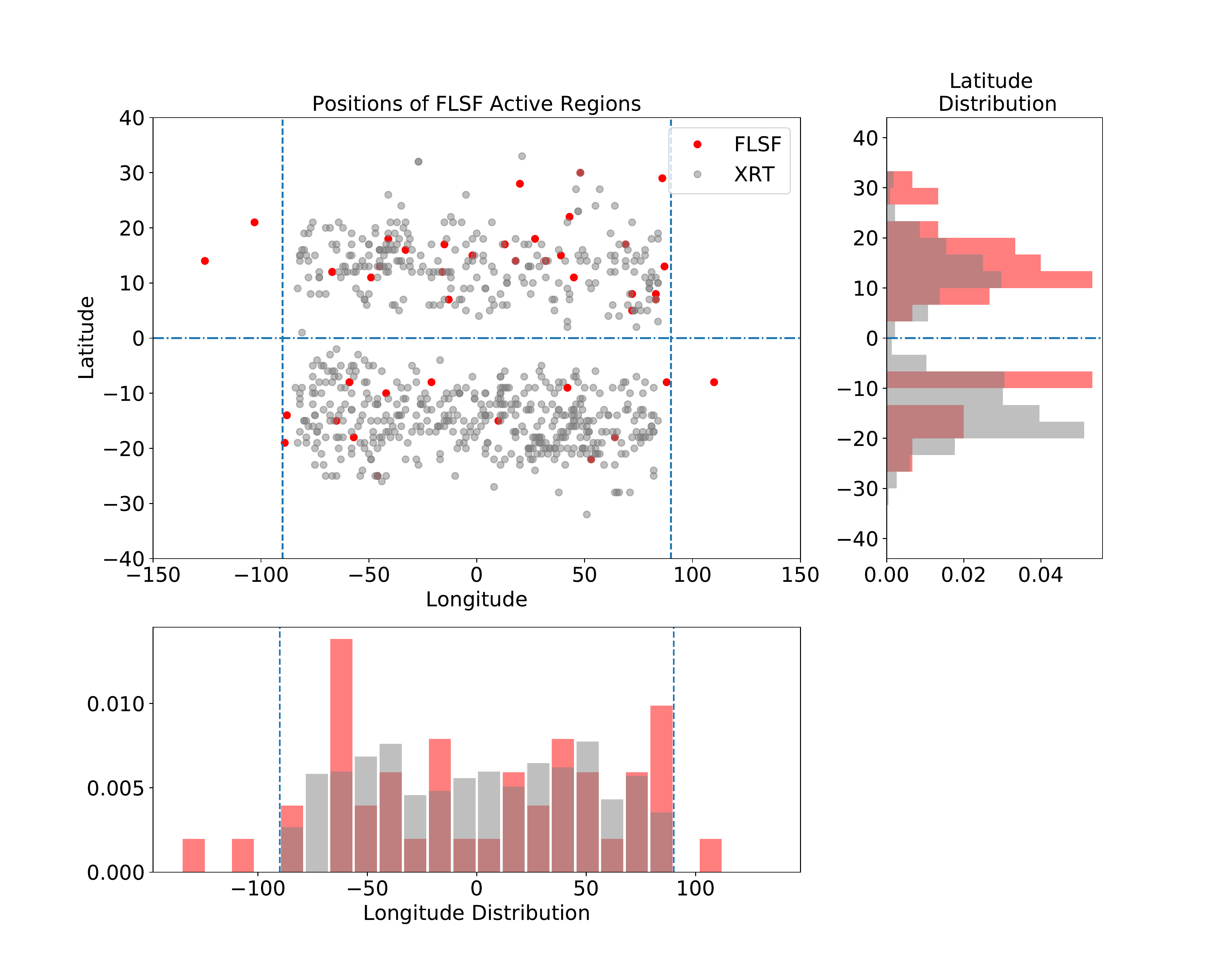} 
\caption{Positions of Active Regions associated with {FLSFs} (red) and M/X-class XRT flares (gray). Longitudes beyond -90\de and +90\de correspond to BTL flares. The right hand panel shows the latitude distribution of the AR positions, illustrating the asymmetry in the population. 64\% of the ARs from which the {FLSFs} originate are located in the northern heliosphere whereas 62\% of the ARs from whic the XRT flares originate are located in the southern heliosphere.}   
\label{fig:AR_position_dist}   
\end{center}   
\end{figure*}

\begin{figure*}[ht!]   
\begin{center}    
\includegraphics[width=\textwidth,trim=10 10 50 50,clip]{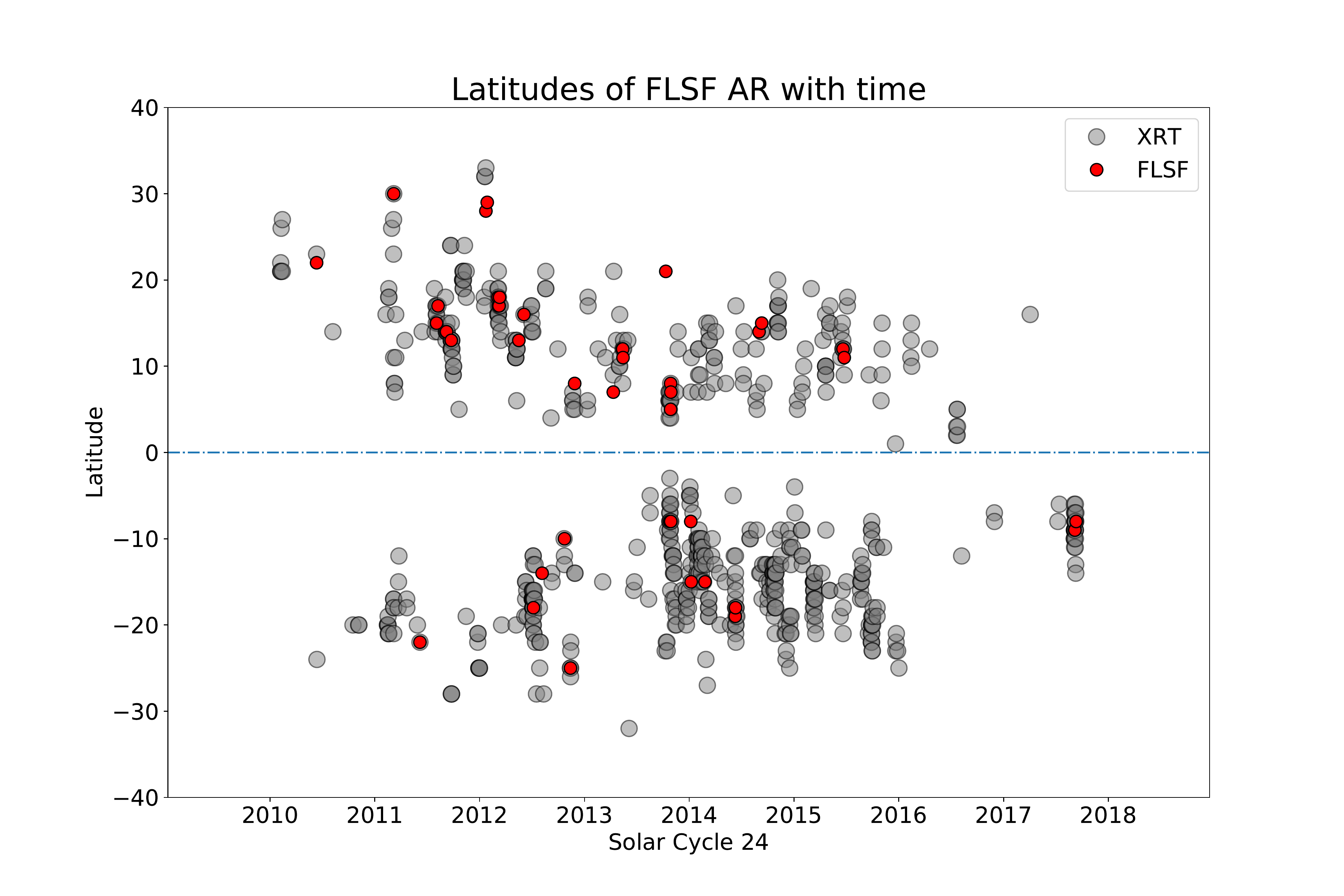} 
\caption{Positions of ARs associated with FLSF (red) and M/X class GOES flare (gray) as a function of time. The distribution of positions follows the so-called butterfly pattern, i.e. at the beginning of a new \solar cycle, sunspots tend to form at high latitudes, but as the cycle reaches its maximum the sunspots tend to form at lower latitudes.}
\label{fig:FLSF_AR_time_dist}   
\end{center}   
\end{figure*}

The positions on the \solar surface of the  ARs associated with the {FLSFs} are plotted together with the M/X class flares detected by Hinodes's XRT~\citep{hinode} in Figure~\ref{fig:AR_position_dist}. {Three BTL flares, whose position was inferred from STEREO, appear with longitudes smaller or greater than -90\de and +90\de.} 
The  distribution {in} longitude is rather uniform, with the same number of flares in positive and negative longitudes between -90\de and +90\de.   
However, there is an asymmetry in the distributions in latitude, with {a preponderance of FLSFs ($\sim$65\%) in the northern hemisphere, while} the opposite is true for the XRT flares. This asymmetry is also evident in Figure~\ref{fig:FLSF_AR_time_dist}, where we plot the positions of FLSF ARs as a function of {time,  illustrating the} so-called Butterfly pattern, with ARs migrating toward the equator as the \solar cycle evolves.

\subsection{Flare Series}

A {notable} feature of the FLSF population is {that more than half (25 out of 45) are part of a cluster of flares originating from the same AR (see Table~\ref{tab:flare_series}).} It is common for an AR to be the source of several flares, but the high {fraction of such clusters} in the FLSF catalog might indicate that some ARs have the right conditions {to be associated with the production of} $\gamma$-rays. 
The most notable series happened {during 2012-03-05 to 2012-03-10 and 2013-03-13 to 2013-03-15 each with four} {FLSFs}. All of these flares were associated with fast CMEs, and both series {produced} strong  and long-lasting SEP events.
They all yielded delayed FLSF $\gamma$-ray emission lasting more than three hours. In addition, {three of the eight flares were identified as having no $>$100 MeV $\gamma$-rays detected during the prompt phase; only delayed emission was detected.} Only one additional flare {behaved this way, FLSF~2013-04-11, that was found to have a short delayed emission and no prompt emission.}
This could indicate that the presence of previous SEP events and multiple fast CMEs is more important  
for the production of long lasting $\gamma$-ray emission than the presence of impulsive HXRs produced by high energy electrons.

\subsection{Gamma-ray localization}
\label{sec:localization}
The \Fermi-LAT is the first 
{telescope capable of determining the centroid of $>$100~MeV emission from \solar flares.}
The position of the emission centroid on the \solar disk can yield valuable information on where on the photosphere the precipitating ions produce 
the high-energy $\gamma$-rays.

For the majority of the {FLSFs} in the catalog, the 68\% error on the emission centroid is larger than 500$\arcsec$ and therefore it becomes difficult to distinguish a specific region on the \solar disk from which the emission is originating. 
{For eight of the {FLSFs}, the 68\% error radius is $\leq$365$\arcsec$ (roughly a third of the \solar disk), providing meaningful constraints on the location of the emission centroid that can then be compared with the lower-energy flare emission sites.}
The localization results for these {eight} flares are {given} in Table~\ref{tab:best_localization}. The first eight columns of Table~\ref{tab:best_localization} report the date and time window of the detection, position of the centroid of the $>$100~MeV emission 
in helioprojective coordinates (X,Y), the 68\% and 95\% uncertainty on the emission centroid, the AR number and position, the angular distance and relative distance 
of the emission centroid from the AR\footnote{The position of the AR at the time of the GOES X-ray flare}. The last column shows the ratio of this distance to the 95\% error radius. 
{We emphasize that the position and the confidence intervals in the table are derived by modeling the high-energy emission as a point source, i.e. with no geometric extent on the \solar surface.}

\begin{figure*}[ht!]   
\begin{center}    
\includegraphics[width=\textwidth]{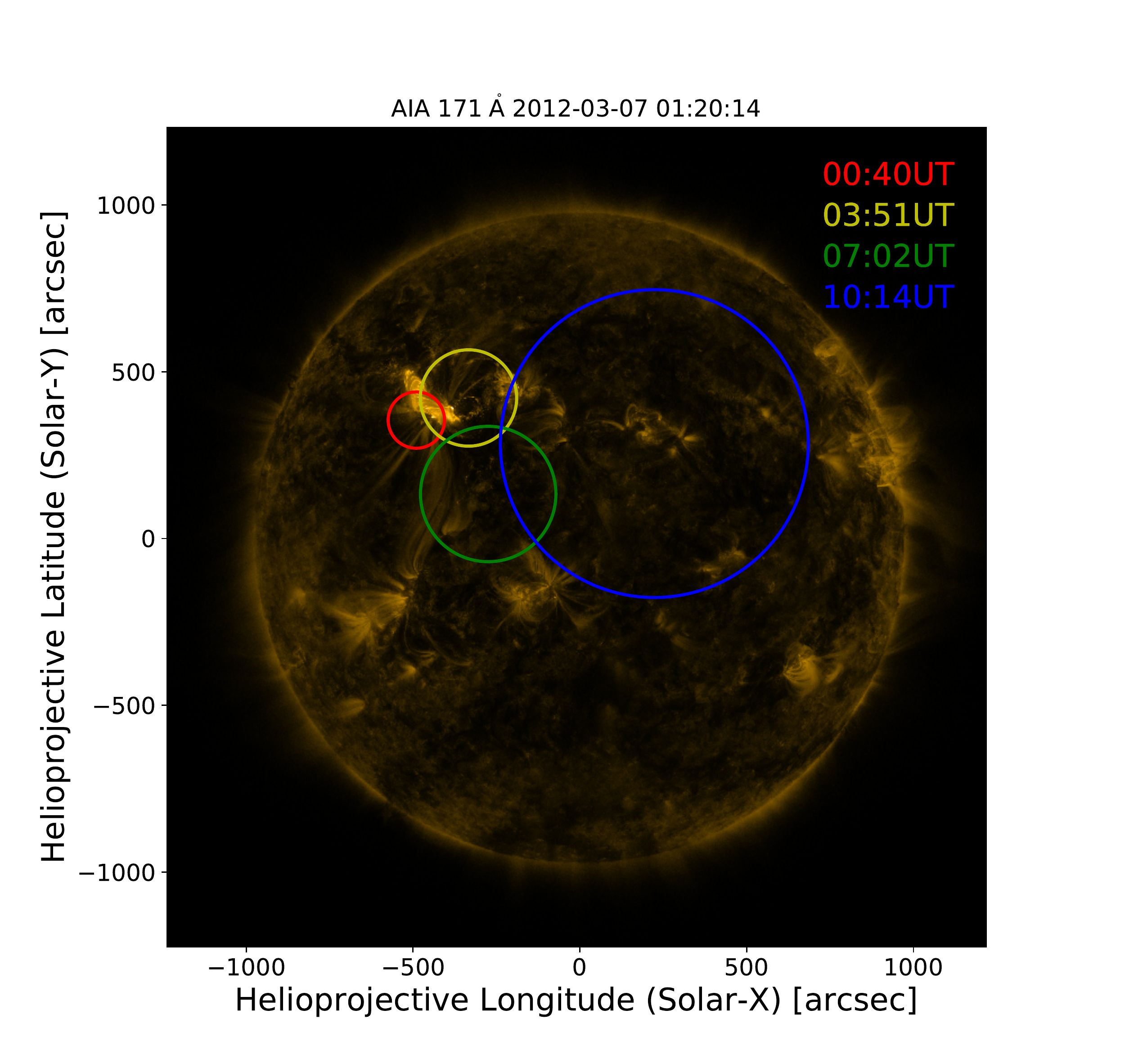} 
\caption{\Fermi-LAT localization of the $>$100~MeV data in multiple time windows from the FLSF~2012-03-07. The error radii correspond to the 95\% confidence region. The start of the time windows is annotated in the upper right corner of the figure. The localization centroid is overplotted on the AIA 171\Angst\ image of the Sun at the time of the flare.}
\label{fig:SOL20120307_multpile_loc}   
\end{center}   
\end{figure*}

\begin{figure*}[ht!]   
\begin{center}    
\includegraphics[width=\textwidth]{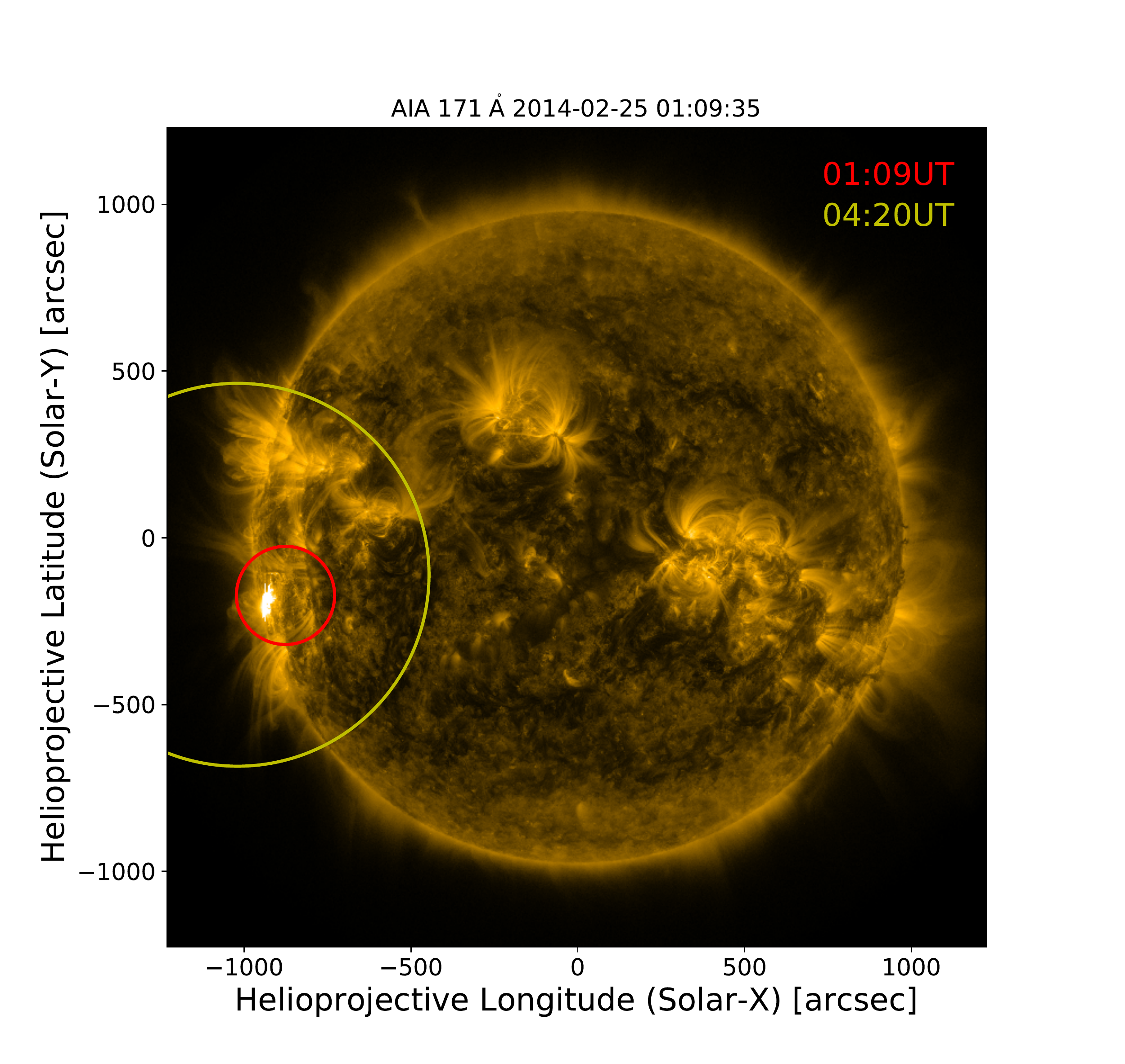} 
\caption{\Fermi-LAT localization of the $>$100~MeV data in multiple time windows from the FLSF~2014-02-25. The error radii correspond to the 95\% confidence region. The start of the time windows is annotated in the upper right corner of the figure. The localization centroid is overplotted on the AIA 171\Angst\ image of the Sun at the time of the flare.}
\label{fig:SOL20140225_multpile_loc}   
\end{center}   
\end{figure*}

\begin{figure*}[ht!]   
\begin{center}    
\includegraphics[width=\textwidth]{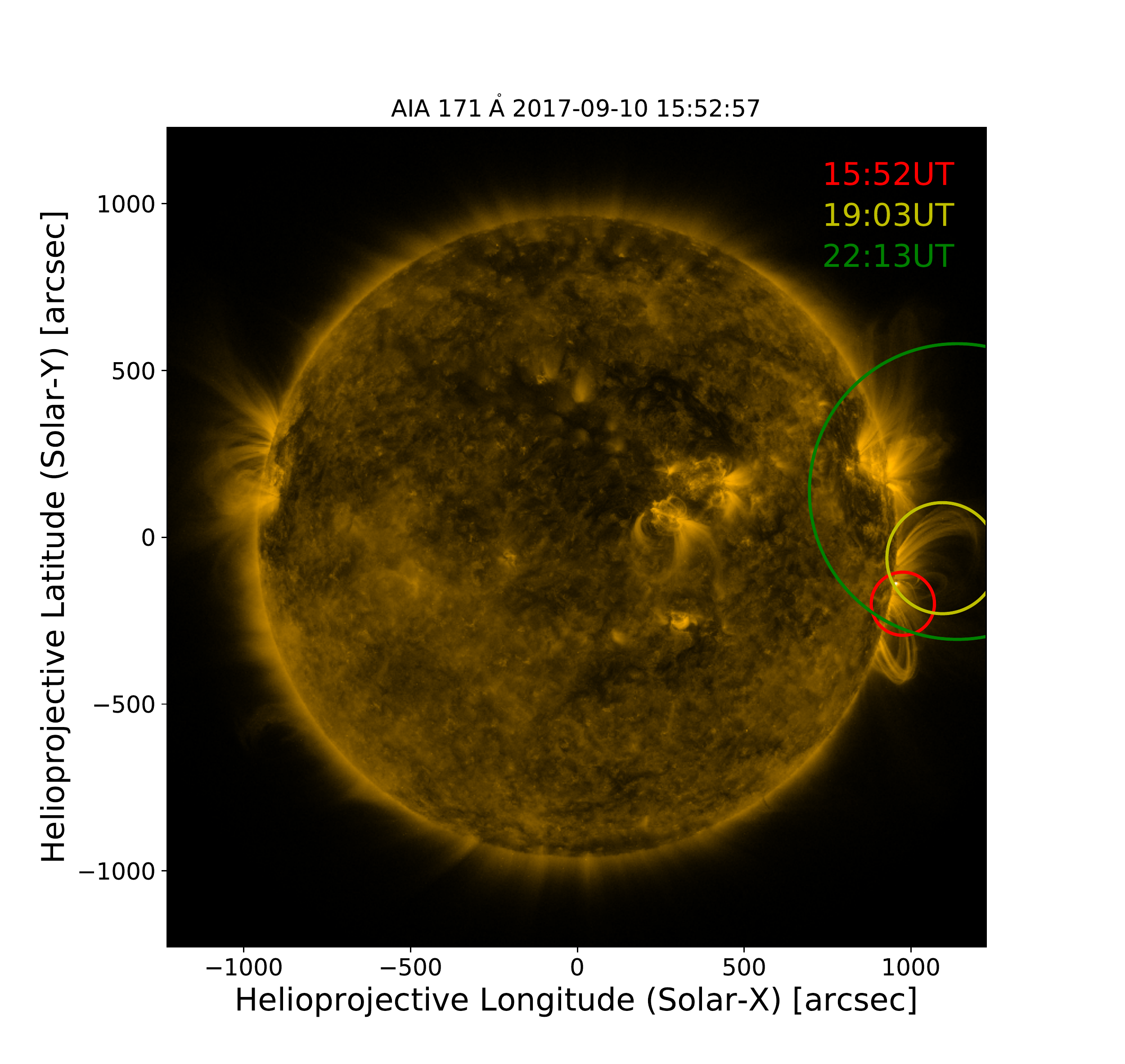} 
\caption{\Fermi-LAT localization of the $100$~MeV data in multiple time windows from the FLSF~2017-09-10. The error radii correspond to the 95\% containment, the start of the time windows are annotated in the upper left-hand corner of the figure. The localization centroid is overplotted on the AIA 171\Angst\ image of the Sun at the time of the flare.}
\label{fig:SOL20170910_multpile_loc}   
\end{center}   
\end{figure*}

Three of the eight flares (FLSF~2012-03-07, FLSF~2014-02-25 and FLSF~2017-09-10) 
were {sufficiently bright and long-lasting to be localized in multiple \texttt{SunMonitor} time windows.} The FLSF~2012-03-07 was an exceptional $\gamma$-ray flare in terms of both duration and brightness. 
{The error radius was smaller than 300$\arcsec$ in four detection windows, and the emission centroid} moved progressively across the \solar disk over the $\sim$10 hours of $\gamma$-ray emission, {as} shown in Figure~\ref{fig:SOL20120307_multpile_loc}.  
{This flare was the first for which this behavior in $>$100~MeV $\gamma$-rays could be observed, and it was interpreted as supporting evidence for the CME driven shock scenario as the particle accelerator \citep{0004-637X-789-1-20}.}
{For FLSF~2014-02-25, the statistics were sufficient to provide meaningful localization in only two time intervals, and the emission centroid remained consistent with the AR position over three hours, as shown in Figure~\ref{fig:SOL20140225_multpile_loc}.} 
Finally, FLSF~2017-09-10 was also an exceptionally bright flare, but, because the AR was located at the very edge of the western limb, it was impossible to observe any progressive motion of the $\gamma$-ray source. Throughout the 7-hour detection, the source centroid remained consistent with the AR position, {as} shown in Figure~\ref{fig:SOL20170910_multpile_loc}. 

Two out of these eight flares originated from {ARs} whose position was located behind the visible \solar disk, {highlighting} how bright these flares were regardless of the position of the AR.
All eight {FLSFs} were classified as GOES X-class flares, with the exception of the BTL FLSF~2013-10-11 whose GOES classification of M4.9 is most likely an underestimation~{\citep{2013SoPh..288..241N,2015ApJ...805L..15P}}. The peak $\gamma$-ray fluxes were all greater than 3$\times$10$^{-5}$ {ph} cm$^{-2}$ s$^{-1}$ and exposure times were {all} greater than 20 minutes{, indicating that they are not impulsive flares}. Five of the {FLSFs} originated from ARs from the eastern quadrant and three from the western quadrant of the \solar disk.

\subsection{GOES X-class flares not detected by the LAT}

\begin{figure*}[ht!]   
\begin{center}   
\includegraphics[width=\textwidth,trim=50 10 80 80,clip]{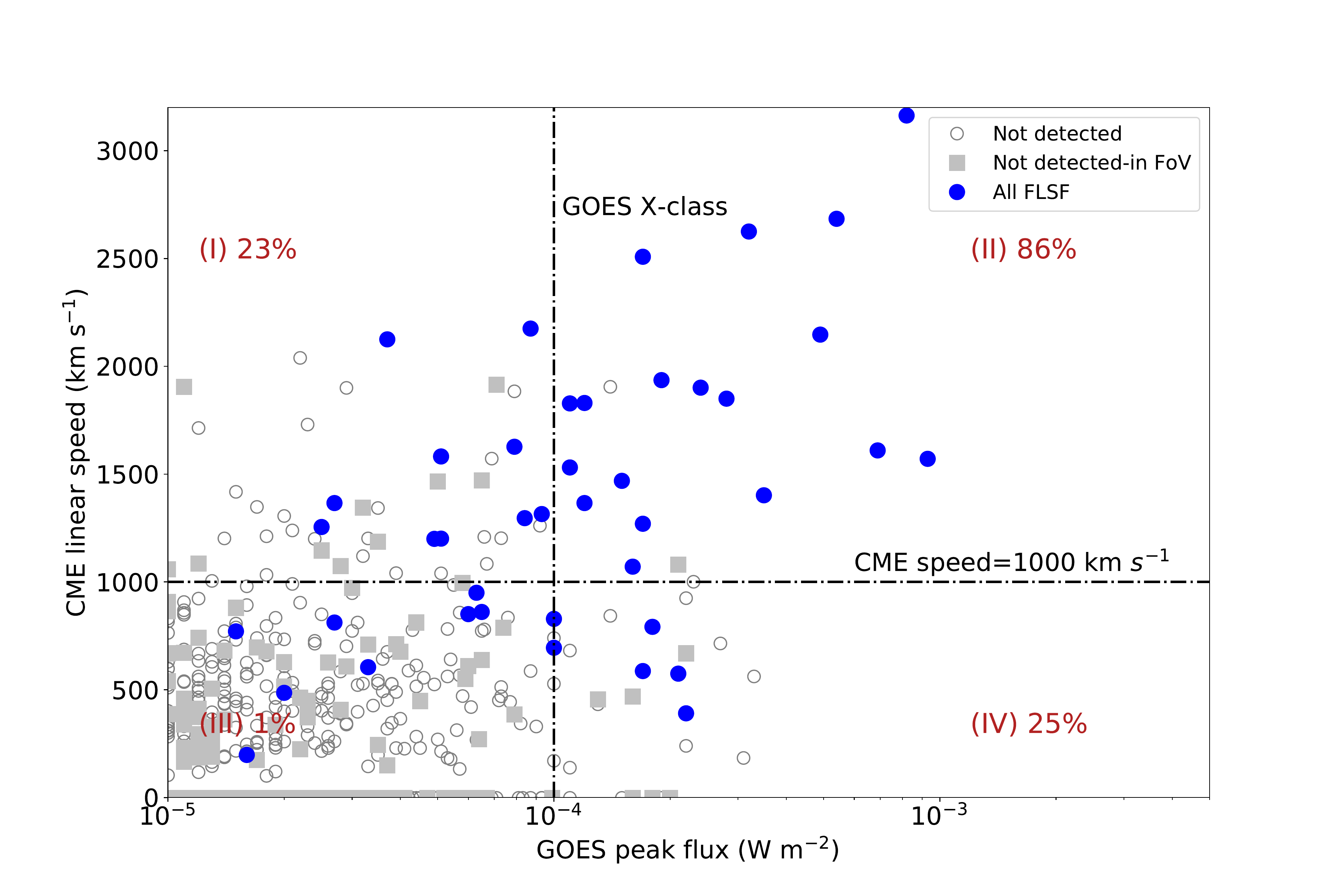}
\caption{{CME linear speed versus GOES peak flux for all the FLSFs (blue points), M/X-class flares not detected  by the \Fermi-LAT outside the LAT {FoV} (gray empty circles) and in the {FoV} (gray filled square) at the time of the GOES X-ray flare. The vertical dashed line indicates the border between M and X class GOES flares. The horizontal  dashed line indicates a 1000 km s$^{-1}$ CME speed. In each of the four quadrants (labeled I-IV) we indicate the fraction of flares detected by the LAT in that quadrant.}} 
\label{fig:non_detections} 
\end{center}   
\end{figure*} 

{In an attempt to characterize the \solar flares associated with $\gamma$-ray detections, we can also examine the population of \solar flares not detected by the \Fermi-LAT above 30 MeV. During the time period considered in this paper, there were a total of 772 M and X class flares (49 were X-class flares and 24 of these were associated with FLSFs)%
\footnote{Here {we include FLSF 2012-03-07, {we associate the $\gamma$-ray emission with the X5.4 X-ray flare and with the CME with linear speed of 2684 km s$^{-1}$.}  Two of the three BTL flares have an estimated GOES class of X3.5 and X2.4, but are not considered in this comparison because we do not have a catalog of X-class flares occurring} behind the limb.}. 
In Table~\ref{tab:Xray_non_detections} we list only the 25 X-class flares not associated with a $\gamma$-ray detection and their possible associations to CMEs and SEP events.}
{Figure~\ref{fig:non_detections} shows a scatter plot of CME speed versus GOES flux for all FLSFs and all the M/X class flares non detected by the LAT.}
{We have labeled the four quadrants (I-IV) that indicate the population of flares classified as M/X class and whether they were associated with a CME with linear speed $>/<$ 1000 km s$^{-1}$. We report the fraction of LAT detected flares over the total number of flares that fall within the quadrant. From this figure it is possible to see that most favourable condition for the LAT to detect $\gamma$-ray emission is for the flare to be of X-class and be associated with a CME with linear speed greater than 1000 km s$^{-1}$ (86\% of the flares detected by the LAT) and that the least favourable condition (1\% of the flares detected by the LAT) is diagonally opposite (i.e. M class and slow CME speed). The conditions in the off diagonal quadrants appear to be equally favourable.}
{Out of the three flares not detected by the LAT and in quadrant IV, the \SunMon picked up a marginal detection in the 3 following observing windows (with a $\sigma$ = 4.5, 4.0, 4.0) for the flare of 2011-09-22 that was associated with a Halo CME with a linear speed of 1905 km\,s$^{-1}$.} 

\section{Summary and Discussion}
\label{sec:discussion}
Continuous monitoring of the Sun by  \Fermi-LAT has led to high-confidence detection of 45 \solar flares with $\gamma$-ray emission above 60 MeV. With such a relatively sizable sample of flares it is now possible to perform population studies of $\gamma$-ray \solar flares.  Based on the temporal characteristics and associations with multiwavelegth flaring activity, we have found that there are at least two 
{distinct types of $\gamma$-ray emission in} \solar flares: {\emph{prompt-impulsive} and \emph{delayed-gradual}}. Within these two broad classes we find a rich and diverse sample of events with a wide variety of characteristics. 
Of the 45 FLSFs discussed in this work, six have {been detected only with}  a prompt-impulsive emission correlated with HXR emission (classified as \emph{prompt-only}), four have no $\gamma$-ray emission detected during the impulsive HXR emission  but were significantly bright after  all other flare emission activities had ceased (classified as \emph{delayed-only}), and ten have both \emph{prompt} and \emph{delayed} emission. For the remaining {25 flares with delayed emission}, we cannot exclude the presence of \emph{prompt} emission because the Sun was not in the {FoV} of the LAT during the impulsive HXR activity phase.  

The most significant results presented in this work  can be summarized as follows:

\begin{enumerate}

\item {Emission above 60 MeV could be due to bremsstrahlung radiation produced by electrons of Lorentz factor $\gamma_e>100$ with relatively hard spectrum is most probably an  unlikely scenario. This is because the acceleration of electrons to such energies is difficult due to  high synchrotron losses. We find that emission due to decay of pions ($\pi ^0, \pi ^\pm$) produced by $>300$ MeV protons and ions, with a power law spectrum of index $\sim 4-5$, extending up to 10s of GeV, produces very good fit to all observed $\gamma$ rays.}

\item All of the FLSFs with LLE \emph{prompt} emission (produced by $>300$ MeV ions) reach their peak within seconds of the 100--300~keV emission peak (produced by $>100$ keV electrons) observed with \Fermi-GBM implying that these ions and electrons are accelerated, transported, and interact with the ambient medium at the same time.  
Similar conclusions for acceleration of lower energy (1 to 30 MeV) ions were reached by ~\cite{Chupp_1987} and \cite{Hurford_2006}  based on the \emph{RHESSI} imaging of the 2.223 MeV neutron-capture $\gamma$-ray line, and by \cite{shih09} who reported a tight correlation between the 2.223 MeV line fluence and the $>$300 keV electron bremsstrahlung fluence. 
    
\item All but three of the flares in the FLSF catalog are associated with CMEs. The \emph{delayed} type flares are associated with faster CMEs (mean speed of 1535 km\,s$^{-1}$ ) whereas the \emph{prompt} type FLSFs are associated with slower CMEs (mean speed of 656 km\,s$^{-1}$ ).
    
\item One of the most important contribution of \fermi-LAT has been its ability to localize the centroids of high-energy $\gamma$-ray emission on the Sun. {In most such cases the initial centroid position is at or near the AR where the flare originated.} In several long lasting strong flares there are clear indications of change of the centroid position with time; often away from the AR. 
This change is best observed in the strong, long-lasting FLSF~2012-03-07, where the centroid of $>$100 MeV emission gradually migrates away from the AR up to tens of degrees. This indicates that the acceleration site of the $\gamma$-ray producing high energy ions is magnetically connected to regions on the photosphere far away from the initial AR. 
    
\item Further evidence for this scenario comes from, for the first time, \Fermi observation of GeV emission from three behind-the-limb flares including two-hour emission from FLSF 2014-09-01 originating 40 degrees behind the limb. Localization of the $\gamma$-ray emission from two of these flares indicates that the emission occurred on the visible disk, again necessitating a way for the ions from the acceleration site to access regions on the visible disk (more than 40 degrees away from the AR) to interact and to produce the observed $\gamma$-rays. Similar conclusions were also reached by~\cite{cliv93} and \cite{1993ApJ...409L..69V} for the observations with CGRO-EGRET of behind-the-limb flares with emission up to 100 MeV.

\item There is an asymmetry in the latitude distribution of the ARs from which the FLSFs originate, with 65\% of the flares coming from the northern heliosphere. The opposite is true for the M/X-class XRT flares detected during the same time interval. \cite{Shrivastava_2005} found that CMEs associated with Forbush decreases also come predominately from the northern heliosphere.

\item More than half of the FLSFs in this catalog are part of a series of flare clusters. The most notable clusters happened from 2012-03-05 to 2012-03-10 and from 2013-05-13 to 2013-05-15 with each consisting of four FLSFs. All of these flares were associated with fast CMEs, and both series produced strong and long-lasting SEP events. They all yielded delayed FLSF $\gamma$-ray emission lasting more than three hours. In addition, three of these eight flares showed no impulsive phase $\gamma$-ray emission (only one other non-series FLSF was found with similar properties). This could suggest that the presence of previous SEP events and multiple fast CMEs is more important for production of long lasting $\gamma$-ray emission than the presence of impulsive HXRs produced by high-energy electrons.
    
\item Seven FLSFs in the catalog are detected with both LLE-prompt and delayed phases, with the average peak flux of the prompt phase 10 times higher than that of the delayed phase. However, the total energy released during the delayed phase is 10--100 times larger than that during the prompt phase. 
    
\end{enumerate}
        
{Solar eruptive events involve two distinct but related phenomena: (1) acceleration of electrons and ions at the reconnection regions in coronal loops that produce the  impulsive nonthermal radiation observed from microwaves to $\gamma$ rays, lasting several minutes, and are observed as impulsive-prompt SEPs, often with substantial enhanced abundances of $^3$He and heavier ions. (2) production of a supersonic CME which drives a shock, where particles are accelerated resulting in long-duration SEPs with normal ionic abundances, with only one radiative signature of type II radio emission 
produced by less numerous SEP electrons. As summarized above, the \Fermi-LAT observations show both prompt-impulsive $\gamma$-ray emission having light curves similar to those of the HXRs, and long-duration delayed emission with temporal behavior similar to SEPs, {and like gradual SEPs,} associated with fast CMEs. These similarities between gradual SEPs and $>60$ MeV gradual-delayed emission, plus the observed drifting of the centroid of $\gamma$-ray emission from the original active region, which is accentuated by the observations of behind-the-limb flares, indicate that the site and mechanism of the acceleration of ions responsible for the long duration $\gamma$ rays is different than that of particles producing the impulsive nonthermal flare radiation, and suggest that long duration $\gamma$ rays are another radiative signature of acceleration in CME-shocks. However, unlike the type II radiation they are produced by ions (accelerated in the CME driven shock) and not in the low density environment of the CME. While SEPs are particles escaping the upstream of the shock, the $\gamma$ rays must be produced by ions escaping from downstream region of the shock back to the high density photosphere of the Sun, and because of complex and changing magnetic connection between the CME and the Sun, sometimes to regions far from the AR from which the eruptions originated. Recent reconstruction of these magnetic connections by~\cite{Jin2018} provide support for this scenario.}

{Alternative scenarios for explaining the gradual-delayed emission observed by \Fermi have been put forth by authors such as \cite{DeNolfo2019} in their comparison between the characteristics of high-energy SEPs observed by PAMELA and those of the \emph{delayed}-type emission $\gamma$-ray flares. One such scenario is that particles are accelerated via the second-order Fermi mechanism and trapped locally within extended coronal loops. These accelerated particles would then diffuse to the denser photosphere to radiate~\citep{1991ApJ...368..316R}. With this approach it is possible to decouple the acceleration of the particles producing $\gamma$ rays from the acceleration and transport of the SEPs, allowing for different energetic particle productivities.}

{Thanks to the increase in sensitivity of the \Fermi-LAT the sample of $>100$~MeV $\gamma$-ray flares has increased by almost a factor of 10 thus allowing to perform population studies on these events for the first time. The observations presented in this work suggest that the particles producing the \emph{prompt}-type emission and those producing the \emph{delayed}-type emission are accelerated via different mechanisms. However, further multiwavelength observations and in-depth simulations are needed in order to come to a definitive answer to which acceleration mechanism is driving the \emph{delayed}-type $\gamma$-ray emission of \solar flares.}

\newpage

\acknowledgements
The \textit{Fermi} LAT Collaboration acknowledges generous ongoing support
from a number of agencies and institutes that have supported both the
development and the operation of the LAT as well as scientific data analysis.
These include the National Aeronautics and Space Administration and the
Department of Energy in the United States, the Commissariat \`a l'Energie Atomique
and the Centre National de la Recherche Scientifique / Institut National de Physique
Nucl\'eaire et de Physique des Particules in France, the Agenzia Spaziale Italiana
and the Istituto Nazionale di Fisica Nucleare in Italy, the Ministry of Education,
Culture, Sports, Science and Technology (MEXT), High Energy Accelerator Research
Organization (KEK) and Japan Aerospace Exploration Agency (JAXA) in Japan, and
the K.~A.~Wallenberg Foundation, the Swedish Research Council and the
Swedish National Space Board in Sweden.
 
Additional support for science analysis during the operations phase is gratefully
acknowledged from the Istituto Nazionale di Astrofisica in Italy and the Centre
National d'\'Etudes Spatiales in France. This work performed in part under DOE
Contract DE-AC02-76SF00515.

MPR and NO acknowledge relevant and helpful discussions with members of the ISSI International Team on Energetic Ions: The Elusive Component of Solar Flares and with participants in the Lorentz Center Workshop on Solar Sources of GeV Gamma-rays, 26 Feb - 2 Mar 2018.

\newgeometry{left = 10mm,right=10mm}  
\begin{table}
\let\center\empty
\let\endcenter\relax
\centering
  \resizebox{0.9\textwidth}{!}{\input{tables/Main_Flare_table_insert.tex}}
\caption{FLSF catalog for flares detected with the \Fermi-LAT \texttt{SunMonitor} and their likely GOES X-ray flare associations. In the {\it GOES}-class column entries with a $*$ identify the BTL flares, whose class 
is estimated based on the STEREO observation, and $^a$ indicate that there is also a LLE detection of the flare. The analysis results for the LLE flares are shown in Table~\ref{tab:lle_main_flare_table}. }
\label{tab:time_integrated_main_table}
\end{table}
\restoregeometry

\begin{landscape}
  \setlength\LTcapwidth{\linewidth}
  \scriptsize
  \begin{longtable}{ccccccccc}
\toprule
            \thead{Date and Time\\(UTC)} & \thead{Exposure \\ (minutes)} &          \thead{Flux\\ ($10^{-5}$ {ph} cm$^{-2}$s$^{-1}$)}  &   \thead{TS} &  \thead{$\Delta$TS} & \thead{Model} &   \thead{Photon Index} &  \thead{Cutoff Energy \\ (MeV)} &  \thead{Proton \\  index}  \\
\midrule
\endhead
\midrule
\multicolumn{3}{r}{{Continued on next page}} \\
\midrule
\endfoot
\midrule 
\endlastfoot
  2011-03-07 20:10 - 20:39 &             29 &   $  2.06    \pm  0.19  $   &  317 &    27 &   Exp &  -0.76 $    \pm  $ 0.45 &    172 $    \pm  $ 55 &     $  4.3    \pm  0.4  $   \\
 2011-03-07 23:21 - 00:05 &             44 &   $  3.04    \pm  0.20  $   &  710 &    70 &   Exp &  -0.31 $    \pm  $ 0.36 &    138 $    \pm  $ 27 &   $  4.13    \pm  0.26  $   \\
 2011-03-08 02:33 - 03:16 &             43 &   $  3.23    \pm  0.22  $   &  621 &    66 &   Exp &  -0.15 $    \pm  $ 0.41 &    110 $    \pm  $ 22 &   $  4.70    \pm  0.32  $   \\
 2011-03-08 05:44 - 06:27 &             44 &   $  1.40    \pm  0.15  $   &  219 &    32 &   Exp &   0.67 $    \pm  $ 0.99 &     63 $    \pm  $ 22 &                  $>$6 \\
 2011-03-08 09:13 - 09:39 &             26 &   $  0.48    \pm  0.11  $   &   46 &  -0.1 &    PL &  -2.55 $    \pm  $ 0.25 &             - &                   - \\
\midrule 
 2011-06-07 07:47 - 08:23 &             36 &   $  3.18    \pm  0.20  $   &  740 &   76 &   Exp &  -0.13 $    \pm  $ 0.37 &    104 $    \pm  $ 19 &   $  4.97    \pm  0.33  $   \\
 2011-06-07 11:16 - 11:34 &             19 &   $  0.32    \pm  0.10  $   &   19 &    5 &    PL &  -2.70 $    \pm  $ 0.35 &             - &                   - \\
\midrule 
 2011-08-04 04:55 - 05:37 &             42 &   $  2.30    \pm  0.18  $   &  413 &   49 &   Exp &  -0.09 $    \pm  $ 0.50 &     95 $    \pm  $ 21 &   $  5.4    \pm  0.4  $   \\
\end{longtable}
\end{landscape}

\newgeometry{left = 10mm,right=10mm}  
\begin{table}
\let\center\empty
\let\endcenter\relax
\centering
  \resizebox{0.87\textwidth}{!}{\input{tables/lle_main_table.tex}}
  \caption{LLE FLSF catalog results with associated {\it GOES} X-ray flare. For the cases where the the curved spectrum is preferred we also list the best inferred proton index. The \SunMon detected column indicates whether the flare was detected by the \SunMon automatic pipeline. The fluxes are in units of 10$^{-5}$ ph s$^{-1}$ cm$^{-2}$.}
\label{tab:lle_main_flare_table}
\end{table}
\restoregeometry

\begin{landscape}
  \setlength\LTcapwidth{\linewidth}
  \scriptsize
  \begin{longtable}{lcccccccc}
\toprule
        \thead{Name} & \thead{Total duration\\ (hrs)}  &                       \thead{Flare Type} &        \thead{GOES\\ Start (UT)} &   \thead{GOES\\ Class} &    \thead{CME Speed\\(km s$^{-1}$)}  & \thead{CME\\First C2 app. (UT)} &    \thead{SEP Emax\\ (MeV)} & \thead{HXR Emax\\ (keV)} \\
        \midrule
\endhead
\midrule
\multicolumn{3}{r}{{Continued on next page}} \\
\midrule
\endfoot
\midrule 
\endlastfoot
  FLSF 2010-06-12 &                        30$^{\star}$ &                LLE-Prompt$^a$ &  2010-06-12 00:30 &         M2.0 &          486 &  2010-06-12 01:31 &          10 &     1000 \\
  FLSF 2011-03-07 &              $  15.8    \pm  3.1  $   &                          Delayed &  2011-03-07 19:43 &  M3.7$^c$ &         2125 &  2011-03-07 20:00 &          50 &     $>$100 \\
  FLSF 2011-06-07 &               $  6.0    \pm  2.2  $   &                          Delayed &  2011-06-07 06:16 &         M2.5 &         1255 &  2011-06-07 06:49 &         100 &      100 \\
  FLSF 2011-08-04 &               $  2.3    \pm  0.7  $   &                          Delayed &  2011-08-04 03:41 &         M9.3 &         1315 &  2011-08-04 04:12 &         100 &      300 \\
  FLSF 2011-08-09 &             $  0.87    \pm  0.34  $   &      Prompt Short-Delayed$^a$ &  2011-08-09 07:48 &         X6.9 &         1610 &  2011-08-09 08:12 &         100 &      300 \\
  FLSF 2011-09-06 &               $  2.0    \pm  1.4  $   &  LLE-Prompt Short-Delayed$^a$ &  2011-09-06 22:12 &         X2.1 &          575 &  2011-09-06 23:05 &         100 &     1000 \\
  FLSF 2011-09-07 &             $  2.02    \pm  0.35  $   &                          Delayed &  2011-09-07 22:32 &         X1.8 &          792 &  2011-09-07 23:05 &   50$^d$ &      500 \\
  FLSF 2011-09-24 &               $  1.2    \pm  0.7  $   &  LLE-Prompt Short-Delayed$^a$ &  2011-09-24 09:21 &         X1.9 &         1936 &  2011-09-24 09:48 &   50$^d$ &     1000 \\
\end{longtable}
\end{landscape}

\begin{landscape}
\begin{table}[hp!]
\begin{center} 
\scriptsize 

\begin{tabular}{cccccccccc}
\toprule
              \thead{Name} & \thead{Flare Type} & \thead{Duration\\ (hrs)}  & \thead{CME Speed\\(km s$^{-1}$)}  & \thead{Width} & \thead{GOES\\ Class} & \thead{SEP Emax\\ (MeV)} & \thead{HXR Emax\\ (keV)} &     \thead{AR} & \thead{AR pos} \\
\midrule
 FLSF~2011-09-06 &  Prompt Delayed &      0.6 &       575 &     H &       X2.1 &      100 &     1000 &  11283 &    N14W18 \\
 FLSF~2011-09-07 &         Delayed &      1.9 &       792 &   290 &       X1.8 &   50$\ddagger$ &      500 &  11283 &    N14W32 \\
 \hline
 FLSF~2012-01-23 &         Delayed &      5.8 &      2175 &     H &       M8.7 &      100 &     $>$100 &  11402 &    N28W20 \\
 FLSF~2012-01-27  &         Delayed &      7.3 &      2508 &     H &       X1.7 &      605 &     $>$100 &  11402 &    N29W86 \\
 \hline
 FLSF~2012-03-05 &         Delayed &      5.4 &      1531 &     H &       X1.1 &   40$\ddagger$ &     $>$100 &  11429 &    N18E41 \\
 FLSF~2012-03-07 &         Delayed &     20.2 &   2684* &     H &    X5.4* &      605 &     1000 &  11429 &    N17E15 \\
 FLSF~2012-03-09 &    Delayed only &      7.3 &       950 &     H &       M6.3 &  100$\ddagger$ &     $>$100 &  11429 &    N17W13 \\
\bottomrule
\end{tabular}
\caption{List of FLSFs from similar Active Regions (* indicates several X-ray classes or CMEs during the duration of the $\gamma$-ray emission. $\ddagger$ indicates the previous presence of SEPs, without this event being an SEP event).} The last column provides the position of the AR in heliographics coordinates.
\label{tab:flare_series}
\end{center}    
\end{table}
\end{landscape}

\begin{table}[h!]
\begin{center} 
\scriptsize 
\begin{tabular}{ccccccccc}
\toprule
 Date and Time &  \thead{Helio X\\($\arcsec$)} & \thead{Helio Y\\($\arcsec$)} & \thead{ERR 68\\($\arcsec$)} & \thead{ERR 95\\($\arcsec$)} &  \thead{AR\\ Number} & \thead{AR\\ Position} & \thead{Angular\\ Dist ($\arcsec$)} & \thead{Relative \\Dist (95)} \\
            \midrule
 2011-09-06 22:11 - 22:47 &     219 &     533 &    139 &    220 &     11283 &      N14W18 &        382 &                    1.7 \\
            \midrule
 2012-03-07 00:40 - 01:20 &    -562 &     231 &     56 &     84 &     11429 &      N17E15 &       45 &                    0.5 \\
 2012-03-07 03:51 - 04:31 &    -300 &     342 &     84 &    144 &     11429 &      N17E15 &       143 &                    1.0 \\
 2012-03-07 07:02 - 07:42 &    -320 &      20 &    126 &    203 &     11429 &      N17E15 &       331 &                    1.6 \\ 
 2012-03-07 10:14 - 10:54 &     207 &     245 &    291 &    462 &     11429 &      N17E15 &       707 &                    1.5 \\
 \midrule
 2012-07-06 23:20 - 00:08 &     530 &    -432 &    362 &    586 &     11515 &      S18W64 &       122 &                    0.2 \\ 
 \midrule
 2013-05-14 02:43 - 03:31 &   -1137 &     333 &    314 &    504 &     11748 &      N12E67 &      279 &                    0.6 \\
\bottomrule
\end{tabular}
\caption{Localization results for the {FLSFs} with 68\% error radius $<$0.1$^{\circ}$. We report the date and detection time window start and stop, LAT $>$100~MeV emission centroid position in Helio X and Y coordinates, the 68\% and 95\% error radius (in arcseconds), the AR number and position, the distance of the centroid from the active region, and the ratio of this distance to the 95\% error radius.}
\label{tab:best_localization}
\end{center}    
\end{table}

\begin{landscape}
\begin{table}[hp!]
\begin{center} 
\scriptsize 
\begin{tabular}{ccccccc}
\toprule
          \thead{GOES \\Start-Stop} & \thead{GOES \\Class} &   \thead{CME First appear. \\ (UT)} & \thead{CME Speed\\($km s^{-1}$)}  &  CME Width (\de)  & LAT Observable & SEP event  \\
\midrule
 2011-02-15 01:44 - 02:06 &       X2.2 &  2011-02-15 02:24 &       669 &      Halo &              X  &      -  \\
 2011-03-09 23:13 - 23:29 &       X1.5 &                -  &        -  &       -   &              X  &       - \\
 2011-09-22 10:29 - 11:44 &       X1.4 &  2011-09-22 10:48 &      1905 &      Halo &              -   &      SEP  \\
 2011-11-03 20:16 - 20:32 &       X1.9 &                -  &        -   &       -    &              -   &        -\\
 2012-07-12 15:37 - 17:30 &       X1.4 &  2012-07-12 16:24 &       843 &        76 &              X  &     SEP   \\
 2013-10-25 14:51 - 15:12 &       X2.1 &  2013-10-25 15:12 &      1081 &      Halo &              -   &     -   \\
 2013-10-29 21:42 - 22:01 &       X2.3 &  2013-10-29 22:00 &      1001 &      Halo &              -   &     -   \\
\bottomrule
\end{tabular}
\caption{X-class GOES flares not associated with any $\gamma$-ray emission above 30~MeV. The \Fermi-LAT observable column indicates whether the prompt phase of the X-ray flare occurred within a \SunMon time window. The SEP event column indicates the presence of this flare in the Major SEP Event list.}
\label{tab:Xray_non_detections}
\end{center}    
\end{table}
\end{landscape}

%
\bibliography{main.bib}
\bibliographystyle{apj}

\end{document}

%% file: tables/Main_Flare_table_insert.tex
\begin{tabular}{lcccccccc}
\toprule
\thead{Name} &   \thead{GOES\\ Class} & \thead{GOES\\ Start-Stop} & \thead{Detection duration\\ (hrs)}  &  \thead{Total duration\\ (hrs)}  & \thead{ Peak Flux\\ ($10^{-5}$cm$^{-2}$s$^{-1}$)}  &    \thead{Fluence $>$ 100 MeV\\ (cm$^{-2}$)}  &                       \thead{Flare Type} &   \thead{Total Protons\\ $>$500 MeV ($10^{27}$)} \\

\midrule
  FLSF 2011-03-07 &  M3.7$^c$ &   19:43 - 20:58 &     13.5 &              $  15.8    \pm  3.1  $   &   $  3.23    \pm  0.22  $   &     $  1.076    \pm  0.029  $   &                          Delayed &                             $  64.4    \pm  1.8  $   \\
  FLSF 2011-06-07 &         M2.5 &   06:16 - 06:59 &      3.8 &               $  6.0    \pm  2.2  $   &   $  3.18    \pm  0.20  $   &     $  0.295    \pm  0.030  $   &                          Delayed &                             $  19.5    \pm  2.0  $   \\
  FLSF 2011-08-04 &         M9.3 &   03:41 - 04:04 &      0.7 &               $  2.3    \pm  0.7  $   &   $  2.30    \pm  0.18  $   &       $  0.13    \pm  0.05  $   &                          Delayed &                                  $  9    \pm  4  $   \\
  FLSF 2011-08-09 &         X6.9 &   07:48 - 08:08 &      0.5 &             $  0.87    \pm  0.34  $   &   $  2.29    \pm  0.23  $   &     $  0.037    \pm  0.018  $   &      Prompt Short-Delayed$^a$ &                              $  2.7    \pm  1.3  $   \\
  FLSF 2011-09-06 &         X2.1 &   22:12 - 22:24 &      0.6 &               $  2.0    \pm  1.4  $   &    $  22.8    \pm  0.4  $   &       $  0.87    \pm  0.17  $   &  LLE-Prompt Short-Delayed$^a$ &                                $  58    \pm  12  $   \\
  FLSF 2011-09-07 &         X1.8 &   22:32 - 22:44 &      0.8 &             $  2.02    \pm  0.35  $   &   $  0.77    \pm  0.08  $   &     $  0.041    \pm  0.014  $   &                          Delayed &                              $  2.3    \pm  0.7  $   \\
\bottomrule
\end{tabular}

%% file: tables/lle_main_table.tex
\begin{tabular}{lccccccc}
 \toprule
\thead{Name} & \thead{Start\\ (UTC)} & \thead{Duration\\(sec)} & \thead{Flux\\(30 MeV -- 10 GeV)} & \thead{Flux\\(100 MeV -- 10 GeV)}  &\thead{Proton \\Index} & \thead{GOES\\ class} & \thead{\SunMon \\ detected}\\ 
 \midrule 
FLSF~2010-06-12 & 2010-06-12 00:55:49 & 30 & $ 446 \pm 35 $  & $ 191 \pm 12 $  & $ 6.0 \pm 0.4 $  & M2.0 & NO \\ 
FLSF~2011-08-09 & 2011-08-09 08:01:51 & 250 & $ 31.20 \pm 0.24 $  & $ 13.02 \pm 0.22 $  & $ 5.68 \pm 0.13 $  & X6.9 & YES \\ 
FLSF~2011-09-06 & 2011-09-06 22:18:07 & 100 & $ 54.0 \pm 1.4 $  & $ 16.6 \pm 1.1 $  & $ 3.2 \pm 0.4 $  & X2.1 & YES \\ 
FLSF~2011-09-24 & 2011-09-24 09:35:53 & 100 & $ 65.2 \pm 1.7 $  & $ 0.43 \pm 0.07 $  & $ 3.2 \pm 0.4 $  & X1.9 & YES \\ 
FLSF~2012-06-03 & 2012-06-03 17:53:20 & 20 & $ 111 \pm 5 $  & $ 50 \pm 5 $  & $ 6.0 \pm 1.5 $  & M3.3 & YES \\ 
FLSF~2012-08-06 & 2012-08-06 04:36:01 & 30 & $ 205 \pm 5 $  & $ 1.79 \pm 0.12 $  & $ 6.0 \pm 1.5 $  & M1.6 & NO \\ 
FLSF~2012-10-23 & 2012-10-23 03:15:33 & 20 & $ \left(3.08 \pm 0.27\right) \times 10^{3} $  & $ 105 \pm 20 $  & $ 6.0 \pm 1.5 $  & X1.8 & YES \\ 
FLSF~2013-10-25b & 2013-10-25 20:56:52 & 10 & $ 38.9 \pm 1.0 $  & $ 1.13 \pm 0.09 $  & $ 6.0 \pm 1.5 $  & M1.9 & NO \\ 
FLSF~2013-10-28a & 2013-10-28 01:59:15 & 70 & $ 0.450 \pm 0.035 $  &   $<3\times10^{-3}$  & $ 6.0 \pm 1.5 $  & X1.0 & NO \\ 
FLSF~2013-10-28b & 2013-10-28 04:37:48 & 50 & $ 25.9 \pm 1.3 $  & $ 0.0029 \pm 0.0016 $  & $ 6.0 \pm 1.5 $  & M5.1 & NO \\ 
FLSF~2013-10-28d & 2013-10-28 20:54:47 & 50 & $ 9.8 \pm 0.6 $  & $ 0.33 \pm 0.05 $  & $ 6.0 \pm 1.5 $  & M1.5 & NO \\ 
FLSF~2014-02-25 & 2014-02-25 00:44:47 & 400 & $ 1407 \pm 25 $  & $ 631 \pm 26 $  & $ 6.0 \pm 0.7 $  & X4.9 & YES \\ 
FLSF~2014-06-10 & 2014-06-10 12:47:18 & 25 & $ 6.7 \pm 1.3 $  & $ 2.9 \pm 1.1 $  & $ 2.2 \pm 1.4 $  & X1.5 & YES \\ 
FLSF~2017-09-10 & 2017-09-10 15:57:47 & 325 & $ 1060 \pm 9 $  & $ 601 \pm 7 $  & $ 3.01 \pm 0.04 $  & X8.2 & YES \\ 
\\ 
\\ 
  \bottomrule
 \end{tabular}

%% file: main.bbl
\newcommand{\noop}[1]{}
\begin{thebibliography}{59}
\expandafter\ifx\csname natexlab\endcsname\relax\def\natexlab#1{#1}\fi

\bibitem[{{Abdo} {et~al.}(2011){Abdo}, {Ackermann}, {Ajello}, {Baldini},
  {Ballet}, {Barbiellini}, {Bastieri}, {Bechtol}, {Bellazzini}, {Berenji},
  {Bonamente}, {Borgland}, {Bouvier}, {Bregeon}, {Brez}, {Brigida}, {Bruel},
  {Buehler}, {Buson}, {Caliandro}, {Cameron}, {Caraveo}, {Casandjian},
  {Cecchi}, {Charles}, {Chekhtman}, {Chiang}, {Ciprini}, {Claus},
  {Cohen-Tanugi}, {Conrad}, {Cutini}, {de Angelis}, {de Palma}, {Dermer},
  {Digel}, {Silva}, {Drell}, {Dubois}, {Favuzzi}, {Fegan}, {Focke}, {Fortin},
  {Frailis}, {Funk}, {Fusco}, {Gargano}, {Gasparrini}, {Gehrels}, {Germani},
  {Giglietto}, {Giordano}, {Giroletti}, {Glanzman}, {Godfrey}, {Grenier},
  {Grillo}, {Guiriec}, {Hadasch}, {Hays}, {Hughes}, {Iafrate},
  {J{\'o}hannesson}, {Johnson}, {Johnson}, {Kamae}, {Katagiri}, {Kataoka},
  {Kn{\"o}dlseder}, {Kuss}, {Lande}, {Latronico}, {Lee}, {Lionetto}, {Longo},
  {Loparco}, {Lott}, {Lovellette}, {Lubrano}, {Makeev}, {Mazziotta}, {McEnery},
  {Mehault}, {Michelson}, {Mitthumsiri}, {Mizuno}, {Moiseev}, {Monte},
  {Monzani}, {Morselli}, {Moskalenko}, {Murgia}, {Nakamori}, {Naumann-Godo},
  {Nolan}, {Norris}, {Nuss}, {Ohsugi}, {Okumura}, {Omodei}, {Orlando}, {Ormes},
  {Ozaki}, {Paneque}, {Pelassa}, {Pesce-Rollins}, {Pierbattista}, {Piron},
  {Porter}, {Rain{\`o}}, {Rando}, {Razzano}, {Reimer}, {Reimer}, {Reposeur},
  {Ritz}, {Sadrozinski}, {Schalk}, {Sgr{\`o}}, {Share}, {Siskind}, {Smith},
  {Spandre}, {Spinelli}, {Strickman}, {Strong}, {Takahashi}, {Tanaka},
  {Thayer}, {Thayer}, {Thompson}, {Tibaldo}, {Torres}, {Tosti}, {Tramacere},
  {Troja}, {Uchiyama}, {Usher}, {Vandenbroucke}, {Vasileiou}, {Vianello},
  {Vilchez}, {Vitale}, {Vladimirov}, {Waite}, {Wang}, {Winer}, {Wood}, {Yang},
  \& {Ziegler}}]{2011ApJ...734..116A}
{Abdo}, A.~A., {Ackermann}, M., {Ajello}, M., {et~al.} 2011, \apj, 734, 116

\bibitem[{{Abeysekara} {et~al.}(2018){Abeysekara}, {Archer}, {Benbow}, {Bird},
  {Brose}, {Buchovecky}, {Buckley}, {Bugaev}, {Chromey}, {Connolly}, {Cui},
  {Daniel}, {Falcone}, {Feng}, {Finley}, {Fortson}, {Furniss}, {H{\"u}tten},
  {Hanna}, {Hervet}, {Holder}, {Hughes}, {Humensky}, {Johnson}, {Kaaret},
  {Kar}, {Kertzman}, {Kieda}, {Krause}, {Krennrich}, {Kumar}, {Lang}, {Lin},
  {McArthur}, {Moriarty}, {Mukherjee}, {O'Brien}, {Ong}, {Otte}, {Park},
  {Petrashyk}, {Pohl}, {Pueschel}, {Quinn}, {Ragan}, {Reynolds}, {Richards},
  {Roache}, {Rulten}, {Sadeh}, {Santander}, {Sembroski}, {Shahinyan}, {Sushch},
  {Tyler}, {Wakely}, {Weinstein}, {Wells}, {Wilcox}, {Wilhelm}, {Williams},
  {Williamson}, {Zitzer}, {VERITAS Collaboration}, {Abdollahi}, {Ajello},
  {Baldini}, {Barbiellini}, {Bastieri}, {Bellazzini}, {Berenji}, {Bissaldi},
  {Bland ford}, {Bonino}, {Bottacini}, {Brandt}, {Bruel}, {Buehler}, {Cameron},
  {Caputo}, {Caraveo}, {Castro}, {Cavazzuti}, {Charles}, {Chiaro}, {Ciprini},
  {Cohen-Tanugi}, {Costantin}, {Cutini}, {D'Ammand o}, {de Palma}, {Di Lalla},
  {Di Mauro}, {Di Venere}, {Dom{\'\i}nguez}, {Favuzzi}, {Fegan}, {Franckowiak},
  {Fukazawa}, {Funk}, {Fusco}, {Gargano}, {Gasparrini}, {Giglietto},
  {Giordano}, {Giroletti}, {Green}, {Grenier}, {Guillemot}, {Guiriec}, {Hays},
  {Hewitt}, {Horan}, {J{\'o}hannesson}, {Kensei}, {Kuss}, {Larsson},
  {Latronico}, {Lemoine-Goumard}, {Li}, {Longo}, {Loparco}, {Lovellette},
  {Lubrano}, {Magill}, {Maldera}, {Mazziotta}, {McEnery}, {Michelson},
  {Mitthumsiri}, {Mizuno}, {Monzani}, {Morselli}, {Moskalenko}, {Negro},
  {Nuss}, {Ojha}, {Omodei}, {Orienti}, {Orlando}, {Palatiello}, {Paliya},
  {Paneque}, {Perkins}, {Persic}, {Pesce-Rollins}, {Petrosian}, {Piron},
  {Porter}, {Principe}, {Rain{\`o}}, {Rando}, {Rani}, {Razzano}, {Razzaque},
  {Reimer}, {Reimer}, {Reposeur}, {Sgr{\`o}}, {Siskind}, {Spandre}, {Spinelli},
  {Suson}, {Tajima}, {Thayer}, {Thompson}, {Torres}, {Tosti}, {Troja},
  {Valverde}, {Vianello}, {Vogel}, {Wood}, {Yassine}, {Fermi-LAT
  Collaboration}, {Alfaro}, {{\'A}lvarez}, {{\'A}lvarez}, {Arceo},
  {Arteaga-Vel{\'a}zquez}, {Avila Rojas}, {Ayala Solares}, {Becerril},
  {Belmont-Moreno}, {BenZvi}, {Bernal}, {Braun}, {Brisbois}, {Caballero-Mora},
  {Capistr{\'a}n}, {Carrami{\~n}ana}, {Casanova}, {Castillo}, {Cotti},
  {Cotzomi}, {Couti{\~n}o de Le{\'o}n}, {De Le{\'o}n}, {De la Fuente},
  {Dichiara}, {Dingus}, {DuVernois}, {D{\'\i}az-V{\'e}lez}, {Engel},
  {Enriquez-Rivera}, {Fiorino}, {Fleischhack}, {Fraija},
  {Garc{\'\i}a-Gonz{\'a}lez}, {Garfias}, {Gonz{\'a}lez Mu{\~n}oz},
  {Gonz{\'a}lez}, {Goodman}, {Hampel-Arias}, {Harding}, {Hernand ez},
  {Hernandez-Almada}, {Hona}, {Hueyotl-Zahuantitla}, {Hui}, {H{\"u}ntemeyer},
  {Iriarte}, {Jardin-Blicq}, {Joshi}, {Kaufmann}, {Lara}, {Lauer}, {Lee},
  {Lennarz}, {Le{\'o}n Vargas}, {Linnemann}, {Longinotti}, {Luis-Raya},
  {Luna-Garc{\'\i}a}, {L{\'o}pez-Coto}, {Malone}, {Marinelli}, {Martinez},
  {Martinez-Castellanos}, {Mart{\'\i}nez-Castro}, {Mart{\'\i}nez-Huerta},
  {Matthews}, {Miranda-Romagnoli}, {Moreno}, {Mostaf{\'a}}, {Nayerhoda},
  {Nellen}, {Newbold}, {Nisa}, {Noriega-Papaqui}, {Pelayo}, {Pretz},
  {P{\'e}rez-P{\'e}rez}, {Ren}, {Rho}, {Rivi{\`e}re}, {Rosa-Gonz{\'a}lez},
  {Rosenberg}, {Ruiz-Velasco}, {Salazar}, {Salesa Greus}, {Sandoval},
  {Schneider}, {Seglar Arroyo}, {Sinnis}, {Smith}, {Springer}, {Surajbali},
  {Taboada}, {Tibolla}, {Tollefson}, {Torres}, {Ukwatta}, {Villase{\~n}or},
  {Weisgarber}, {Westerhoff}, {Wisher}, {Wood}, {Yapici}, {Yodh}, {Zepeda},
  {Zhou}, \& {HAWC Collaboration}}]{2018ApJ...866...24A}
{Abeysekara}, A.~U., {Archer}, A., {Benbow}, W., {et~al.} 2018, \apj, 866, 24

\bibitem[{{Ackermann} {et~al.}(2012{\natexlab{a}}){Ackermann}, {Ajello},
  {Allafort}, {Atwood}, {Baldini}, {Barbiellini}, {Bastieri}, {Bechtol},
  {Bellazzini}, {Bhat}, {Blandford}, {Bonamente}, {Borgland}, {Bregeon},
  {Briggs}, {Brigida}, {Bruel}, {Buehler}, {Burgess}, {Buson}, {Caliandro},
  {Cameron}, {Casandjian}, {Cecchi}, {Charles}, {Chekhtman}, {Chiang},
  {Ciprini}, {Claus}, {Cohen-Tanugi}, {Connaughton}, {Conrad}, {Cutini},
  {Dennis}, {de Palma}, {Dermer}, {Digel}, {Silva}, {Drell}, {Drlica-Wagner},
  {Dubois}, {Favuzzi}, {Fegan}, {Ferrara}, {Fortin}, {Fukazawa}, {Fusco},
  {Gargano}, {Germani}, {Giglietto}, {Giordano}, {Giroletti}, {Glanzman},
  {Godfrey}, {Grillo}, {Grove}, {Gruber}, {Guiriec}, {Hadasch}, {Hayashida},
  {Hays}, {Horan}, {Iafrate}, {J{\'o}hannesson}, {Johnson}, {Johnson}, {Kamae},
  {Kippen}, {Kn{\"o}dlseder}, {Kuss}, {Lande}, {Latronico}, {Longo}, {Loparco},
  {Lott}, {Lovellette}, {Lubrano}, {Mazziotta}, {McEnery}, {Meegan}, {Mehault},
  {Michelson}, {Mitthumsiri}, {Monte}, {Monzani}, {Morselli}, {Moskalenko},
  {Murgia}, {Murphy}, {Naumann-Godo}, {Nuss}, {Nymark}, {Ohno}, {Ohsugi},
  {Okumura}, {Omodei}, {Orlando}, {Paciesas}, {Panetta}, {Parent},
  {Pesce-Rollins}, {Petrosian}, {Pierbattista}, {Piron}, {Pivato}, {Poon},
  {Porter}, {Preece}, {Rain{\`o}}, {Rando}, {Razzano}, {Razzaque}, {Reimer},
  {Reimer}, {Ritz}, {Sbarra}, {Schwartz}, {Sgr{\`o}}, {Share}, {Siskind},
  {Spinelli}, {Takahashi}, {Tanaka}, {Tanaka}, {Thayer}, {Tibaldo},
  {Tinivella}, {Tolbert}, {Tosti}, {Troja}, {Uchiyama}, {Usher},
  {Vandenbroucke}, {Vasileiou}, {Vianello}, {Vitale}, {von Kienlin}, {Waite},
  {Wilson-Hodge}, {Wood}, {Wood}, \& {Yang}}]{2012ApJ...745..144A}
{Ackermann}, M., {Ajello}, M., {Allafort}, A., {et~al.} 2012{\natexlab{a}},
  \apj, 745, 144

\bibitem[{{Ackermann} {et~al.}(2012{\natexlab{b}}){Ackermann}, {Ajello},
  {Albert}, {Allafort}, {Atwood}, {Axelsson}, {Baldini}, {Ballet},
  {Barbiellini}, {Bastieri}, {Bechtol}, {Bellazzini}, {Bissaldi}, {Blandford},
  {Bloom}, {Bogart}, {Bonamente}, {Borgland}, {Bottacini}, {Bouvier}, {Brandt},
  {Bregeon}, {Brigida}, {Bruel}, {Buehler}, {Burnett}, {Buson}, {Caliandro},
  {Cameron}, {Caraveo}, {Casandjian}, {Cavazzuti}, {Cecchi}, {{\c C}elik},
  {Charles}, {Chaves}, {Chekhtman}, {Cheung}, {Chiang}, {Ciprini}, {Claus},
  {Cohen-Tanugi}, {Conrad}, {Corbet}, {Cutini}, {D'Ammando}, {Davis}, {de
  Angelis}, {DeKlotz}, {de Palma}, {Dermer}, {Digel}, {Silva}, {Drell},
  {Drlica-Wagner}, {Dubois}, {Favuzzi}, {Fegan}, {Ferrara}, {Focke}, {Fortin},
  {Fukazawa}, {Funk}, {Fusco}, {Gargano}, {Gasparrini}, {Gehrels}, {Giebels},
  {Giglietto}, {Giordano}, {Giroletti}, {Glanzman}, {Godfrey}, {Grenier},
  {Grove}, {Guiriec}, {Hadasch}, {Hayashida}, {Hays}, {Horan}, {Hou}, {Hughes},
  {Jackson}, {Jogler}, {J{\'o}hannesson}, {Johnson}, {Johnson}, {Johnson},
  {Kamae}, {Katagiri}, {Kataoka}, {Kerr}, {Kn{\"o}dlseder}, {Kuss}, {Lande},
  {Larsson}, {Latronico}, {Lavalley}, {Lemoine-Goumard}, {Longo}, {Loparco},
  {Lott}, {Lovellette}, {Lubrano}, {Mazziotta}, {McConville}, {McEnery},
  {Mehault}, {Michelson}, {Mitthumsiri}, {Mizuno}, {Moiseev}, {Monte},
  {Monzani}, {Morselli}, {Moskalenko}, {Murgia}, {Naumann-Godo}, {Nemmen},
  {Nishino}, {Norris}, {Nuss}, {Ohno}, {Ohsugi}, {Okumura}, {Omodei},
  {Orienti}, {Orlando}, {Ormes}, {Paneque}, {Panetta}, {Perkins},
  {Pesce-Rollins}, {Pierbattista}, {Piron}, {Pivato}, {Porter}, {Racusin},
  {Rain{\`o}}, {Rando}, {Razzano}, {Razzaque}, {Reimer}, {Reimer}, {Reposeur},
  {Reyes}, {Ritz}, {Rochester}, {Romoli}, {Roth}, {Sadrozinski}, {Sanchez},
  {Saz Parkinson}, {Sbarra}, {Scargle}, {Sgr{\`o}}, {Siegal-Gaskins},
  {Siskind}, {Spandre}, {Spinelli}, {Stephens}, {Suson}, {Tajima}, {Takahashi},
  {Tanaka}, {Thayer}, {Thayer}, {Thompson}, {Tibaldo}, {Tinivella}, {Tosti},
  {Troja}, {Usher}, {Vandenbroucke}, {Van Klaveren}, {Vasileiou}, {Vianello},
  {Vitale}, {Waite}, {Wallace}, {Winer}, {Wood}, {Wood}, {Wood}, {Yang}, \&
  {Zimmer}}]{2012ApJS..203....4A}
{Ackermann}, M., {Ajello}, M., {Albert}, A., {et~al.} 2012{\natexlab{b}},
  \apjs, 203, 4

\bibitem[{{Ackermann} {et~al.}(2014{\natexlab{a}}){Ackermann}, Ajello, Albert,
  Allafort, Baldini, Barbiellini, Bastieri, Bechtol, Bellazzini, Bissaldi,
  Bonamente, Bottacini, Bouvier, Brandt, Bregeon, Brigida, Bruel, Buehler,
  Buson, Caliandro, Cameron, Caraveo, Cecchi, Charles, Chekhtman, Chen, Chiang,
  Chiaro, Ciprini, Claus, Cohen-Tanugi, Conrad, Cutini, D'Ammando, de~Angelis,
  de~Palma, Dermer, Desiante, Digel, Venere, do~Couto~e Silva, Drell,
  Drlica-Wagner, Favuzzi, Fegan, Focke, Franckowiak, Fukazawa, Funk, Fusco,
  Gargano, Gasparrini, Germani, Giglietto, Giordano, Giroletti, Glanzman,
  Godfrey, Grenier, Grove, Guiriec, Hadasch, Hayashida, Hays, Horan, Hughes,
  Inoue, Jackson, Jogler, J{\'o}hannesson, Johnson, Kamae, Kawano,
  Kn{\"o}dlseder, Kuss, Lande, Larsson, Latronico, Lemoine-Goumard, Longo,
  Loparco, Lott, Lovellette, Lubrano, Mayer, Mazziotta, McEnery, Michelson,
  Mizuno, Moiseev, Monte, Monzani, Moretti, Morselli, Moskalenko, Murgia,
  Murphy, Nemmen, Nuss, Ohno, Ohsugi, Okumura, Omodei, Orienti, Orlando, Ormes,
  Paneque, Panetta, Perkins, Pesce-Rollins, Petrosian, Piron, Pivato, Porter,
  Rain{\`o}, Rando, Razzano, Reimer, Reimer, Ritz, Schulz, Sgr{\`o}, Siskind,
  Spandre, Spinelli, Takahashi, Takeuchi, Tanaka, Thayer, Thayer, Thompson,
  Tibaldo, Tinivella, Tosti, Troja, Tronconi, Usher, Vandenbroucke, Vasileiou,
  Vianello, Vitale, Werner, Winer, Wood, Wood, Wood, \&
  Yang}]{0004-637X-787-1-15}
{Ackermann}, M., Ajello, M., Albert, A., {et~al.} 2014{\natexlab{a}}, \apj,
  787, 15

\bibitem[{{Ackermann} {et~al.}(2014{\natexlab{b}}){Ackermann}, {Ajello},
  {Albert}, {Allafort}, {Baldini}, {Barbiellini}, {Bastieri}, {Bechtol},
  {Bellazzini}, {Bissaldi}, {Bonamente}, {Bottacini}, {Bouvier}, {Brandt},
  {Bregeon}, {Brigida}, {Bruel}, {Buehler}, {Buson}, {Caliandro}, {Cameron},
  {Caraveo}, {Cecchi}, {Charles}, {Chekhtman}, {Chen}, {Chiang}, {Chiaro},
  {Ciprini}, {Claus}, {Cohen-Tanugi}, {Conrad}, {Cutini}, {D'Ammando}, {de
  Angelis}, {de Palma}, {Dermer}, {Desiante}, {Digel}, {Di Venere}, {Silva},
  {Drell}, {Drlica-Wagner}, {Favuzzi}, {Fegan}, {Focke}, {Franckowiak},
  {Fukazawa}, {Funk}, {Fusco}, {Gargano}, {Gasparrini}, {Germani}, {Giglietto},
  {Giordano}, {Giroletti}, {Glanzman}, {Godfrey}, {Grenier}, {Grove},
  {Guiriec}, {Hadasch}, {Hayashida}, {Hays}, {Horan}, {Hughes}, {Inoue},
  {Jackson}, {Jogler}, {J{\'o}hannesson}, {Johnson}, {Kamae}, {Kawano},
  {Kn{\"o}dlseder}, {Kuss}, {Lande}, {Larsson}, {Latronico}, {Lemoine-Goumard},
  {Longo}, {Loparco}, {Lott}, {Lovellette}, {Lubrano}, {Mayer}, {Mazziotta},
  {McEnery}, {Michelson}, {Mizuno}, {Moiseev}, {Monte}, {Monzani}, {Moretti},
  {Morselli}, {Moskalenko}, {Murgia}, {Murphy}, {Nemmen}, {Nuss}, {Ohno},
  {Ohsugi}, {Okumura}, {Omodei}, {Orienti}, {Orlando}, {Ormes}, {Paneque},
  {Panetta}, {Perkins}, {Pesce-Rollins}, {Petrosian}, {Piron}, {Pivato},
  {Porter}, {Rain{\`o}}, {Rando}, {Razzano}, {Reimer}, {Reimer}, {Ritz},
  {Schulz}, {Sgr{\`o}}, {Siskind}, {Spandre}, {Spinelli}, {Takahashi},
  {Takeuchi}, {Tanaka}, {Thayer}, {Thayer}, {Thompson}, {Tibaldo}, {Tinivella},
  {Tosti}, {Troja}, {Tronconi}, {Usher}, {Vandenbroucke}, {Vasileiou},
  {Vianello}, {Vitale}, {Werner}, {Winer}, {Wood}, {Wood}, {Wood}, \&
  {Yang}}]{2014ApJ...787...15A}
{Ackermann}, M., {Ajello}, M., {Albert}, A., {et~al.} 2014{\natexlab{b}}, \apj,
  787, 15

\bibitem[{{Ackermann} {et~al.}(2017){Ackermann}, {Allafort}, {Baldini},
  {Barbiellini}, {Bastieri}, {Bellazzini}, {Bissaldi}, {Bonino}, {Bottacini},
  {Bregeon}, {Bruel}, {Buehler}, {Cameron}, {Caragiulo}, {Caraveo},
  {Cavazzuti}, {Cecchi}, {Charles}, {Ciprini}, {Costanza}, {Cutini},
  {D'Ammando}, {de Palma}, {Desiante}, {Digel}, {Di Lalla}, {Di Mauro}, {Di
  Venere}, {Drell}, {Favuzzi}, {Fukazawa}, {Fusco}, {Gargano}, {Giglietto},
  {Giordano}, {Giroletti}, {Grenier}, {Guillemot}, {Guiriec}, {Jogler},
  {J{\'o}hannesson}, {Kashapova}, {Krucker}, {Kuss}, {La Mura}, {Larsson},
  {Latronico}, {Li}, {Liu}, {Longo}, {Loparco}, {Lubrano}, {Magill}, {Maldera},
  {Manfreda}, {Mazziotta}, {Mitthumsiri}, {Mizuno}, {Monzani}, {Morselli},
  {Moskalenko}, {Negro}, {Nuss}, {Ohsugi}, {Omodei}, {Orlando}, {Pal'shin},
  {Paneque}, {Perkins}, {Pesce-Rollins}, {Petrosian}, {Piron}, {Principe},
  {Rain{\`o}}, {Rando}, {Razzano}, {Reimer}, {Rubio da Costa}, {Sgr{\`o}},
  {Simone}, {Siskind}, {Spada}, {Spandre}, {Spinelli}, {Tajima}, {Thayer},
  {Torres}, {Troja}, \& {Vianello}}]{2017ApJ...835..219A}
{Ackermann}, M., {Allafort}, A., {Baldini}, L., {et~al.} 2017, \apj, 835, 219

\bibitem[{{Ackermann} {et~al.}(2018){Ackermann}, {Ajello}, {Baldini}, {Ballet},
  {Barbiellini}, {Bastieri}, {Bellazzini}, {Bissaldi}, {Blandford}, {Bloom},
  {Bonino}, {Bottacini}, {Brandt}, {Bregeon}, {Bruel}, {Buehler}, {Cameron},
  {Caputo}, {Caraveo}, {Castro}, {Cavazzuti}, {Charles}, {Cheung}, {Chiaro},
  {Ciprini}, {Cohen-Tanugi}, {Costantin}, {Cutini}, {D'Ammand o}, {de Palma},
  {Desai}, {Di Lalla}, {Di Mauro}, {Di Venere}, {Favuzzi}, {Finke},
  {Franckowiak}, {Fukazawa}, {Funk}, {Fusco}, {Gargano}, {Gasparrini},
  {Giglietto}, {Giordano}, {Giroletti}, {Green}, {Grenier}, {Guillemot},
  {Guiriec}, {Hays}, {Hewitt}, {Horan}, {J{\'o}hannesson}, {Kensei}, {Kuss},
  {Larsson}, {Latronico}, {Lemoine-Goumard}, {Li}, {Longo}, {Loparco},
  {Lovellette}, {Lubrano}, {Magill}, {Maldera}, {Manfreda}, {Mazziotta},
  {McEnery}, {Meyer}, {Mizuno}, {Monzani}, {Morselli}, {Moskalenko}, {Negro},
  {Nuss}, {Omodei}, {Orienti}, {Orlando}, {Ormes}, {Palatiello}, {Paliya},
  {Paneque}, {Perkins}, {Persic}, {Pesce-Rollins}, {Piron}, {Porter},
  {Principe}, {Rain{\`o}}, {Rando}, {Rani}, {Razzaque}, {Reimer}, {Reimer},
  {Reposeur}, {Sgr{\`o}}, {Siskind}, {Spandre}, {Spinelli}, {Suson}, {Tajima},
  {Thayer}, {Tibaldo}, {Torres}, {Tosti}, {Valverde}, {Venters}, {Vogel},
  {Wood}, {Wood}, {Zaharijas}, {Fermi-LAT Collaboration}, \&
  {Biteau}}]{2018ApJS..237...32A}
{Ackermann}, M., {Ajello}, M., {Baldini}, L., {et~al.} 2018, \apjs, 237, 32

\bibitem[{{Ahnen} {et~al.}(2019){Ahnen}, {Ansoldi}, {Antonelli}, {Arcaro},
  {Baack}, {Babi{\'c}}, {}, {Banerjee}, {Bangale}, {Barres de Almeida},
  {Barrio}, {Becerra Gonz{\'a}lez}, {Bednarek}, {Bernardini}, {Berse}, {Berti},
  {Bhattacharyya}, {Biland}, {Blanch}, {Bonnoli}, {Carosi}, {Carosi},
  {Ceribella}, {Chatterjee}, {Colak}, {Colin}, {Colombo}, {Contreras},
  {Cortina}, {Covino}, {Cumani}, {da Vela}, {Dazzi}, {de Angelis}, {de Lotto},
  {Delfino}, {Delgado}, {di Pierro}, {Dom{\'\i}nguez}, {Dominis Prester},
  {Dorner}, {Doro}, {Einecke}, {Elsaesser}, {Fallah Ramazani},
  {Fern{\'a}ndez-Barral}, {Fidalgo}, {Fonseca}, {Font}, {Fruck}, {Galindo},
  {Garc{\'\i}a L{\'o}pez}, {Garczarczyk}, {Gaug}, {Giammaria}, {Godinovi{\'c}},
  {}, {Gora}, {Guberman}, {Hadasch}, {Hahn}, {Hassan}, {Hayashida}, {Herrera},
  {Hose}, {Hrupec}, {Ishio}, {Konno}, {Kubo}, {Kushida}, {Kuve{\v{z}}di{\'c}},
  {}, {Lelas}, {Lindfors}, {Lombardi}, {Longo}, {L{\'o}pez}, {Maggio},
  {Majumdar}, {Makariev}, {Maneva}, {Manganaro}, {Mannheim}, {Maraschi},
  {Mariotti}, {Mart{\'\i}nez}, {Masuda}, {Mazin}, {Mielke}, {Minev}, {Miranda},
  {Mirzoyan}, {Moralejo}, {Moreno}, {Moretti}, {Nagayoshi}, {Neustroev},
  {Niedzwiecki}, {Nievas Rosillo}, {Nigro}, {Nilsson}, {Ninci}, {Nishijima},
  {Noda}, {Nogu{\'e}s}, {Paiano}, {Palacio}, {Paneque}, {Paoletti}, {Paredes},
  {Pedaletti}, {Peresano}, {Persic}, {Prada Moroni}, {Prand ini}, {Puljak},
  {Garcia}, {Reichardt}, {Rhode}, {Rib{\'o}}, {Rico}, {Righi}, {Rugliancich},
  {Saito}, {Satalecka}, {Schweizer}, {Sitarek}, {{\v{S}}nidari{\'c}}, {},
  {Sobczynska}, {Stamerra}, {Strzys}, {Suri{\'c}}, {}, {Takahashi}, {Takalo},
  {Tavecchio}, {Temnikov}, {Terzi{\'c}}, {}, {Teshima}, {Torres-Alb{\`a}},
  {Treves}, {Tsujimoto}, {Vanzo}, {Vazquez Acosta}, {Vovk}, {Ward}, {Will},
  {Zari{\'c}}, {}, {MAGIC Collaboration}, {Albert}, {Alfaro}, {Alvarez},
  {Arceo}, {Arteaga-Vel{\'a}zquez}, {Avila Rojas}, {Ayala Solares}, {Becerril},
  {Belmont-Moreno}, {Benzvi}, {Bernal}, {Braun}, {Caballero-Mora},
  {Capistr{\'a}n}, {Carrami{\~n}ana}, {Casanova}, {Castillo}, {Cotti},
  {Cotzomi}, {Couti{\~n}o de Le{\'o}n}, {de Le{\'o}n}, {de La Fuente}, {Diaz
  Hernandez}, {Dichiara}, {Dingus}, {Duvernois}, {D{\'\i}az-V{\'e}lez},
  {Ellsworth}, {Engel}, {Enriquez-Rivera}, {Fiorino}, {Fleischhack}, {Fraija},
  {Garc{\'\i}a-Gonz{\'a}lez}, {Garfias}, {Gonz{\'a}lez-Mu{\~n}oz},
  {Gonz{\'a}lez}, {Goodman}, {Hampel-Arias}, {Harding}, {Hernand ez},
  {Hueyotl-Zahuantitla}, {Hui}, {H{\"u}ntemeyer}, {Iriarte}, {Jardin-Blicq},
  {Joshi}, {Kaufmann}, {Lara}, {Lauer}, {Lee}, {Lennarz}, {Le{\'o}n Vargas},
  {Linnemann}, {Longinotti}, {Luis-Raya}, {Luna-Garc{\'\i}a}, {L{\'o}pez-Coto},
  {Malone}, {Marinelli}, {Martinez}, {Martinez-Castellanos},
  {Mart{\'\i}nez-Castro}, {Mart{\'\i}nez-Huerta}, {Matthews},
  {Miranda-Romagnoli}, {Moreno}, {Mostaf{\'a}}, {Nayerhoda}, {Nellen},
  {Newbold}, {Nisa}, {Noriega-Papaqui}, {Pelayo}, {Pretz},
  {P{\'e}rez-P{\'e}rez}, {Ren}, {Rho}, {Rivi{\`e}re}, {Rosa-Gonz{\'a}lez},
  {Rosenberg}, {Ruiz-Velasco}, {Salesa Greus}, {Sandoval}, {Schneider}, {Seglar
  Arroyo}, {Sinnis}, {Smith}, {Springer}, {Surajbali}, {Taboada}, {Tibolla},
  {Tollefson}, {Torres}, {Ukwatta}, {Vianello}, {Villase{\~n}or}, {Werner},
  {Westerhoff}, {Wood}, {Yapici}, {Yodh}, {Zepeda}, {Zhou}, {{\'A}lvarez},
  {Hawc Collaboration}, {Ajello}, {Baldini}, {Barbiellini}, {Berenji},
  {Bissaldi}, {Bland ford}, {Bonino}, {Bottacini}, {Brandt}, {Bregeon},
  {Bruel}, {Cameron}, {Caputo}, {Caraveo}, {Castro}, {Cavazzuti}, {Chiaro},
  {Ciprini}, {Costantin}, {D'Ammando}, {de Palma}, {Desai}, {di Lalla}, {di
  Mauro}, {di Venere}, {Dom{\'\i}nguez}, {Favuzzi}, {Fukazawa}, {Funk},
  {Fusco}, {Gargano}, {Gasparrini}, {Giglietto}, {Giordano}, {Giroletti},
  {Glanzman}, {Green}, {Grenier}, {Guiriec}, {Harding}, {Hays}, {Hewitt},
  {Horan}, {J{\'o}hannesson}, {Kuss}, {Larsson}, {Liodakis}, {Longo},
  {Loparco}, {Lubrano}, {Magill}, {Maldera}, {Manfreda}, {Mazziotta}, {Mereu},
  {Michelson}, {Mizuno}, {Monzani}, {Morselli}, {Moskalenko}, {Negro}, {Nuss},
  {Omodei}, {Orienti}, {Orlando}, {Ormes}, {Palatiello}, {Paliya}, {Persic},
  {Pesce-Rollins}, {Petrosian}, {Piron}, {Porter}, {Principe}, {Rain{\`o}},
  {Rani}, {Razzano}, {Razzaque}, {Reimer}, {Reimer}, {Sgr{\`o}}, {Siskind},
  {Spandre}, {Spinelli}, {Tajima}, {Takahashi}, {Thayer}, {Thompson}, {Torres},
  {Torresi}, {Troja}, {Valverde}, {Wood}, {Yassine}, \& {Fermi-Lat
  Collaboration}}]{2019MNRAS.485..356A}
{Ahnen}, M.~L., {Ansoldi}, S., {Antonelli}, L.~A., {et~al.} 2019, \mnras, 485,
  356

\bibitem[{{Ajello} {et~al.}(2014){Ajello}, {Albert}, {Allafort}, {Baldini},
  {Barbiellini}, {Bastieri}, {Bellazzini}, {Bissaldi}, {Bonamente}, {Brandt},
  {Bregeon}, {Brigida}, {Bruel}, {Buehler}, {Buson}, {Caliandro}, {Cameron},
  {Caraveo}, {Cecchi}, {Charles}, {Chekhtman}, {Chiang}, {Chiaro}, {Ciprini},
  {Claus}, {Cohen-Tanugi}, {Cominsky}, {Conrad}, {Cutini}, {D'Ammando}, {de
  Palma}, {Dermer}, {Desiante}, {Digel}, {Silva}, {Drell}, {Drlica-Wagner},
  {Favuzzi}, {Focke}, {Franckowiak}, {Fukazawa}, {Fusco}, {Gargano},
  {Gasparrini}, {Germani}, {Giglietto}, {Giommi}, {Giordano}, {Giroletti},
  {Glanzman}, {Godfrey}, {Grenier}, {Grove}, {Guiriec}, {Hadasch}, {Hayashida},
  {Hays}, {Horan}, {Hou}, {Hughes}, {Inoue}, {Jackson}, {Jogler},
  {J{\'o}hannesson}, {Johnson}, {Johnson}, {Kamae}, {Kn{\"o}dlseder},
  {Kocevski}, {Kuss}, {Lande}, {Larsson}, {Latronico}, {Longo}, {Loparco},
  {Lott}, {Lovellette}, {Lubrano}, {Mayer}, {Mazziotta}, {McEnery},
  {Michelson}, {Mizuno}, {Moiseev}, {Monte}, {Monzani}, {Morselli},
  {Moskalenko}, {Murgia}, {Murphy}, {Nakamori}, {Nemmen}, {Nuss}, {Ohno},
  {Ohsugi}, {Omodei}, {Orienti}, {Orland o}, {Ormes}, {Paneque}, {Panetta},
  {Perkins}, {Pesce-Rollins}, {Petrosian}, {Piron}, {Pivato}, {Porter},
  {Rain{\`o}}, {Rando}, {Razzano}, {Reimer}, {Reimer}, {Roth}, {Schulz},
  {Sgr{\`o}}, {Siskind}, {Spandre}, {Spinelli}, {Takahashi}, {Thayer},
  {Thayer}, {Thompson}, {Tibaldo}, {Tinivella}, {Tosti}, {Troja}, {Usher},
  {Vandenbroucke}, {Vasileiou}, {Vianello}, {Vitale}, {Werner}, {Winer},
  {Wood}, {Wood}, \& {Yang}}]{0004-637X-789-1-20}
{Ajello}, M., {Albert}, A., {Allafort}, A., {et~al.} 2014, \apj, 789, 20

\bibitem[{Ajello {et~al.}(2020)Ajello, Baldini, Bastieri, Bellazzini, Berretta,
  Bissaldi, Blandford, Bonino, Bruel, Cameron, Caputo, Cavazzuti, Cheung,
  Chiaro, Costantin, Cutini, D'Ammando, De~Palma, Di~Lalla, Dirirsa, Di~Venere,
  Fegan, Fukazawa, Funk, Fusco, Gargano, Gasparrini, Giordano, Giroletti,
  Green, Guirec, Hayes, Hewitt, Horan, Johannesson, Kovac'evic, Kuss, Larsson,
  Latronico, Li, Longo, Lovellette, Lubrano, Maldera, Manfreda, Marti-Devesa,
  Mazziotta, Mereu, Michelson, Mizuno, Monzani, Morselli, Moskalenko, Negro,
  Omodei, Orienti, Orlando, Paneque, Pei, Persic, Pesce-Rollins, Petrosian,
  Piron, Porter, Principe, Racusin, Raino', Rando, Rani, Razzano, Razzaque,
  Reimer, Reimer, Serini, Sgro', Siskind, Spandre, Spinelli, Tak, Troja,
  Valverde, Wood, \& Zaharijas}]{ajello_m_2020_4311157}
Ajello, M., Baldini, L., Bastieri, D., {et~al.} 2020, The First Fermi-LAT Solar
  Flare Catalog Appendix

\bibitem[{{Arnaud}(1996)}]{XSPEC}
{Arnaud}, K.~A. 1996, in Astronomical Society of the Pacific Conference Series,
  Vol. 101, Astronomical Data Analysis Software and Systems V, ed. G.~H.
  {Jacoby} \& J.~{Barnes}, 17

\bibitem[{{Atwood} {et~al.}(2013){Atwood}, {Albert}, {Baldini}, {Tinivella},
  {Bregeon}, {Pesce-Rollins}, {Sgr{\`o}}, {Bruel}, {Charles}, {Drlica-Wagner},
  {Franckowiak}, {Jogler}, {Rochester}, {Usher}, {Wood}, {Cohen-Tanugi}, \&
  {Zimmer}}]{Pass8}
{Atwood}, W., {Albert}, A., {Baldini}, L., {et~al.} 2013, arXiv e-prints,
  arXiv:1303.3514

\bibitem[{{Atwood} {et~al.}(2009){Atwood}, {Abdo}, {Ackermann}, {Ajello},
  {Baldini}, {Ballet}, {Barbiellini}, {Baring}, {Bastieri}, {Bechtol},
  {Bellazzini}, {Berenji}, {Bhat}, {Bissaldi}, {Blandford}, {Bonamente},
  {Bonnell}, {Borgland}, {Bouvier}, {Bregeon}, {Brigida}, {Bruel}, {Buehler},
  {Buson}, {Caliandro}, {Cameron}, {Caraveo}, {Casandjian}, {Cecchi},
  {Charles}, {Chekhtman}, {Chiang}, {Ciprini}, {Claus}, {Connaughton},
  {Conrad}, {Cutini}, {de Angelis}, {de Palma}, {Dermer}, {Silva}, {Drell},
  {Dubois}, {Favuzzi}, {Fukazawa}, {Fusco}, {Gargano}, {Gehrels}, {Germani},
  {Giglietto}, {Giommi}, {Giordano}, {Giroletti}, {Glanzman}, { Godfrey},
  {Granot}, {Grenier}, {Guiriec}, {Hadasch}, {Hanabata}, {Hughes},
  {J{\'o}hannesson}, {Johnson}, {Kamae}, {Katagiri}, {Kataoka}, {Kerr},
  {Kn{\"o}dlseder}, {Kuss}, {Lande}, {Latronico}, {Lee}, {Longo}, {Loparco},
  {Lott}, {Lubrano}, {Mazziotta}, {McEnery}, {M{\'e}sz{\'a}ros}, {Michelson},
  {Mizuno}, {Moiseev}, {Monzani}, {Morselli}, {Moskalenko}, {Murgia},
  {Nakamori}, {Naumann-Godo}, {Nolan}, {Norris}, {Nuss}, {Ohsugi}, {Okumura},
  {Omodei}, {Orlando}, {Paciesas}, {Pelassa}, {Pesce-Rollins}, {Pierbattista},
  {Piron}, {Porter}, {Racusin}, {Rain{\`o}}, {Razzano}, {Razzaque}, {Reimer},
  {Reimer}, { Reyes}, {Roth}, {Sadrozinski}, {Sgr{\`o}}, {Siskind}, {Smith},
  {Sonbas}, {Spandre}, {Spinelli}, {Stamatikos}, {Strickman}, {Takahashi},
  {Tanaka}, {Tanaka}, {Thayer}, {Thayer}, {Torres}, {Tosti}, {Troja}, {Uehara},
  {Usher}, {Vandenbroucke}, {Vasileiou}, {Vianello}, {Vilchez}, {Vitale}, {von
  Kienlin}, {Waite}, {Wang}, {Winer}, {Wood}, {Yamazaki}, {Yang}, {Ziegler},
  {Piro}, \& {Fermi Collaboration}}]{LATPaper}
{Atwood}, W.~B., {Abdo}, A.~A., {Ackermann}, M., {et~al.} 2009, \apj, 697, 1071

\bibitem[{{Barat} {et~al.}(1994){Barat}, {Trottet}, {Vilmer}, {Dezalay},
  {Talon}, {Sunyaev}, {Terekhov}, \& {Kuznetsov}}]{1994ApJ...425L.109B}
{Barat}, C., {Trottet}, G., {Vilmer}, N., {et~al.} 1994, \apjl, 425, L109

\bibitem[{Bruel {et~al.}(2018)Bruel, Burnett, Digel, Johannesson, Omodei, \&
  Wood}]{bruel2018fermilat}
Bruel, P., Burnett, T.~H., Digel, S.~W., {et~al.} 2018, Fermi-LAT improved
  Pass~8 event selection

\bibitem[{{Chupp}(1987)}]{Chupp_1987}
{Chupp}, E.~L. 1987, Physica Scripta Volume T, 18, 5

\bibitem[{{Chupp} \& {Ryan}(2009)}]{chup09}
{Chupp}, E.~L., \& {Ryan}, J.~M. 2009, Research in Astronomy and Astrophysics,
  9, 11

\bibitem[{{Chupp} {et~al.}(1982){Chupp}, {Forrest}, {Ryan}, {Heslin}, {Reppin},
  {Pinkau}, {Kanbach}, {Rieger}, \& {Share}}]{chup82}
{Chupp}, E.~L., {Forrest}, D.~J., {Ryan}, J.~M., {et~al.} 1982, \apjl, 263, L95

\bibitem[{{Cliver} {et~al.}(1993){Cliver}, {Kahler}, \& {Vestrand}}]{cliv93}
{Cliver}, E.~W., {Kahler}, S.~W., \& {Vestrand}, W.~T. 1993, in International
  Cosmic Ray Conference, Vol.~3, 23rd International Cosmic Ray Conference
  (ICRC23), Volume 3, 91

\bibitem[{De~Nolfo {et~al.}(2019)De~Nolfo, Bruno, Ryan, Dalla, Giacalone,
  Richardson, Christian, Stochaj, Bazilevskaya, Boezio, Martucci, Mikhailov, \&
  Munini}]{DeNolfo2019}
De~Nolfo, G., Bruno, A., Ryan, J., {et~al.} 2019, Astrophysical Journal, 879,
  cited By 7

\bibitem[{{Dennis}(1988)}]{1988SoPh..118...49D}
{Dennis}, B.~R. 1988, \solphys, 118, 49

\bibitem[{{Di Mauro} {et~al.}(2018){Di Mauro}, {Manconi}, {Zechlin}, {Ajello},
  {Charles}, \& {Donato}}]{2018ApJ...856..106D}
{Di Mauro}, M., {Manconi}, S., {Zechlin}, H.~S., {et~al.} 2018, \apj, 856, 106

\bibitem[{Dolan \& Fazio(1965)}]{doi:10.1029/RG003i002p00319}
Dolan, J.~F., \& Fazio, G.~G. 1965, Reviews of Geophysics, 3, 319

\bibitem[{{Forrest} {et~al.}(1986){Forrest}, {Vestrand}, {Chupp}, {Rieger}, \&
  {Cooper}}]{forr86}
{Forrest}, D.~J., {Vestrand}, W.~T., {Chupp}, E.~L., {Rieger}, E., \& {Cooper},
  J. 1986, Advances in Space Research, 6, 115

\bibitem[{{Forrest} {et~al.}(1985){Forrest}, {Vestrand}, {Chupp}, {Rieger},
  {Cooper}, \& {Share}}]{forr85}
{Forrest}, D.~J., {Vestrand}, W.~T., {Chupp}, E.~L., {et~al.} 1985, in
  International Cosmic Ray Conference, Vol.~4, 19th International Cosmic Ray
  Conference (ICRC19), Volume 4, 146

\bibitem[{Hurford {et~al.}(2006)Hurford, Krucker, Lin, Schwartz, Share, \&
  Smith}]{Hurford_2006}
Hurford, G.~J., Krucker, S., Lin, R.~P., {et~al.} 2006, \apj, 644, L93

\bibitem[{{Jin} {et~al.}(2018){Jin}, {Petrosian}, {Liu}, {Nitta}, {Omodei},
  {Rubio da Costa}, {Effenberger}, {Li}, {Pesce-Rollins}, {Allafort}, \&
  {Manchester}}]{Jin2018}
{Jin}, M., {Petrosian}, V., {Liu}, W., {et~al.} 2018, \apj, 867, 122

\bibitem[{{Kanbach} {et~al.}(1993){Kanbach}, {Bertsch}, {Fichtel}, {Hartman},
  {Hunter}, {Kniffen}, {Kwok}, {Lin}, {Mattox}, \&
  {Mayer-Hasselwander}}]{kanb93}
{Kanbach}, G., {Bertsch}, D.~L., {Fichtel}, C.~E., {et~al.} 1993, \aaps, 97,
  349

\bibitem[{Kocharov {et~al.}(2020)Kocharov, Pesce-Rollins, Laitinen, Mishev,
  K{\"u}hl, Klassen, Jin, Omodei, Longo, Webb, Cane, Heber, Vainio, \&
  Usoskin}]{Kocharov_2020}
Kocharov, L., Pesce-Rollins, M., Laitinen, T., {et~al.} 2020, \apj, 890, 13

\bibitem[{{Lin} \& {Rhessi Team}(2003)}]{LIN20031001}
{Lin}, R.~P., \& {Rhessi Team}. 2003, Advances in Space Research, 32, 1001

\bibitem[{Linden {et~al.}(2018)Linden, Zhou, Beacom, Peter, Ng, \&
  Tang}]{PhysRevLett.121.131103}
Linden, T., Zhou, B., Beacom, J.~F., {et~al.} 2018, Phys. Rev. Lett., 121,
  131103

\bibitem[{Mattox {et~al.}(1996)Mattox, Bertsch, Chiang, Dingus, Digel,
  Esposito, Fierro, Hartman, Hunter, Kanbach, Kniffen, Lin, Macomb,
  Mayer-Hasselwander, Michelson, von Montigny, Mukherjee, Nolan, Ramanamurthy,
  Schneid, Sreekumar, Thompson, \& Willis}]{Mattox:96}
Mattox, J.~R., Bertsch, D.~L., Chiang, J., {et~al.} 1996, \apj, 461, 396

\bibitem[{Mazziotta {et~al.}(2020)Mazziotta, Luque, Di~Venere, Fass\`o,
  Ferrari, Loparco, Sala, \& Serini}]{PhysRevD.101.083011}
Mazziotta, M.~N., Luque, P. D. L.~T., Di~Venere, L., {et~al.} 2020, Phys. Rev.
  D, 101, 083011

\bibitem[{{Meegan} {et~al.}(2009){Meegan}, {Lichti}, {Bhat}, {Bissaldi},
  {Briggs}, {Connaughton}, {Diehl}, {Fishman}, {Greiner}, {Hoover}, {van der
  Horst}, {von Kienlin}, {Kippen}, {Kouveliotou}, {McBreen}, {Paciesas},
  {Preece}, {Steinle}, {Wallace}, {Wilson}, \& {Wilson-Hodge}}]{meeg09}
{Meegan}, C., {Lichti}, G., {Bhat}, P.~N., {et~al.} 2009, \apj, 702, 791

\bibitem[{{Moskalenko} {et~al.}(2006){Moskalenko}, {Porter}, \&
  {Digel}}]{2006ApJ...652L..65M}
{Moskalenko}, I.~V., {Porter}, T.~A., \& {Digel}, S.~W. 2006, \apjl, 652, L65

\bibitem[{{Murphy} {et~al.}(1987){Murphy}, {Dermer}, \& {Ramaty}}]{murp87}
{Murphy}, R.~J., {Dermer}, C.~D., \& {Ramaty}, R. 1987, \apjs, 63, 721

\bibitem[{{Nitta} {et~al.}(2013){Nitta}, {Aschwanden}, {Boerner}, {Freeland},
  {Lemen}, \& {Wuelser}}]{2013SoPh..288..241N}
{Nitta}, N.~V., {Aschwanden}, M.~J., {Boerner}, P.~F., {et~al.} 2013, \solphys,
  288, 241

\bibitem[{{Omodei} {et~al.}(2018){Omodei}, {Pesce-Rollins}, {Longo},
  {Allafort}, \& {Krucker}}]{2018ApJ...865L...7O}
{Omodei}, N., {Pesce-Rollins}, M., {Longo}, F., {Allafort}, A., \& {Krucker},
  S. 2018, \apjl, 865, L7

\bibitem[{{Orlando} \& {Strong}(2007)}]{2007Ap&SS.309..359O}
{Orlando}, E., \& {Strong}, A.~W. 2007, \apss, 309, 359

\bibitem[{{Orlando} \& {Strong}(2008)}]{2008A&A...480..847O}
---. 2008, \aap, 480, 847

\bibitem[{{Pesce-Rollins} {et~al.}(2015){Pesce-Rollins}, {Omodei}, {Petrosian},
  {Liu}, {Rubio da Costa}, {Allafort}, \& {Chen}}]{2015ApJ...805L..15P}
{Pesce-Rollins}, M., {Omodei}, N., {Petrosian}, V., {et~al.} 2015, \apjl, 805,
  L15

\bibitem[{Poluianov {et~al.}(2017)Poluianov, Usoskin, Mishev, Shea, \&
  Smart}]{Poluianov2017}
Poluianov, S.~V., Usoskin, I.~G., Mishev, A.~L., Shea, M.~A., \& Smart, D.~F.
  2017, Solar Physics, 292, 176

\bibitem[{{Reames}(1995)}]{ream95}
{Reames}, D.~V. 1995, Advances in Space Research, 15, 41

\bibitem[{{Ryan}(2000)}]{ryan00}
{Ryan}, J.~M. 2000, \ssr, 93, 581

\bibitem[{{Ryan} \& {Lee}(1991)}]{1991ApJ...368..316R}
{Ryan}, J.~M., \& {Lee}, M.~A. 1991, \apj, 368, 316

\bibitem[{Sakurai(2008)}]{hinode}
Sakurai, T. 2008, The Hinode Mission (Springer)

\bibitem[{{Seckel} {et~al.}(1991){Seckel}, {Stanev}, \&
  {Gaisser}}]{1991ApJ...382..652S}
{Seckel}, D., {Stanev}, T., \& {Gaisser}, T.~K. 1991, \apj, 382, 652

\bibitem[{{Shih} {et~al.}(2009){Shih}, {Lin}, \& {Smith}}]{shih09}
{Shih}, A.~Y., {Lin}, R.~P., \& {Smith}, D.~M. 2009, \apjl, 698, L152

\bibitem[{Shrivastava \& Singh(2005)}]{Shrivastava_2005}
Shrivastava, P.~K., \& Singh, N. 2005, Chinese Journal of Astronomy and
  Astrophysics, 5, 198

\bibitem[{{Thompson} {et~al.}(1993){Thompson}, , {et~al.}}]{thompson93}
{Thompson}, D.~J., , {et~al.} 1993, \apjs, 86, 629

\bibitem[{{Trottet} {et~al.}(1998){Trottet}, {Vilmer}, {Barat}, {Benz},
  {Magun}, {Kuznetsov}, {Sunyaev}, \& {Terekhov}}]{1998A&A...334.1099T}
{Trottet}, G., {Vilmer}, N., {Barat}, C., {et~al.} 1998, \aap, 334, 1099

\bibitem[{{Vestrand} \& {Forrest}(1993)}]{1993ApJ...409L..69V}
{Vestrand}, W.~T., \& {Forrest}, D.~J. 1993, \apjl, 409, L69

\bibitem[{{Vianello} {et~al.}(2015){Vianello}, {Lauer}, {Younk}, {Tibaldo},
  {Burgess}, {Ayala}, {Harding}, {Hui}, {Omodei}, \&
  {Zhou}}]{2015arXiv150708343V}
{Vianello}, G., {Lauer}, R.~J., {Younk}, P., {et~al.} 2015, arXiv e-prints,
  arXiv:1507.08343

\bibitem[{Vilmer(1987)}]{Vilmer1987}
Vilmer, N. 1987, Hard X-ray Emission Processes in Solar Flares (Dordrecht:
  Springer Netherlands), 207--223

\bibitem[{{Vilmer} {et~al.}(2011){Vilmer}, {MacKinnon}, \&
  {Hurford}}]{2011SSRv..159..167V}
{Vilmer}, N., {MacKinnon}, A.~L., \& {Hurford}, G.~J. 2011, \ssr, 159, 167

\bibitem[{{Vilmer} {et~al.}(2003){Vilmer}, {MacKinnon}, {Trottet}, \&
  {Barat}}]{vilm03}
{Vilmer}, N., {MacKinnon}, A.~L., {Trottet}, G., \& {Barat}, C. 2003, \aap,
  412, 865

\bibitem[{{Vilmer} {et~al.}(1999){Vilmer}, {Trottet}, {Barat}, {Schwartz},
  {Enome}, {Kuznetsov}, {Sunyaev}, \& {Terekhov}}]{1999A&A...342..575V}
{Vilmer}, N., {Trottet}, G., {Barat}, C., {et~al.} 1999, \aap, 342, 575

\bibitem[{{Wood} {et~al.}(2017){Wood}, {Caputo}, {Charles}, {Di Mauro},
  {Magill}, {Perkins}, \& {Fermi-LAT Collaboration}}]{2017ICRC...35..824W}
{Wood}, M., {Caputo}, R., {Charles}, E., {et~al.} 2017, in International Cosmic
  Ray Conference, Vol. 301, 35th International Cosmic Ray Conference
  (ICRC2017), 824

\end{thebibliography}
